\documentclass{JHEP}
\usepackage{amsmath,amssymb,epsfig}

\newcommand{\be}{\begin{equation}}
\newcommand{\ee}{\end{equation}}
\newcommand{\beq}{\begin{equation}}
\newcommand{\eeq}{\end{equation}}
\newcommand{\ba}{\begin{eqnarray}}
\newcommand{\ea}{\end{eqnarray}}
\newcommand{\nn}{\nonumber}

\def\IC {\mathbb{C}}

\def\II {\mathbb{I}}

\def\IP {\mathbb{P}}

\def\IR {\mathbb{R}}

\def\IZ {\mathbb{Z}}

\def\IGa{\relax\hbox{${\rm I}\kern-.18em\Gamma$}}

\def\CC {\mathcal{C}}
\def\CD {\mathcal{D}}

\def\CL {\mathcal{L}}
\def\CM {\mathcal{M}}
\def\CN {\mathcal{N}}
\def\CO {\mathcal{O}}

\def\CR {\mathcal{R}}
\def\CS {\mathcal{W}}

\def\CX {\mathcal{X}}
\def\CY {\mathcal{Y}}

\def\half {\frac{1}{2}}

\def\ihalf {\frac{i}{2}}

\def\pa {\partial}

\def\rhalf {\frac{1}{\sqrt{2}}}
\def\Tr {\rm Tr}

\def\Vol {{\rm Vol}}
\def\we {{\wedge}}

\def\a {\alpha}
\def\b {\beta}
\def\d {\delta}
\def\e {\epsilon}
\def\g {\gamma}
\def\G {\Gamma}
\def\hG {\hat\Gamma}
\def\k {\kappa}
\def\l {\lambda}
\def\L {\Lambda}
\def\m {\mu}
\def\n {\nu}
\def\o {\omega}
\def\p {\phi}
\def\P {\Phi}
\def\s {\sigma}
\def\t {\theta}

\def\tw {\tilde w}
\def\tS {\tilde S}

\def\gs {g_s}
\def\gym {g_{YM}}

\def\ls {l_s}

\def\tym {\t_{YM}}
\def\vt#1#2#3 {{\vartheta[{#1 \atop  #2}](#3\vert \tau)}}

\title{Gauge/String Duality in Confining Theories}
\author{Jos\'e D. Edelstein \\
Departamento de F\'\i sica de Part\'\i culas, Universidade de
Santiago de Compostela and Instituto Galego de F\'\i sica de
Altas Enerx\'\i as (IGFAE), E-15782 Santiago de Compostela, Spain
\\ \vskip2mm
Instituto de F\'\i sica de La Plata (IFLP), Universidad Nacional de
La Plata, La Plata, Argentina
\\ \vskip2mm Centro de Estudios Cient\'\i ficos (CECS), Casilla
1469, Valdivia, Chile \\ \vskip3mm E-mail:
\email{jedels-at-usc-dot-es}}
\author{Rub\'en Portugues \\
Centro de Estudios Cient\'\i ficos (CECS), Casilla
1469, Valdivia, Chile \\ \vskip3mm E-mail: \email{rp-at-cecs-dot-cl}}

\abstract{This is the content of a set of lectures given at the ``XIII Jorge
Andr\'e Swieca Summer School on Particles and Fields'', Campos do
Jord\~ao, Brazil in January 2005. They intend to be a basic introduction to the topic
of gauge/gravity duality in confining theories. We start by reviewing some
key aspects of the low energy physics of non-Abelian gauge theories. Then,
we present the basics of the AdS/CFT correspondence and its extension both
to gauge theories in different spacetime dimensions with sixteen supercharges and to more realistic situations with less supersymmetry. We discuss the different options of interest: placing D--branes at singularities and
wrapping D--branes in calibrated cycles of special holonomy manifolds. We finally present an outline of a number of non-perturbative phenomena in
non-Abelian gauge theories as seen from supergravity.}

\preprint{CECS-PHY-05/14 \\ {\tt hep-th/0602021}}

\begin{document}

\noindent {\bf Note:}

\noindent This set of lectures intends to be a basic introduction to the topic
of gauge/gravity duality in confining theories. There are other sources for a
nice introduction to these topics \cite{berto,div,hko,ime}. We nevertheless attempted to keep our own perspective on the subject and hope that the final result could be seen as a valuable addition to the existing literature.

\section{Lecture I: Strongly Coupled Gauge Theories}

Quantum Chromodynamics (QCD) is the theory that governs the strong interaction of quarks and gluons. It is a non-Abelian quantum field theory with (color) gauge group $SU(3)$. The fundamental degrees of freedom are gluons
$A_\mu^{k\bar k}$ $\in F \otimes \bar F$ and quarks $q_\a^k$ $\in F$, where
$\a$ is a flavor index (whose values
are customarily taken to be up, down, strange, charm, bottom and
top) and $k, \bar k$ are color indices. The elementary fields are
evidently charged under $SU(3)$, whose {\it local} action on the
fields is
\be q_\a \to U ~q_\a ~, ~~~~~
A_\mu \to U A_\mu ~U^{-1}+ i ~(\pa_\mu U) ~U^{-1} ~.
\label{trlaws}
\ee
Mesons and baryons,
instead, which are made up of quarks and gluons, are singlets under
the color gauge group. The $SU(3)$ invariant Lagrangian is
\be
{\CL}_{QCD} = \frac{1}{4 g_{YM,0}^2} ~Tr\ F_{\m\n} F^{\m\n} +
\frac{\t_{YM,0}}{32\pi^2} ~Tr\ F_{\m\n} {}^*F^{\m\n}  +
\sum_{\a=1}^{N_f} \bar q_\a ( -i \g^\m D_\m + m_\a ) q_\a ~,
\label{qcdlag}
\ee
where $g_{YM,0}$ and $\t_{YM,0}$ are,
respectively, the bare coupling constant and $\t$--angle, $m_\a$ the
bare mass matrix of the quarks and $N_f$ the number of Weyl
fermions. If we set $m_\a$ to zero, QCD is also invariant under \be
U(N_f)_L \times U(N_f)_R = SU(N_f)_L \times SU(N_f)_R \times U(1)_V
\times U(1)_A ~, \label{symqcd} \ee at the classical level, where it
also seems conformal (there are no dimensionful parameters in the
Lagrangian). At the quantum level, however, instantons break the
axial symmetry $U(1)_A$ and a quark condensate emerges $<\bar q_{L
k} ~q_R^{~k}> \neq 0$ which breaks one of the $SU(N_f)$ factors. We
now review some aspects of QCD, including the striking feature that
its elementary constituents, the quarks, appear to be weakly coupled
at short distances and strongly coupled at long distances.

\subsection{Running Coupling Constant and Asymptotic Freedom}

Consider the one-loop effective action expanded around a classical QCD background. We will set in what follows $\t_{YM} = 0$. \footnote{Present
bounds on the value of the theta angle tell us that $\t_{YM} \leq 10^{-9}$.
It is actually an important open problem to understand why it is so small
or possibly null.} Its computation amounts to the calculus of some
functional determinants, as can be immediately seen in the path
integral formulation of the theory. After expansion, using the fact
that a classical configuration satisfies the Euler--Lagrange
equations, $S^{(1)}(\bar A_\mu, \bar q_\a) = 0$, linear terms vanish
(the bar over a quantity refers to its classical value). The
resulting integration over the fluctuating fields is Gaussian. This
implies, by means of the operatorial identity $\Tr \log \mathcal{O}
= \log \det \mathcal{O}$, that the result can be written in terms of
functional determinants
$$
S^{eff}_{1-loop} = \int d^4x ~\Bigg\{ \frac{1}{4 g_{YM,0}^2} Tr
\bar{F}_{\m\n} \bar{F}^{\m\n} + \half Tr \log \Delta_1 - \frac{N_f}{2}
Tr \log \Delta_{1/2}
\Bigg\} ~,
$$
where
$$
\left(\Delta_1\right)^{ab}_{\m\n} = -\left(\bar{D}^2\right)^{ab}
\delta_{\mu\nu} + 2 f^{ab}_{~~c} \bar{F}^c_{\mu\nu}\,,\quad\,,\quad
\Delta_{1/2} = i \g^\m \bar D_\m ~,
$$
$D_{\mu}$ is the covariant derivative and $f^{ab}_{~~c}$ are the
structure constants which appear in the gauge group algebra
$$ [T^a,T^b] = f^{ab}_{~~c} ~T^c ~. $$
Since determinants are gauge invariant functionals of the fields a
series expansion will begin with quadratic contributions in the
gauge field strength, which happens to be proportional to the terms
appearing in the Yang--Mills Lagrangian. This amounts to a quantum
mechanical correction to the gauge coupling. Arising from a one-loop
computation, we expect to get a logarithmic divergence \be S^{eff
~(quad)}_{1-loop} = \Bigg( \frac{1}{4 g_{YM,0}^2} - \frac{b_0}{32
\pi^2}\log \frac{\mu}{\L} \Bigg) \int d^4x ~Tr \bar{F}_{\m\n}
\bar{F}^{\m\n} ~, \label{actionq} \ee where $\mu$ is the energy
characterizing the variation of the background field and $\L$ a
regulator (the scale at which we impose renormalization), while
$b_0$ results to be the following number \be  b_0 = -\frac{11}{3}
c_v + \frac{4N_f}{3} c_f = - \left( 11 - \frac{2}{3} N_f \right) ~,
\label{betafunc} \ee where $c_v = N_c = 3$ and $c_f = 1/2$ are
Casimir values corresponding to the vector and fermion
representation of the gauge group,
$$ Tr_{\,r}\ ( T^a T^b ) = c_r ~\delta^{ab} ~. $$
We immediately extract from Eq.(\ref{actionq}) an important result:
{\it Quantum effects render $g_{YM}$ scale--dependent} \be
\frac{1}{g_{YM}^2(\mu)} = \frac{1}{g_{YM,0}^2} - \frac{b_0}{8\pi^2}
\log \frac{\mu}{\L} ~. \label{running} \ee When there are few
fermions ($N_f<17$), this results in a {\it negative}
$\beta$--function, \be \b\ (g_{YM}) = \frac{\partial
g_{YM}}{\partial\log\frac{\mu}{\L}} = - \frac{g_{YM}^3}{(4\pi)^2}
\left( 11 - \frac{2}{3} N_f \right) ~. \label{bef} \ee Strikingly,
if we take $\mu \to \infty$, {\it i.e.} at high energies, $g_{YM}
\to 0$. QCD is a {\it perturbative} theory at sufficiently {\it high
energies} or {\it short distances}. The theory is analytically
tractable in that regime. This is the remarkable phenomenon known as
{\it asymptotic freedom} \cite{gwp}. Notice that, as we reduce the
energy at which the theory is probed, the coupling increases. This
is shocking and in complete contrast to what happens, for example,
in Quantum Electrodynamics (QED).

\subsection{Dimensional Transmutation}

Let us now compute the renormalization of the gauge coupling assuming
$g_{YM}^2 << 1$ for external gluons of momentum $Q$. The Renormalization
Group (RG) equation can be read from the leading Feynman diagrams
(Fig.\ref{fig:vertices}):

\FIGURE[ph]{\centerline{\epsfig{file=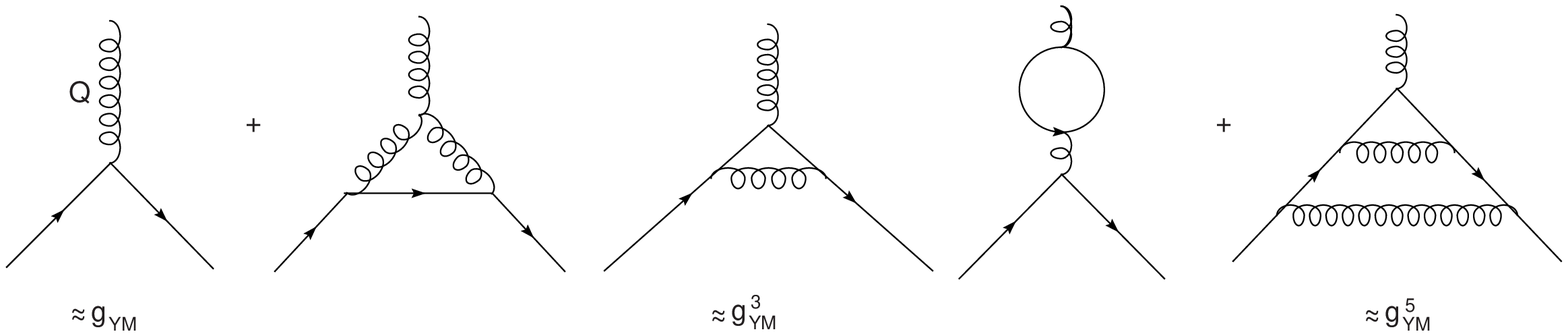,width=15cm}}
\caption{Leading, next-to-leading and next-to-next-to-leading perturbative contributions to the Yang--Mills coupling constant for an external gluon of
momentum $Q$.}
\label{fig:vertices}}

\be \frac{dg_{YM}}{dt} = \frac{b_0}{(4\pi)^2}
~g_{YM}^3 + \frac{2 b_1}{(4\pi)^4} ~g_{YM}^5 + ~\dots ~~~~~~~ t =
\log \frac{Q}{\L} ~, \label{rgeq} \ee
where $b_0$ is the value
computed above in (\ref{betafunc}) and $b_1$ is the number
$$ b_1 = - \left( 51 - \frac{19}{3} N_f \right) ~, $$
(negative for $N_f<9$) that will not be of great relevance to us.
The Renormalization Group equation (\ref{rgeq}), including the next
to leading term, can be rewritten as \be \frac{dg_{YM}^{-2}}{dt} = -
\frac{1}{8\pi^2} \left( b_0 + \frac{b_1}{8\pi^2} ~g_{YM}^2 + ~\dots
\right) ~, \label{step} \ee which, in the regime where $g_{YM}^2
\sim - \frac{8\pi^2}{b_0} t^{-1}$ is small, can be integrated with
the following result: \be \frac{1}{g_{YM}^2} + \frac{b_0}{8\pi^2}
\left( t - \frac{b_1}{b_0^2} \log t \right) + ~\dots = {\rm constant} ~.
\label{step2} \ee From this expression, it is straightforward to see
an extremely significant consequence: we can construct a
dimensionful quantity that survives the removal of the regulator:
\be Q ~\exp \left( \frac{8\pi^2}{b_0 ~g_{YM}^2(Q)} \right) \left(
\log \frac{Q}{\L} \right)^{- \frac{b_1}{b_0^2}} \longrightarrow
{\rm constant} ~\L ~, \label{dimtr} \ee when $t, Q \to \infty$. An {\it
energy scale} appears quantum mechanically in an otherwise massless
theory! This is the phenomenon known as {\it dimensional
transmutation} \cite{cw}. The dynamically generated scale is
customarily called ~$\L_{QCD}$. It is an unexpected built-in energy
scale of QCD, which is experimentally found to be $\L_{QCD} \approx
200 MeV$.

\subsection{QCD at Low Energy}

As we have seen earlier, at energies high enough (higher than $\L_{QCD}$),
the theory is perturbative and we can use Feynman diagrams to compute
particle scattering processes. When we decrease the energy, ~$g_{YM}$
increases and, at some point it should be of order one or larger. The
theory ceases to be well-defined perturbatively. This is in a sense welcome! Indeed, classical non-Abelian gauge theory is very different from the observed world. For QCD to successfully describe the strong force, it must have at
the quantum level properties which are dramatically different from the
behavior of the classical theory or perturbation theory built from it. This
is telling us that the theory is strongly coupled at low energies. In particular, the following phenomena have been observed in the laboratory:
\begin{itemize}
\item {\it Mass Gap}:~ there is a $\Delta > 0$ such that every excitation of
the vacuum has energy at least $\Delta$.
\item {\it Quark Confinement}:~ even though the theory is described in terms
of elementary fields, the physical particle states (baryons, mesons,
glueballs) are $SU(3)$ invariant. In a sense, the $SU(3)$ symmetry
{\it disappears} in the infrared (IR).
\item {\it Chiral Symmetry Breaking}:~ when the quark bare masses vanish the
vacuum is invariant only under a certain subgroup of the full symmetry group
acting on the quark fields. ~The $SU(N_f) \times SU(N_f)$ chiral flavor
symmetry of the QCD Lagrangian is broken, as explained above, in the vacuum
of the theory. ~Besides, let us remind that there is also a UV breaking of
a global $U(1)$ --the one that we called $U(1)_A$ in (\ref{symqcd})--,
$$ q_\a ~\to~ e^{i \varphi \Gamma_5} q_\a ~, $$
to a discrete subgroup, due to the contribution of instantons.
\end{itemize}
These are conjectured to be non-perturbative phenomena that govern
the low-energy physics of QCD. There are other aspects of this
nature that we will mention below. To show that this phenomenology
can be actually derived from the theory of quantum chromodynamics is
still an open problem. These phenomena take place in a regime of the
theory which is not tractable by analytic means. {\it How do we
study strongly coupled gauge theories?}.

\subsection{Strings in QCD}

It is natural to think that, as soon as the theory becomes strongly
coupled, the relevant degrees of freedom cease to be the elementary
point--like excitations of the original quantum fields. The
phenomenon of color confinement might be telling us that the
appropriate objects to be described are not the building blocks of
classical QCD, quarks and gluons, but, for example, the confining
strings formed by the collimated lines of chromoelectric flux. The
underlying idea is that at low energies magnetic monopoles might
become relevant, condense, and squeeze the otherwise spread
chromoelectric flux into vortices (in analogy with Abrikosov
vortices formed in ordinary superconductors). Thus, when studying
QCD at low energies, we should change the theory to, say, a
(perturbative) theory of strings describing the dynamics of these
collimated flux tubes. \footnote{Notice that, while these strings are open
having quarks attached at their ends, there are also closed strings in QCD
forming complicated loops. These are purely gluonic objects known as
glueballs.} What kind of strings are these? First of all,
they are thick strings. Their thickness should be of order
$1/\L_{QCD}$. This seems very different from the fundamental object
studied in string theory, which has zero thickness. Remarkably, it
is possible to argue that thickness might be a holographic
phenomenon due to the existence of extra dimensions where the
strings can propagate \cite{polsus}.

There is another coincidence between QCD and string theory: they have a
similar particle spectrum given by an infinite set of resonances, mesons
and hadrons, with masses on a Regge trajectory
\be
M_J^2 \sim M_0^2 + \frac{J}{\alpha^\prime} ~,
\label{regge}
\ee
where $J$ is the spin and $\alpha^\prime$ sets the length scale (in the string
picture, $1/\alpha^\prime$ is the string tension), that is, $\alpha^\prime
\sim \L_{QCD}^{-2}$.
 These were actually the reasons
behind the original ideas about strings in the context of the theory of
strong interactions.

Once we accept that a non--Abelian gauge theory as QCD admits a
description in terms of closed strings --that is, quantum gravity--
in a higher dimensional background, another important issue comes to
mind. It was shown by Bekenstein \cite{jb} and Hawking \cite{sh}
that the entropy of a black hole is proportional to the area of its
event horizon. If gravity behaves as a local field theory we would
expect, from elementary notions in statistical mechanics, the
entropy of a gravitational system to be proportional to its volume.
Somehow, this suggests that quantum gravity is related, to local
field theories in lower dimensions in a holographic sense \cite{tho}.

Gauge theory and string theory are related in a deep way which we
are starting to understand. As a final piece of evidence, we shall
recall that string theory has some solitonic objects called
D--branes, on whose worldvolume a particular class of gauge theories
live \cite{pol} which, on the other hand, are sources for the
gravitational field. We will come back to this point later on.
Before, let us see an interesting framework where the relation
between gauge theory and string theory can be established on a solid
footing.

\subsection{The Large $N_c$ limit of Gauge Theory}

A remarkable proposal was made thirty years ago by 't Hooft
\cite{thooft}. Take the number of colors, $N_c$, from three to
infinity, ~$N_c \to \infty$ and expand in powers of $1/N_c$. The
so--called 't Hooft coupling, \be \l \equiv g_{YM}^2 N_c ~,
\label{thcoupl} \ee has to be held fixed in the limiting procedure.
This framework provides a feasible perturbative method in terms of
the 't Hooft coupling. It is worth noticing that lattice numerical
studies indicate that large $N_c$ is a reasonably good approximation
to $N_c = 3$ \cite{teper}.

Using double line notation for gluons $A_\mu^{k\bar k}$ and quarks $q_\a^k$,
where each line represents a gauge group index of the field, Feynman diagrams
become ribbon graphs. Thus, Feynman diagrams can be drawn on closed Riemann
surfaces ~$\Sigma_g$. Vertices, propagators and loops contribute, respectively,
with factors of $g_{YM}^{-2}$, $g_{YM}^2$ and $N_c$ to the diagrams. Then,
a Feynman diagram including $V$ vertices, $P$ propagators and $L$ loops,
contributes as $N_c^\chi$, ~$\chi = V - P + L = 2-2g = Eul(\Sigma_g)$. In the
large $N_c$ limit, a given amplitude $\mathcal{A}$ can be written as a sum
over topologies
\be
\mathcal{A} = \sum_{g=0}^\infty c_g(\l) ~N_c^{2-2g} ~.
\label{sumovert}
\ee
When $N_c \to \infty$, with $\l$ fixed, {\it planar} Feynman diagrams
dominate (see Fig.\ref{fig:1}) and we have two well distinct regimes:
\begin{itemize}
\item If $\l << 1$, also $g_{YM} << 1$, ~thus, perturbative gauge theory
applies and we can use Feynman diagrams.
\vskip1mm
\item If $\l >> 1$, ~$\mathcal{A}$ can be rearranged in a form that reminds
us of a familiar expression in string theory,
\be
\mathcal{A} = \sum_{g=0}^\infty g_s^{2g - 2} \mathcal{A}_g (\l) ~,
\label{sumovtt}
\ee
$\mathcal{A}_g (\l)$ being a perturbative closed string amplitude on
$\Sigma_g$, ~where $g_s \equiv 1/N_c << 1$ is the string coupling, and
$\l = g_s N_c$ is a {\em modulus} of the target space.
\end{itemize}

\FIGURE[ph]{\centerline{\epsfig{file=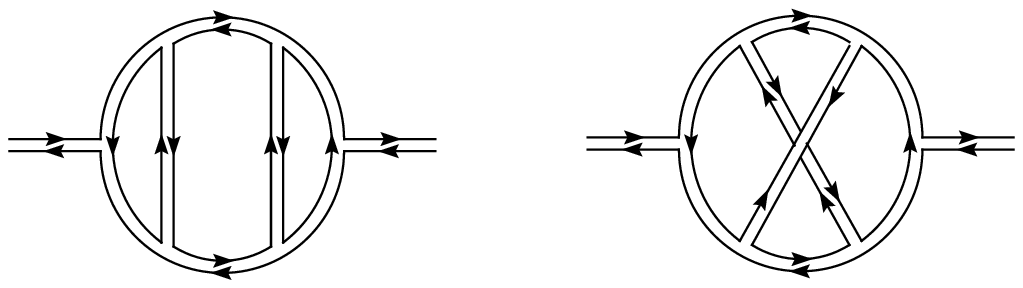,width=10cm}}
\caption{A plot showing two diagrams that contribute to quantum corrections
of a meson mass to the same order in the coupling constant. The diagram on
the left has 3 loops and will be proportional to $N_c^3$, whereas the single
loop diagram on the right just contributes a factor of $N_c$.}
\label{fig:1}}

\noindent
Two comments are in order at this point. First, if we add quarks in
the fundamental representation of the gauge group the resulting
diagrams are drawn on Riemann surfaces with {\it boundaries}. This
amounts to the introduction of open strings --that can be thought of
as corresponding to the flux lines that bind a quark and an
antiquark to form a meson-- into the computation of a given
amplitude. Second, it should be noticed that Feynman diagrams would
span {\it discretized} Riemann surfaces that do not seem to
correspond to a continuous world-sheet as that of string theory.
Only very large Feynman graphs may provide a good approximation to
continuous world sheets --notice that the string theory description
is valid at strong coupling, where it is reasonable to assume that
large Feynman graphs indeed dominate--. The idea is that a full
non-perturbative description of the gauge theory will fill the
holes. The results that will be discussed in these lectures point
encouragingly in this direction.

In summary, strongly coupled large $N_c$ gauge theories are dual to
(better described in terms of) perturbative closed string theories.
Again, we should ask: {\it what kind of string theories are those
appearing here?} Early examples of large $N_c$ duality, involving
bosonic strings on various backgrounds that are dual to zero
dimensional gauge theories (matrix models), were constructed some
ten years ago \cite{kdv}. In the last few years our understanding of
the relation between gauge theory and string theory has been
dramatically enhanced after it was realized that the former can be
made to live on multi-dimensional solitonic extended objects
appearing in the latter: the D--branes we mentioned earlier
\cite{pol}. These gauge theories are typically supersymmetric. Thus,
in order to be able to deal with these examples, let us introduce an
extra ingredient: supersymmetry.

\subsection{Supersymmetry and Gauge Theory}

Supersymmetry relates fermions and bosons, matter and interactions.
It has been proposed as a symmetry beyond the Standard Model mostly
because gauge couplings unify under this principle. Indeed, if we
consider a minimal supersymmetric generalization of the Standard
Model where all superpartners of the known elementary particles have
masses above the effective supersymmetry scale $M_{SUSY} \approx 1\,
TeV$, then unification is achieved at the so-called GUT scale.
Besides, supersymmetry provides a mechanism for understanding the
large hierarchy in scale between the masses of the particles at the
electroweak scale $M_{EW} \approx 100 \,GeV$ and the GUT scale
$M_{GUT} \approx 10^{16} \, GeV$. To preserve a small number such as
$M_{EW}/M_{GUT} \approx 10^{-14}$ requires a fine tuning that would
hardly survive quantum corrections. Due to the magic of
supersymmetry, dangerous loop corrections to masses cancel to all
orders in perturbation theory. In short, if unification is assumed,
then it seems clear that there has to be new physics between the
electroweak scale and the Planck scale and supersymmetry is a
promising candidate to fill the gap.

On the other hand, from a purely theoretical point of view,
supersymmetry is also extremely interesting because it allows a
deeper look into the {\it study of gauge dynamics}. Supersymmetry
severely constrains the theory and this leads to drastic
simplifications that make it possible to deal with some of its
non-perturbative aspects in many cases in an exact way. It is a
symmetry that implies the appearance of robust mathematical
structures on top of the gauge theory such as special geometry,
holomorphicity and integrability. Furthermore, some supersymmetric
systems belong to the same {\it universality class} as
non-supersymmetric ones, thus opening a window to study {\it
non-perturbative phenomena of gauge theories} in closely related
systems.

A four--dimensional Lorentz--invariant theory can have different amounts
of supersymmetry. We shall briefly recall in what follows the most
interesting cases. Those aspects which are specially relevant to the
content of these lectures will be discussed in greater detail later.
\begin{itemize}
\item $\mathcal{N}=8$ ~Supergravity: This is the maximal amount of
supersymmetry that an interacting theory in four dimensions can possess.
It is related by Kaluza--Klein reduction to the 11d Cremmer--Julia--Scherk
theory \cite{cjs}. It will be the subject of Dan Freedman's lectures at
this School.
\vskip1mm
\item $\mathcal{N}=4$ ~Supersymmetric Yang--Mills Theory: It is a unique,
exact and finite theory (no quantum corrections). It will be briefly discussed
in the next lecture and will be considered in much greater detail by Carlos
N\'u\~nez in his course. This is the gauge theory where Maldacena's
correspondence \cite{malda} applies in its full glory.
\vskip1mm
\item $\mathcal{N}=2$ ~Supersymmetric Yang--Mills Theory: This is a theory
whose low-energy dynamics has been exactly solved through the so-called
Seiberg--Witten solution \cite{sw}. This solution relies heavily on special
geometry and has been shown to be part of a broader framework given by the
Whitham hierarchy of a given integrable system (in the absence of matter,
for example, it is the periodic Toda lattice) \cite{GMMM,emm}.
\vskip1mm
\item $\mathcal{N}=1$ ~Supersymmetric Gauge Theory: ~This theory is the
closest relative of QCD: they share a number of key features that
can be more easily explored in the former. Holomorphic quantities
are protected by means of non-renormalization theorems \cite{sei}.
This leads to a variety of exact results that we would like to keep
in mind for later reference:

\textbf{1. } {\it Chiral symmetry breaking}: The $U(1)$
chiral symmetry is broken by instantons to a discrete subgroup, in
this case $\IZ_{2N_c}$. This is a UV effect. Moreover, an IR
phenomenon, gaugino condensation, further breaks the residual chiral
symmetry $\IZ_{2N_c}$ to $\IZ_2$.

\textbf{2. } {\it Gluino condensation}: there is a composite chiral field
whose lowest component is $S = \Tr ~\l\l$, $\l$ being the supersymmetric
partner of the gluon,
that condenses due to the existence of an IR effective superpotential
$W_{eff}(S)$. This leads to $N_c$ isolated and degenerate vacua labeled
by the value of the condensate in each, $\langle S \rangle\vert_k =
e^{2\pi i \frac{k}{N_c}} N_c~ \Lambda^3_{QCD}$, ~$k = 0, \cdots,
N_c-1$. The coefficient $N_c$ is exactly calculable by instanton
methods \cite{HKLM}.

\textbf{3. } {\it $\beta$--function}: The (so-called NSVZ)
$\b$--function can be computed to all orders in perturbation theory
resulting in a remarkably simple expression \cite{nsvz}: \be
\b_{NSVZ}(g_{YM}) = - \frac{3N_c g_{YM}^3}{16\pi^2} \left( 1 -
\frac{N_c g_{YM}^2}{8\pi^2} \right) ~. \label{nsvzbeta} \ee

\textbf{4. } {\it Confinement and screening of magnetic monopoles}: A linear
quark-antiquark potential appears in the IR. A nice mechanism that is usually
advocated to implement quark confinement is the vortex formation in an Abelian Higgs model where the scalar field comes from a magnetic monopole chiral field
that condenses. Indeed, the monopole-antimonopole potential can be computed
and it can be checked that, instead of being confined, monopoles are screened.

\textbf{5. } {\it Confining strings}: Confinement occurs when
chromoelectric flux cannot spread out in space over regions larger
than $\Lambda_{QCD}^{-1}$ in radius and hence forms flux tubes. The
tension of these confining strings depends on the {\it n-ality} of
the representation under $SU(N_c)$ of the sources for the flux at
either end of the tube.  This tension has been computed in several
supersymmetric approximations to QCD as well as in the lattice.

\textbf{6. } {\it Domain walls}: Due to the appearance of $N_c$ isolated
and degenerate vacua, BPS domain walls (invariant under some supersymmetry
transformations) should exist interpolating between any two such vacua.
Flat domain walls preserve $\frac{1}{2}$ of the supersymmetries \cite{DvSh}.
Being BPS objects, one can compute a quantum mechanically exact formula
for the tension of the wall separating the $n$-th and the $n+k$-th vacua,
\be
T_k = \frac{N_c^2 \Lambda_{QCD}^3}{4\pi^2} ~\sin \frac{k\pi}{N_c} ~.
\label{dwtension}
\ee
Since $T_k + T_{\tilde k} > T_{k + \tilde k}$, it is clear that these
domain walls tend to form a bound state. The gluing particles that carry the
attractive interactions are the previously alluded glueballs.

\textbf{7. } {\it Wilson loops}: Wilson loops can be used to calculate the
quark anti-quark potential and determine whether a theory is confining or
not. They are given by the following expression:
\be
W[\CC] = \frac{1}{N_c} \Tr ~P\left[ e^{i \oint A}\right] ~,
\label{wilson1}
\ee
where $\CC$ is some contour. If we take the contour $\CC$ to be a rectangle
(the quarks being a distance $L$ apart and the rectangle extending for a time
$T$) we can read off the quark anti-quark potential $E(L)$ in the infinite
strip limit, $T \to \infty$, via
\be
<W[\CC]> = A(L)\, e^{- T E(L)} ~.
\label{wilson2}
\ee
It will turn out that the calculation of the Wilson loop is simple in the
context of the gauge/gravity duality, and a formula will be derived that determines whether a dual supergravity background is confining or not in
terms of just two components of the metric.

\textbf{8. } {\it Instantons}: As mentioned above, instantons are responsible
for the symmetry breaking $U(1)_R \to \IZ_{2N_c}$ in $\CN = 1$ supersymmetric
Yang--Mills theory. They are characterized by a value of the Euclidean action
\be S_{\rm inst} = \frac{8\pi^2}{g_{YM}^2(\mu)} + i \tym ~. \ee

\textbf{9. } {\it Finite temperature effects}: Interesting phenomena
arise when the system has nonzero temperature. Thermodynamical
properties of $\CN = 1$ supersymmetric Yang--Mills theory include an
expected hydrodynamical behaviour near equilibrium. In this context,
transport properties such as the shear viscosity, $\eta$, can be
studied. Of particular interest is the ratio of $\eta$ and the
entropy density $s$ of the associated plasma. In several supersymmetric
gauge theories, this ratio has a universal value of $\frac{1}{4\pi}$.

\textbf{10. } {\it Adding flavors}: The theory of QCD has quarks. In this
sense, the addition of dynamical quarks into our analysis is clearly relevant.
In particular, the introduction of flavors leads to composite states as
mesons and hadrons, whose spectrum is an important observable.

This list does not exhaust at all the plethora of non-perturbative phenomena
of interest. For example, the mass spectrum of composite objects entirely
built out of gluons --known as {\it glueballs}-- is an interesting observable
in the physics of confinement.

\item $\mathcal{N}=0$ ~Gauge Theory: ~Where we started from. Where we should
eventually come back at low energies. Non-perturbative physics is hard to
tackle analytically. It is, however, possible to study non-supersymmetric
gauge theories by considering supersymmetric systems in higher dimensions and breaking supersymmetry upon compactification. An interesting case is given
by a five dimensional theory compactified on a circle with antiperiodic
boundary conditions for the fermionic fields \cite{wittent}.

\item It is also possible to turn on a marginal deformation in the $\CN = 4$
theory that renders it less supersymmetric in the IR. In particular, this provides a window to study non-supersymmetric theories of
a particular kind known as defect conformal field theories \cite{BGH,FNSS}. There are moreover the so--called $\mathcal{N}=2^*$ or $\mathcal{N}=1^*$,
which are softly broken $\mathcal{N}=4$ theories. We mention them for completeness even though we will not say anything else about these theories. Interested readers should take a look, for example, at \cite{polstr}.
\end{itemize}

\subsection{The Large $N_c$ limit of Supersymmetric Gauge Theory}

The large $N_c$ string duals of many of these theories have been
discovered in the last few years. These lectures will be devoted to
dig deeper in this subject but let us end this section by giving a
brief outline of what is coming. The best understood case is, by
far, $\mathcal{N} = 4$ super Yang--Mills theory in four dimensions,
whose dual is type IIB superstring theory on $AdS_5 \times S^5$
\cite{malda,gkp,witten1}. The (conformal) gauge theory is realized
on the world--volume of flat D3--branes. It is also worth mentioning
that there is an interesting limit of this system whose study was
extremely fruitful. From the gauge theory point of view, it consists
in focusing on a sector of the theory given by states with large
angular momentum, while on the string theory side it amounts to
taking the Penrose limit \cite{penrose,BFOFHP} which leads to type
IIB superstring theory
on a so called pp-wave background \cite{bmn}. This conjectured
duality was also extended to theories with sixteen supercharges that
correspond to the low--energy dynamics of flat Dp--branes, $p \neq
3$ \cite{imsy}. These are, in general, non--conformal, and the
gravity/gauge theory correspondence provides a powerful tool to
study the phase structure of the resulting RG flows \cite{imsy}. We
will discuss some aspects of these duals in the upcoming lecture.

After Maldacena's proposal, an impressive amount of work was
undertaken that allowed to understand how to extend these ideas to a
variety of systems with the aim of eventually arriving some day to
QCD itself. In particular, analogue results were obtained in the
context of topological strings (these are given by topologically
twisted sigma models \cite{vonk}; we will explain more on the
topological twist in lecture III). Let us state some of these
results with the hope that they will become clear by the end of this
set of lectures (a prudent person might prefer to avoid the
following two paragraphs in a first reading of the manuscript). The
so-called A-model topological string in the background of the
resolved conifold, for example, is dual to 3d Chern--Simons gauge
theory on $S^3$ \cite{gv,ov,lm}. There is also a mirrored (that is,
mapped under mirror symmetry) version of this: the B--model
topological string on local Calabi--Yau threefolds being dual to a
holomorphic matrix model \cite{dijv}. Now, topological string
amplitudes are known to compute $F$--terms in related $\CN=1$
gauge/string theories \cite{ov,bcov}. This allowed Vafa \cite{vafa}
to conjecture that $\CN = 1$ supersymmetric Yang--Mills theory in
four dimensions must be dual to either type IIA superstrings on
$\CO(-1) + \CO(-1) \to \IP^1$ with RR fluxes through the exceptional
$\IP^1$ or, passing through the looking glass, to type IIB
superstrings on the deformed conifold with RR fluxes threading the
blown--up $S^3$. The type IIA scenario can be uplifted to M--theory
on a manifold with $G_2$ holonomy, where it was found that both
sides of the duality are smoothly connected through a geometric
transition called the {\it flop} \cite{amv}. In type IIB, instead,
an interesting setup was developed by Maldacena and N\'u\~nez
\cite{mn2} by studying D5--branes wrapped on a supersymmetric
2--cycle in a local Calabi--Yau manifold. This opened the route to
the study of a huge family of theories that can be obtained from
Dp--branes wrapped on supersymmetric cycles of special holonomy
manifolds. We will attempt to clarify some of these claims in the
following lectures.

Finally, another kind of theories where Maldacena's conjecture was
successfully applied are constructed on D3--branes placed at the
apex of Calabi--Yau cones \cite{kw,ms} as well as by considering
orbifolds of type II theories. We will discuss in some detail the
former case (see lecture IV), whose interest was recently revived
after the discovery of an infinite family of metrics corresponding
to Calabi--Yau cones on Sasaki--Einstein manifolds labeled by a set
of coprime integers $Y^{p,q}$ \cite{ms}, as well as $L^{a,b,c}$.
These theories are superconformal in four dimensions. We will also
introduce {\it fractional
branes} which are responsible for destroying conformal invariance,
and discuss several interesting non-perturbative physical phenomena
arising in these scenarios. Most notably, the appearance of duality
cascades in quiver supersymmetric gauge theories (see \cite{strass}
for a recent review).
\newpage

\section{Lecture II: AdS/CFT Correspondence}

The AdS/CFT correspondence establishes that there is a complete
equivalence between $\mathcal{N}= 4$ super Yang--Mills theory and
type IIB string theory on $AdS_5 \times S^5$
\cite{malda,gkp,witten1}. There is by now an impressive amount of
evidence supporting this assertion (see \cite{agmoo} for a review).
A quick attempt to motivate the subject would start by considering a
stack of $N_c$ flat parallel D3--branes. This configuration
preserves sixteen supercharges. The spectrum of massless open string
states whose endpoints are located on the D3--branes is, at low
energies, that of $\mathcal{N}= 4$ $U(N_c)$ super Yang--Mills theory
in four dimensions, whose main features will be shortly addressed
later on this lecture. On the other hand, from the closed string
theory point of view, these D3--branes source some fields that
satisfy the field equations of type IIB supergravity \footnote{We
collect some useful formulas of type IIB supergravity in Appendix
B.}. The solution reads, in the string frame,
\be
ds^2 = H_3^{-\half} dx_{1,3}^2 + H_3^{\half} (dr^2 + r^2 d\Omega_{5}^2) ~,
\label{d3branesol1}
\ee
\be
e^{2\phi} = e^{2\phi_{\infty}} = const ~,
\label{d3branesol2}
\ee
\be
F_{[5]} = (1 + \star) ~dH_3^{-1} \wedge dx^0 \wedge dx^1 \wedge dx^2 \wedge
dx^3  ~,
\label{d3branesol3}
\ee
where $d\Omega_5^2$ is the round metric on $S^5$, $\star$ stands for the
Hodge dual, and $H_3$ is a harmonic function of the transverse coordinates
\be
H_3 = 1 + 4\pi g_s N_c \frac{{\alpha^\prime}^2}{r^4} ~.
\label{d3armf}
\ee
The idea now is to realize that it is possible to decouple the open and
closed string massive modes by taking the limit $\alpha^\prime \to 0$. As
the Planck length is given by $l^2_P = g_s^{1/2} \alpha^\prime$ and
$g_s$ is constant, we see that this limit also decouples the
open/closed interactions, $l_P \to 0$. An important point is that
the Yang--Mills coupling, $g^2_{YM}$, remains finite, \be g^2_{YM} =
4\pi g_s ~, \label{d3ymcoup} \ee in the low energy effective action.
The limit $\alpha^\prime \to 0$ leads, in the string theory side, to
type IIB supergravity. The right limiting procedure also involves a
{\it near-horizon limit}, $r \to 0$, such that \be U \equiv
\frac{r}{\alpha^\prime} = \mbox{fixed} ~, ~~~~~ r, \alpha^\prime \to
0 ~, \label{d3nhlimit} \ee where we would like the reader to notice
that $U$ has dimensions of energy. Performing such a limit in the
supergravity solution (\ref{d3branesol1}), we obtain
\be
ds^2 = {\alpha^\prime}\; \Bigg[ \frac{{\alpha^\prime}\, U^2}{L^2}
~dx_{1,3}^2 + \frac{L^2}{{\alpha^\prime}\, U^2} ~\Bigg( dU^2 + U^2\,
d\Omega_{5}^2 \Bigg) \Bigg] ~,
\label{nhd3branesol} \ee where we have introduced
the characteristic scale parameter $L$, \be L^4 = 4\pi g_s N_c
{\alpha^\prime}^2 ~. \label{quR} \ee
It is not difficult to check that the metric above
corresponds to the direct product of $AdS_5 \times S^5$ with both spaces
having the same radius, $L$, and opposite constant curvature, $\mathcal{R}
\sim \pm\, L^{-2}$. The dilaton is also constant. If we take the limit $g_s \to
0$, $N_c \to \infty$ with $g_s N_c$ constant and large enough, we
see that $L$ is also large and the type IIB supergravity description is perfectly valid
for any value of $U$. Let us finally notice that the curvature and
the 't Hooft coupling are inversely proportional, $\alpha^\prime \CR
\sim \l^{-1/2}$. This means that {\it the gauge theory description
and the string/gravity one are complementary and do not overlap}.
The AdS/CFT correspondence is an example of a strong/weak coupling
duality. That is, the system is well described by $\mathcal{N}= 4$
$U(N_c)$ super Yang--Mills theory for small values of the 't Hooft
coupling while it is better described by type IIB string/gravity
theory whenever $\lambda$ gets large. In spite of the fact that this
issue makes it extremely hard to prove or disprove the AdS/CFT
correspondence, there are several quantities that are known to be
independent of the coupling and whose computation using both
descriptions coincides.

There are arguments that will not be discussed in these lectures
pointing towards the fact that the Maldacena conjecture might be
stronger than what we have presented here. We might state three
different versions of the conjecture which, in increasing level of
conservatism, can be named as the strong, the mild and the weak.
They are described in Table 1, where the appropriate
limits that are assumed on the parameters of the theory are also
mentioned.

\TABLE[ph]{\centerline {\small \begin{tabular}{@{}ccc@{}}
\hline
{} &{} &{} \\[-1.5ex]
Gravity side & Gauge Theory side & \\[1ex]
\hline
{} &{} &{} \\[-1.5ex]
Type IIB string theory & $\mathcal{N}=4$ $SU(N_c)$ super & \\[1ex]
on $AdS_5 \times S^5$ & Yang--Mills theory & Strong \\[1ex]
$\forall g_s, ~L^2/\alpha^\prime$ & $\forall g_{YM}, ~N_c$ & \\[1ex]
\hline
{} &{} &{} \\[-1.5ex]
Classical type IIB strings & 't Hooft large $N_c$ limit of & \\[1ex]
on $AdS_5 \times S^5$ & $\mathcal{N}=4$ $SU(N_c)$ SYM & Mild \\[1ex]
$g_s \to 0, ~L^2/\alpha^\prime$ fixed & $N_c\to\infty\,, ~\l=\gym^2
N_c$
fixed & \\[1ex]
\hline
{} &{} &{} \\[-1.5ex]
Classical type IIB supergravity & Large 't Hooft coupling limit & \\[1ex]
on $AdS_5 \times S^5$ & of $\mathcal{N}=4$ $SU(N_c)$ SYM & Weak \\[1ex]
$g_s \to 0, ~L^2/\alpha^\prime \to\infty\,$ & $N_c\to\infty\,,
~\l\to\infty\,$ & \\[1ex]
\hline
\end{tabular}
\caption{Three versions of the Maldacena conjecture: the strong,
the mild and the weak.}
\label{maldaconj}}}

Let us briefly discuss some basic aspects of the conjecture that
will be important in the rest of the lectures. We start by presenting the
basics of $\mathcal{N}= 4$ supersymmetric Yang--Mills theory.

\subsection{$\CN =4$ Supersymmetric Yang--Mills Theory}

Four dimensional $\CN =4$ supersymmetric theory is special in that
it has a unique superfield in a given representation of the gauge
group. Let us take it to be the adjoint representation. The field
content of the supermultiplet consists of a vector field $A_\mu$,
six scalars $\phi^I$ $I = 1, ... ,6$, and four fermions
$\chi_{\alpha}^i$, $\chi_{\dot\alpha}^{\bar i}$, where $\alpha$ and
$\dot \alpha$ are chiral indices while $i =1,2,3,4$ and $\bar i$ are
in the ${\bf 4}$ and ${\bf\bar 4}$ of the $R$-symmetry group $SU(4)
= SO(6)$. It is a global internal symmetry of $\CN=4$ supersymmetric
Yang--Mills theory. The supercharges also belong to these
representations of $SO(6)$. Notice that it acts as a chiral
symmetry. The Lagrangian contains two parameters, the coupling
constant $g_{YM}$ and the theta angle $\theta_{YM}$, \ba \CL_{\CN=4}
& = & Tr \left[ \frac{1}{g_{YM}^2} \bigg( F \wedge \star F +
(D\phi)^2 + \sum_{IJ} [\phi^I,\phi^J]^2 \bigg) + \theta_{YM} ~F
\wedge F \right] \cr && ~~~~~ + \frac{1}{g_{YM}^2} Tr \left[
\bar\chi \not\!\! D \chi + \bar\chi \Gamma^I \phi^I \chi \right] ~.
\label{actionn4} \ea The theory is scale invariant quantum
mechanically. Indeed, just by counting the degrees of freedom of the
supermultiplet, it is immediate to see that the $\beta$--function is
zero to all orders. Thus, it is conformal invariant. This gives
raise to sixteen conformal supercharges (in any conformal theory,
the supersymmetries are doubled). The theory is maximally
supersymmetric. It displays a (strong/weak) S--duality under which
the complexified coupling constant \be \tau_{YM} =
\frac{\theta_{YM}}{2\pi} + \frac{4\pi i}{g^2_{YM}} ~,
\label{complcoup} \ee transforms into $-1/\tau_{YM}$. This combines
with shifts in $\theta_{YM}$ to complete the group $SL(2,\IZ)$. As
we saw in (\ref{d3ymcoup}), there is a relation between the
Yang--Mills coupling, on the gauge theory side, and the string
coupling. This relation has to be supplemented by another one that
links the $\theta$--angle with the vacuum expectation value of the
RR scalar $\chi$, \be \theta_{YM} = 2\pi \chi ~, \label{d3thchi} \ee
such that the $SL(2,\IZ)$ symmetry is clearly connected with the
usual S--duality symmetry in type IIB string theory.

\subsection{Symmetries and Isometries}

The $AdS_5$ factor can be thought of, after a simple change of variables to
adimensional coordinates $U = \frac{L}{\alpha^\prime}\, e^r$ and $dx_{1,3}^2 \to L^2\, dx_{1,3}^2$, as a warped codimension one 4d Minkowski
space --this actually guarantees the Poincar\'e invariance of the
gauge theory, as reflected in the appearance of a Minkowskian 4d
factor--, whose line element reads \be ds^2 = L^2 ( dr^2 + e^{2r}
dx_{1,3}^2 ) ~. \label{adsmet} \ee The limit where the radial
coordinate goes to infinity, $r \to \infty$, and the exponential
factor blows up is called the boundary of $AdS_5$. String
theory excitations extend all the way to the boundary where the dual
gauge theory is interpreted to live \cite{dbort}. When this space
is written as a hypersurface embedded in $\IR^{2,4}$, \be - w_1^2 -
w_2^2 + x_1^2 + x_2^2 + x_3^2 + x_4^2 = - L^2 ~; \label{adsemb} \ee
it is manifest that it possesses an $SO(2,4)$ isometry group. The
remaining $S^5$ factor of the background provides an extra $SO(6)$
isometry. Now, it is not a coincidence that $SO(2,4)$ is the
conformal group in 4d, while $SO(6) \approx SU(4)$ is exactly the
$R$--symmetry group of $\CN = 4$ supersymmetric Yang--Mills theory.
The bosonic global symmetries match perfectly. There are also
fermionic symmetries which, together with the bosonic ones, form the
supergroup $PSU(2,2|4)$. It is possible to verify that both the
massless fields in string theory and the BPS operators of
supersymmetric Yang--Mills theory, whose precise matching is not
discussed here, are classified in multiplets of this supergroup
\cite{witten1}.

\subsection{Correlators}

Consider a free massive scalar field propagating in $AdS_5$. Its
equation of motion is the Klein--Gordon equation, \be (\Box + m^2)
~\phi = 0 ~, \label{freesc} \ee which has two linearly independent
solutions that behave asymptotically as exponentials \be \phi
\approx e^{(\Delta - 4) r} ~, ~~~ {\rm and}~~~~~ \phi \approx e^{-
\Delta r} ~, \label{asymsc} \ee with $\Delta = 2 + \sqrt{4 + m^2
L^2}$. The first solution dominates near the boundary of the space
while the second one is suppressed. Now consider a solution of the
supergravity equations of motion whose asymptotic behavior is \be
\Phi_i(r,x^\mu) \sim \varphi_i(x^\mu) ~e^{(\Delta_i - 4) r} ~,
\label{asyads} \ee where the $x^\mu$ are gauge theory coordinates
living at the boundary. An operator $\mathcal{O}_i$ in the gauge
theory is associated with fluctuations of its dual supergravity
field (this identification is non-trivial). More precisely, the
generating functional for correlators in the field theory is related
to the type IIB string theory partition function by
\cite{gkp,witten1}
\be
\mathcal{Z}_{\rm string}\left[ \Phi_i \right] = \bigg< \exp \bigg( \int
d^4x ~\varphi_i \mathcal{O}_i \bigg) \bigg> ~, \label{adscft} \ee
subject to the relevant boundary
conditions (\ref{asyads}). This is a deep and extremely relevant
entry in the AdS/CFT dictionary. For most applications, we really
have to deal with the saddle-point approximation of this formula,
\be \exp \left( -\Gamma_{\rm sugra}[\Phi_i] \right) \approx \bigg<
\exp \bigg( \int d^4x ~\varphi_i \mathcal{O}_i \bigg) \bigg> ~,
\label{adscft2} \ee where $\Gamma_{\rm sugra}[\Phi_i]$ is the
supergravity action evaluated on the classical solution. It turns
out that the operator $\mathcal{O}_i$ has conformal dimension
$\Delta_i$. This is a nontrivial prediction of AdS/CFT. The brief
discussion here has only involved scalar fields. It naturally
extends to all degrees of freedom in $AdS_5$: this dictionary
implies that for every gauge invariant operator in the gauge theory
there exists a corresponding closed string field on $AdS_5 \times
S^5$ whose mass is related to the scaling dimension of the operator
and vice-versa. We will comment in the last lecture on some
additional features of the AdS/CFT correspondence that arise when
studying the gravity duals of confining gauge theories.

\subsection{Theories with Sixteen Supercharges}

Consider now a stack of $N_c$ flat Dp--branes. This configuration
preserves sixteen supercharges. The light open string spectrum is
that of a $U(N_c)$ super Yang--Mills theory in p+1 dimensions. We
would thus expect maximally supersymmetric Yang--Mills theory in p+1
dimensions to be {\it dual} to type IIA/IIB string theory in the
near horizon limit of the p-brane supergravity solution. This
solution reads (in the string frame): \be ds^2 = H_p^{-\half}
dx_{1,p}^2 + H_p^{\half} (dr^2 + r^2 d\Omega_{8-p}^2) ~,
\label{dbranesol1} \ee \be e^{-2(\phi-\phi_{\infty})} =
H_p^{\frac{p-3}{2}} ~, \label{dbranesol2} \ee \be A_{[p+1]} = -
\half (H_p^{-1}-1) ~dx^0 \wedge \dots \wedge dx^p ~,
\label{dbranesol3} \ee where $d\Omega_{8-p}^2$ is the round metric
on $S^{8-p}$ and $H_p$ is a harmonic function of the transverse
coordinates \be H_p= 1 + 2^{5-p} \pi^{\frac{5-p}{2}} g_s N_c\,
\Gamma\left(\frac{7-p}{2}\right)
\frac{{\alpha^\prime}^\frac{7-p}{2}}{r^{7-p}} ~. \label{armf} \ee
There are however new problems as soon as we abandon ~$p=3$. In
order to generalize the Maldacena correspondence, we should keep in
mind that it is necessary to decouple both the open and closed
string massive modes (that is, $\alpha^\prime \rightarrow 0$) as
well as the open/closed interactions (correspondingly, the Planck
length $l^2_P = g_s^{1/2} \,\alpha^\prime \rightarrow 0$), while
maintaining a finite $g^2_{YM}$, \be g^2_{YM} = (2\pi)^{p-2} g_s
\,(\alpha^\prime)^{\frac{p-3}{2}} ~, \label{ymcoup} \ee in the low
energy effective action. (We will derive this relation by the end of
the present lecture.) The latter condition implies that $g_s \sim
(\alpha^\prime)^{\frac{3-p}{2}}$ in the field theory limit which,
combined with the other requirements, forces the string coupling to
diverge for $p>3$. Moreover, plugging this expression into the
condition on the Planck length, we get ~$l_P \sim
(\alpha^\prime)^{\frac{7-p}{8}} \rightarrow 0$ ~as~ $\alpha^\prime
\to 0$. Thus, we should restrict our attention to $p<7$. Naively, we
would conclude that the decoupling is not possible for Dp--branes
with $p\geq 7$ ~and the dual strong coupling description is needed
for D4, D5 and D6--branes. However, the situation is more subtle
than this due to the fact that the string coupling depends on the
radial variable. Indeed, the near-horizon limit, \be U \equiv
\frac{r}{\alpha^\prime} = \mbox{fixed} ~, ~~~~~ r, \alpha^\prime \to
0 ~, \label{nhlimit} \ee of the supergravity solution
(\ref{dbranesol1})--(\ref{dbranesol3}) corresponding to $N_c$
Dp--branes reads \be ds^2 = \alpha^\prime \bigg[ \bigg( \frac{d_p\,
g^2_{YM} N_c}{U^{7-p}} \bigg)^{-\frac{1}{2}} dx_{1,p}^2 + \bigg(
\frac{d_p\, g^2_{YM} N_c}{U^{7-p}} \bigg)^{\frac{1}{2}} \left( dU^2
+ U^2 d\Omega_{8-p}^2 \right) \bigg] ~, \label{nhdbranesol1} \ee \be
e^{\phi} = (2\pi)^{2-p} g_{YM}^2 \left( \frac{d_p\, g^2_{YM}
N_c}{U^{7-p}} \right)^{\frac{3-p}{4}} ~, \label{nhdbranesol2} \ee
where
$$ d_p = 2^{7-2p} \pi^{\frac{9-3p}{2}} \Gamma\left( \frac{7-p}{2} \right)
~. $$ From the field theory point of view, $U$ is an energy scale.
The near horizon limit of Dp--brane solutions has non-constant
curvature for $p \neq 3$, \be \alpha^\prime ~\mathcal{R} \sim
\frac{1}{g_{eff}} \sim \bigg( \frac{U^{3-p}}{d_p\, g^2_{YM} N_c}
\bigg)^{\frac{1}{2}} ~. \label{curvat} \ee We have introduced in
this equation a dimensionless coupling, $g^2_{eff} \sim \gym^2 N_c
U^{p-3}$, which is particularly useful as long as the Yang--Mills
coupling is dimensionful for gauge theories in dimensions other than
four. It is proportional to the inverse of the curvature, which
means that the gauge theory description and the string/gravity one
are complementary and do not apply at the same energy scale. Notice
that the dilaton is not constant either, so the ranges of validity
of the various descriptions become more complicated here. The
decoupling limit does not work so cleanly. The supergravity solution
is valid in regions where ~$\alpha^\prime \mathcal{R} \ll 1$ and
$e^{\phi} \ll 1$. These conditions result to be energy dependent. In
particular, we see that for $p < 3$ the effective coupling is small
at large $U$ and the theory becomes UV free. Instead, for $p > 3$
the situation is the opposite, the effecting coupling increases at
high energies and we have to move to the dual string/gravity
description. This reflects the fact that for $p > 3$ supersymmetric
Yang--Mills theories are non-renormalizable and hence, at short
distances, new degrees of freedom appear \cite{imsy}.

The isometry group of the resulting metric is $ISO(1,p) \times
SO(9-p)$ for $p \neq 3$. There is no $AdS$ factor, something that
reflects the fact that the dual supersymmetric theories are not
conformally invariant. Again, there is a matching of symmetries at
work. The $ISO(1,p)$ factor clearly corresponds to the Poincar\'e
symmetry of the supersymmetric Yang--Mills theory, while $SO(9-p)$
is the $R$--symmetry. It is an $R$--symmetry since spinors and
scalars on the worldvolume of the Dp--brane transform respectively
as spinors and a vector in the directions transverse to the brane;
hence, under $SO(9-p)$. Let us consider a couple of examples that
will be useful in the following lectures.

\subsubsection{The case of Flat D5--branes}

In this subsection we use the formulas presented above to analyze
the decoupling limit, $r, \alpha^\prime \to 0$, for $N_c$
D5--branes. The Yang--Mills coupling has to be held fixed in this
limit, $g^2_{YM} = (2\pi)^3 g_s \,\alpha^\prime$ and the resulting
solution is \be ds^2 = \alpha^\prime \bigg[ \bigg( \frac{8\pi^3\,
U^2}{g^2_{YM} N_c} \bigg)^{\frac{1}{2}} dx_{1,5}^2 + \bigg(
\frac{g^2_{YM} N_c}{8\pi^3\, U^2} \bigg)^{\frac{1}{2}} \left( dU^2 +
U^2 d\Omega_{3}^2 \right) \bigg] ~, \label{nhd5branesol1} \ee \be
e^{\phi} = \frac{g_{YM} U}{(2\pi)^{3/2}\sqrt{N_c}} ~.
\label{nhd5branesol2} \ee We can see that there are different energy
scales where the system is described in different ways. Perturbative
super Yang--Mills is valid in the deep IR region, \be g^2_{eff} =
g^2_{YM} N_c U^2 \ll 1 ~~~\Rightarrow~~~ g_{YM} U \ll
\frac{1}{\sqrt{N_c}} ~. \label{d5one} \ee Let us now consider what
happens as we increase the energy. The conditions for supergravity
to be a valid approximation, namely that the curvature in string
units and the string coupling remain small are: \be \alpha^\prime
~\mathcal{R} \ll 1 ~~~~~ {\rm and} ~~~~~ e^{\phi} \ll 1 ~,
\label{d5two} \ee and we see that this happens within an energy
window \be \frac{1}{\sqrt{N_c}} \ll g_{YM} U \ll \sqrt{N_c} ~.
\label{d5three} \ee In this range of energies we can trust the type
IIB D5--brane supergravity solution and it provides an appropriate
description of the system. At higher energies, $\gym U \gg
\sqrt{N_c}$, the string coupling becomes large and we need to
perform an S--duality transformation to describe the system in terms
of the supergravity solution corresponding to flat NS5--branes. In
particular, $\tilde\alpha^\prime = g_s \alpha^\prime =
g^2_{YM}/(2\pi)^3$, and after S--duality is applied, the solution
reads, \be ds^2 = dx_{1,5}^2 + \tilde\alpha^\prime \bigg(
\frac{N_c}{U^2}\, dU^2 + N_c\, d\Omega_{3}^2 \bigg) ~,
\label{nhns5branesol1} \ee and the string coupling gets inverted.
The curvature in string units of this solution, $\tilde\alpha^\prime
\CR \sim N_c^{-1}$, is still small for large $N_c$ and the
supergravity solution is therefore valid.

\subsubsection{The case of Flat D6--branes}

After taking the decoupling limit for $N_c$ D6--branes, while
keeping as usual the Yang--Mills coupling fixed, $g^2_{YM} =
(2\pi)^4 g_s \,(\alpha^\prime)^{3/2}$,
\be
ds^2 = \alpha^\prime \bigg[ \frac{(2\pi)^2}{g_{YM}} \bigg( \frac{2U}{N_c}
\bigg)^{\frac{1}{2}} dx_{1,6}^2 + \frac{g_{YM}}{(2\pi)^2} \bigg(
\frac{N_c}{2U} \bigg)^{\frac{1}{2}} \left( dU^2 + U^2 d\Omega_{2}^2
\right) \bigg] ~,
\label{nhd6branesol1}
\ee
\be
e^{\phi} = \frac{g_{YM}^2}{2\pi} \left( \frac{2U}{g^2_{YM} N_c}
\right)^{\frac{3}{4}} ~,
\label{nhd6branesol2}
\ee
we see that the perturbative super Yang--Mills description is valid in the
deep IR region,
\be
g^2_{eff} = g^2_{YM} N_c U^3 \ll 1 ~~~\Rightarrow~~~ U \ll \frac{1}{(g_{YM}^2 N_c)^{\frac{1}{3}}} ~.
\label{d6one}
\ee
For higher energies, the type IIA supergravity solution is valid to
describe the system at energy scales in the interval
\be
\frac{1}{(g_{YM}^2 N_c)^{\frac{1}{3}}} \ll U \ll \frac{N_c}{g_{YM}^{2/3}} ~. \label{d6two}
\ee
In the UV, that is for energies above $U \gg N_c/g_{YM}^{2/3}$, the dilaton grows and the 11th dimensional circle opens up with radius $R_{11}(U) \gg
l_p$: we have to uplift the IIA solution to M--theory. This uplift results
in a purely gravitational background, the Ramond-Ramond potential
$(C_{[1]})_\mu$ to which the D6--brane couples magnetically lifts to the
metric components $g^{(11)}_{\natural\mu}$, where $\natural$ denotes
the 11th dimension. The actual solution in 11d reads, after a change
of variable of the form $y^2 = 2 N_c \gym^2 U/(2\pi)^4$, \be
ds^2_{11d} = dx_{1,6}^2 + dy^2 + y^2 \left(d \theta ^2 + \sin^2
\theta d \varphi^2 + \cos^2\theta d\phi^2 \right) ~, \label{d6in11d}
\ee where the angular variables are identified upon shifts \be
(\varphi,\phi) \sim \left( \varphi + \frac{2\pi}{N_c},\phi +
\frac{2\pi}{N_c}\right) ~, ~~~~~ 0 \leq \theta < \frac{\pi}{2} ~,
~~~ 0 \leq \varphi,\phi < 2\pi ~, \label{angd6} \ee thus describing
an asymptotically locally Euclidean (ALE) space with an $SU(N_c)$
singularity. This amounts to the fact that the $S^3$ metric in
(\ref{d6in11d}--\ref{angd6}) is written as a $U(1)$ bundle over
$S^2$ with monopole charge $N_c$. An interesting point to keep in
mind is that D6--branes provide a {\it purely gravitational
background} as seen from eleven dimensional M--theory. The metric is
locally flat, so that the curvature vanishes everywhere except at
the singularities. At very large values of $y$ (that is, in the far
UV), the proper length of circles is of order $R_{circles} \sim \gym
N_c^{-1/2} U^{1/2}$ while $l_P^{(11)} = g_s^{1/3} l_s$: an
everywhere flat 11d background as long as $R_{circles} \gg
l_P^{(11)} ~\Rightarrow~ U \gg \frac{N_c}{\gym^{2/3}}$. As a result
of this, it is not necessary to be in the large $N_c$ limit to trust
that 11d supergravity provides a good description in the UV.
However, when a massive radial geodesic in the IIA near-horizon
D6--brane background runs away from the small $U$ region, it starts
seeing the extra 11th dimension and the geometry becomes flat, so
that it can easily escape to infinity. Decoupling is spoiled. The
proper description should then be the whole M-theory (and not just
supergravity excitations) in the ALE space background.

\subsection{The D--brane world-volume action}

Let us end this lecture by rapidly mentioning some aspects of the
world-volume low-energy description of Dp--branes. We have already
mentioned that the massless string states on a Dp--brane form the
vector multiplet of a $(p+1)$--dimensional gauge theory with sixteen
supercharges. Consider a single Dp--brane with world-volume
$\mathcal{M}_{p+1}$, whose coordinates are $\xi^a$. The bosonic part
of its world-volume action is given by the sum of the
Dirac--Born--Infeld part and the Wess--Zumino part
\ba
S_{Dp} & = & - \tau_p \int_{\mathcal{M}_{p+1}} d^{p+1}\xi\ e^{-\Phi}
\sqrt{-\det \left(\hat{G}_{ab} + \hat{B}_{ab} + 2\pi\alpha^\prime F_{ab}
\right)}\cr & & ~~~~~ + \tau_p \int_{\mathcal{M}_{p+1}} \sum_q
\hat{C}_{[q]} \wedge e^{\hat{B}_{[2]} + 2\pi\alpha^\prime F} ~,
\label{SwvDp}
\ea
where hats denote pull-backs of bulk fields onto the
world-volume of the Dp--brane, $C_{[q]}$ are the RR potentials, and $F$
is the world-volume gauge field strength. The tension of the
Dp--brane is $\tau_p$, and the fact that the RR couplings are equal
to the tension is telling us that these are BPS objects. Let us study
the low-energy dynamics of this action. Consider the static gauge
$x^a = \xi^a$ and $x^I = x^I (\xi^a)$, which can always be fixed
because of diffeomorphism invariance. The DBI action can be expanded
at low energies in powers of $\alpha^\prime$, \ba && \sqrt{- \det
\left( \hat{G}_{ab} + \hat{B}_{ab} + 2\pi\alpha^\prime F_{ab}
\right)} = \sqrt{- \det G_{ab} } ~\times \cr && ~~~~~~~\times
\left\{ 1 + \frac{1}{2} G^{ab} G_{IJ} \partial_a x^I
\partial_b x^J + \frac{(2\pi\alpha^\prime)^2}{4} G^{ac} G^{bd} F_{ab}
F_{cd} \right\} ~. \ea Now, plugging this expansion in to
(\ref{SwvDp}), and using the explicit solution
(\ref{dbranesol1})--(\ref{dbranesol3}), the action becomes \ba
S_{Dp} & \simeq & - \tau_p \int d^{p+1}x\ \left\{ H_p^{-1} +
\frac{1}{2}
\partial_a x^I \partial_a x^I + \frac{(2\pi\ls^2)^2}{4}
F_{ab}F_{ab}\right\} \cr && ~~~~~ + \tau_p \int d^{p+1}\xi\ (
H_p^{-1} - 1 ) ~. \ea If we neglect a constant term, and define $x^I
= 2\pi\alpha^\prime \Phi^I$, the action simplifies to
\be
S_{Dp} = - \tau_p \frac{(2\pi\alpha^\prime)^2}{2} \int d^{p+1}x\ \bigg\{
\frac{1}{2} \partial_a \Phi^I \partial_a \Phi^I + \frac{1}{4} F_{ab} F_{ab}
\bigg\} ~,
\label{dbiym}
\ee
which is precisely the kinetic (bosonic) part of the $(p+1)$--dimensional supersymmetric gauge theory action, after the identification
\be
g_{YM}^2 = \frac{2}{(2\pi\alpha^\prime)^2 \tau_p} = (2\pi)^{p-2} g_s
\,(\alpha^\prime)^{\frac{p-3}{2}} ~,
\label{gymgs}
\ee
where we have replaced the value of the tension of a Dp--brane, and this is
the origin of eq.(\ref{ymcoup}) which we used above.

If we put $N_c$ Dp--branes on top of each other, there are some
difficulties and ambiguities in writing the corresponding
world-volume action. However, all attempts that have been considered
in the literature have resulted in the same low-energy dynamics: the
natural extension of the formulas above to the non--Abelian case.
\newpage

\section{Lecture III: Reducing Supersymmetry and Breaking Conformal Invariance}

The scaling arguments behind decoupling gravity from gauge theories
living in D--branes do not rely on maximal supersymmetry. As such,
one might be reasonably interested in generalizing the AdS/CFT
correspondence to supersymmetric gauge theories with less than
sixteen supercharges. In order to make contact with Nature, we need
to reduce the number of supersymmetries and to spoil conformal
invariance.

We might proceed by deforming $\CN = 4$ supersymmetric Yang--Mills
theory by adding a relevant or marginal operator which breaks
conformal invariance and supersymmetry, for example a mass term $M$
or a superpotential term, into the action and exploring the
corresponding deformation in the gravity dual. In this way, the
theories would be conformal in the UV (asymptotically $AdS$
configurations). In general, these scenarios lead to a RG invariant
scale given by \be \L_{QCD} \sim M e^{-\frac{1}{g_{YM}^2 N_c}} ~,
\label{rgiscale} \ee which means that the decoupling can only be
attained if we let $M \to \infty$ together with $\lambda = g_{YM}^2
N_c \to 0$, so that $\L_{QCD}$ is kept fixed. As $\l \to 0$, there
is little hope to apply this deformation method to study strongly
coupled QCD via supergravity. The masses $M$ of the unwanted degrees
of freedom do not decouple from the theory before the
strong-coupling phase is reached. We will not further discuss this
line of thought and the interested reader should take a look at
references \cite{aha,polstr}.

We will instead follow a different approach. We will engineer
D--brane configurations of type II string theory (or M--Theory)
which accomplish one or both of the following features: \vskip3mm
\begin{itemize}
\item {\it reduce supersymmetry} by considering an appropriate {\it
closed string background} known to preserve a specific fraction of
the supersymmetries of flat space or \vskip1mm
\item {\it break conformal symmetry} by engineering particular {\it
configurations of D--branes} in the above closed string backgrounds.
\end{itemize}
Roughly speaking, what we need is to modify the flatness of the D--branes,
the flatness of the target spaces, or both simultaneously.

\subsection{Closed String Backgrounds with Reduced Supersymmetry}

Consider a bosonic background.  We say that it is supersymmetric or
that it preserves supersymmetry if it is invariant under
supersymmetry transformations with spinor parameter $\epsilon$.  The
number of linearly independent solutions for $\epsilon$ determines
the amount of supersymmetry it preserves.  We are therefore looking
for a supersymmetric parameter $\e$ that fullfils \be \d_{\e} ~
\Phi_{b} |_{sol} \sim \Psi_{f} |_{sol} = 0 ~, \label{susy1} \ee \be
\d_{\e} ~ \Psi_{f} |_{sol} \sim \Phi_{b} |_{sol} = 0 ~,
\label{susy2} \ee for bosonic ($\Phi_{b}$) and fermionic
($\Psi_{f}$) fields. For a bosonic solution the latter one is the
non-trivial requirement. Indeed, it amounts to first order
differential equations that the bosonic fields must obey, the
so-called BPS equations. Among them, the transformation law of the
gravitino, in the absence of RR fields, reads $\d_{\e} \Psi_\mu =
D_\mu \e$. Therefore, supersymmetric backgrounds are given by
Ricci-flat manifolds with covariantly constant spinors
(Ricci-flatness arises as the integrability condition of the above
transformation law set to zero). Let us assume the existence of one
such spinor $\e$ and consider its parallel transport along closed
curves of the manifold. All matrices $M \in SO(d)$ relating $\e$
before and after the parallel transport along the loop define a
group, the holonomy group, which can be shown to be independent of
the base point. In order to leave the spinor unchanged, the holonomy
group $H$ of the manifold must admit an invariant subspace, and must
therefore be a proper subgroup of $SO(d)$. This means that a generic
manifold, with holonomy group $SO(d)$, breaks supersymmetry fully.
Therefore, in order to preserve some supersymmetry, a d-dimensional
manifold must have a holonomy group $H$, such that the group of
rotations $SO(d)$, must admit at least one singlet under the
decomposition of its spinor representation in irreducible
representations of $H \subset SO(d)$. Such a manifold is called a
{\it reduced} or {\it special holonomy} manifold. For any of these
manifolds, we can construct closed forms of various degrees, \be
\o_n ~=~ \frac{1}{n!} ~\bar{\e} \,\G_{i_1 ... i_n} \e ~\, dx^{i_1}
\we ... \we dx^{i_n} ~, \label{cforms} \ee out of the covariantly
constant spinor. However, just a few values of $n$ are such that
$\o_n \neq 0$ lie in a non-trivial cohomology class, and their
corresponding homology $n$-cycles are therefore non-trivial. The
list of all possible special holonomy Riemannian manifolds and
their corresponding non-trivial closed forms is summarized in Table 2.

\TABLE[ph]{\centerline {\small \begin{tabular}{@{}cccc@{}}
\hline
{} &{} &{} &{} \\[-1.5ex]
dim & Holonomy group & Fraction of SUSY & Forms \\[1ex]
\hline
{} &{} &{} &{} \\[-1.5ex]
4 & SU(2) & $\half$ & $\o_2$ \\[1ex]
6 & SU(3) & $\frac{1}{4}$  & $\o_2, \o_3, \o_4$ \\[1ex]
7 & $G_2$ & $\frac{1}{8}$ & $\o_3, \o_4$ \\[1ex]
8 & SU(2) $\times$ SU(2) & $\frac{1}{4}$  & $\o_2, \o_4, \o_6$ \\[1ex]
8 & Sp(2) & $\frac{3}{16}$  & $\o_2, \o_4, \o_6$ \\[1ex]
8 & SU(4) & $\frac{1}{8}$  & $\o_2, \o_4, \o_6$ \\[1ex]
8 & Spin(7) & $\frac{1}{16}$ & $\o_4$ \\[1ex]
10 & SU(3) $\times$ SU(2) & $\frac{1}{16}$ & $\o_2, \o_3, \o_4, \o_6, \o_8$
\\[1ex]
10 & SU(5) & $\frac{1}{16}$ & $\o_2, \o_4,\o_5, \o_6, \o_8$ \\[1ex]
\hline
\end{tabular}
\caption{Manifolds with special holonomy, their dimensionality, holonomy
group, the fraction of supersymmetry preserved when compactifying on them
and the list of closed forms.}
\label{sholonomy}}}

The available possibilities are not exhausted by the listed special
holonomy manifolds (which, as we will see later, include {\it
non-compact} and even {\it singular} cases). It is also interesting
to consider {\it orbifolds} of type II theories, namely quotients of
a part of space-time by a discrete symmetry. These spaces can be
seen as geometrically singular points in the moduli space of
Calabi--Yau manifolds, which however admit a sensible conformal
field theory description in terms of perturbative strings. In
particular, $\IC^2/\Gamma$ orbifolds, where $\Gamma$ is a discrete
subgroup of $SU(2)$, preserve one half of the supersymmetries of
flat space, while one quarter of the supercharges are preserved by
$\IC^3/\Gamma$ orbifolds, where $\Gamma\subset SU(3)$. Another
closed string theory background that provides an interesting setup
is given by parallel NS5 branes in flat space. This is a BPS
configuration breaking one half of the supersymmetries. In these
lectures, we will mainly focus on closed string backgrounds given by
a special holonomy manifold.

\subsection{Removing Conformal Invariance}

It is natural to ask at this point if it is possible to embed a
D--brane in such a way that some of its directions span a
submanifold of a special holonomy manifold while preserving some
supersymmetry. There is a hint telling us that we should if we think
of the D--brane as a fixed surfaces arising from orientifolding an
oriented closed string theory \cite{dlp,hor}. Nothing tells us
within this framework that the D--branes' worldvolume has to be
flat. However, this may be puzzling at a first glance, because a
curved worldvolume does not generally support a covariantly constant
spinor. At first sight, due to the non-trivial spin connection
$\omega_\mu^{ab}$, a D--brane whose world-volume $\mathcal{M}$
extends along a curved manifold will not preserve any supersymmetry,
\be \left( \partial_\mu + \half \omega_\mu \right) \epsilon \neq 0
~. \label{nospi} \ee This is simply the old problem of writing a
global supersymmetric gauge theory in curved space-time. It was
addressed by Witten long ago \cite{witw}, who introduced a subtle
procedure called the {\it topological twist} by means of which the
Lorentz group is combined with the (global) R--symmetry into a
twisted Lorentz group. This affects the irreducible representations
under which every field transforms. After the twist, supersymmetry
is not realized in the standard form but it is partially twisted on
the worldvolume theory of curved D--branes \cite{bsv}. It is
important to notice that supersymmetries corresponding to flat
directions of the D--branes are not twisted at all. Thus, we would
like to explore the possibility of a higher dimensional D--brane
with $(3+1)$ non-compact dimensions and whose remaining directions
wrap a curved submanifold within a special holonomy manifold.

Not any submanifold admits wrapped D--branes preserving some
supersymmetry. Submanifolds which do are called
\textbf{supersymmetric} or \textbf{calibrated cycles} and are
defined by the condition that the worldvolume theory is
supersymmetric. In other words, a global supersymmetry
transformation might be undone by a $\kappa$--transformation, which
is a fermionic gauge symmetry of the worldvolume theory. Table 3
displays a list of these cycles in special holonomy manifolds.

\TABLE[ph]{\centerline {\small \begin{tabular}{@{}cccccc@{}}
\hline
{} &{} &{} &{} &{} &{} \\[-1.5ex]
dim & $SU(2)$ & $SU(3)$ & $G_2$ & $SU(4)$ & Spin(7) \\[1ex]
\hline
{} &{} &{} &{} &{} &{} \\[-1.5ex]
2 & {\rm div./sLag} & {\rm holom.} & - & {\rm holom.} & \\[1ex]
3 & - & {\rm sLag} & {\rm assoc.} & - & - \\[1ex]
4 & {\rm manifold} & {\rm div.} & {\rm coassoc.} & {\rm Cayley} &
{\rm Cayley} \\[1ex]
5 & x & - & - & - & - \\[1ex]
6 & x & {\rm manifold} & - & {\rm div.} & - \\[1ex]
7 & x & x & {\rm manifold} & - & - \\[1ex]
8 & x & x & x & {\rm manifold} & {\rm manifold} \\[1ex]
\hline
\end{tabular}
\caption{Supersymmetric cycles in Manifolds with special holonomy.}
\label{scycles}}}

The low--energy dynamics of a collection of D--branes wrapping
supersymmetric cycles is governed, when the size of the cycle is
taken to zero, by a lower dimensional supersymmetric gauge theory
with less than sixteen supercharges. As discussed above, the
non--trivial geometry of the worldvolume leads to a gauge theory in
which supersymmetry is appropriately twisted. The amount of
supersymmetry preserved has to do with the way in which the cycle is
embedded in the higher dimensional space. When the number of branes
is taken to be large, the near horizon limit of the corresponding
supergravity solution provides a gravity dual of the field theory
living on their world--volume. The gravitational description of the
strong coupling regime of these gauge theories allows for a
geometrical approach to the study of such important aspects of their
infrared dynamics such as, for example, chiral symmetry breaking,
gaugino condensation, domain walls, confinement and the existence of
a mass gap.

We have to engineer configurations of D--branes whose world-volume
is, roughly speaking, topologically non-trivial.  There are two ways
of doing this:
\begin{itemize}
\item {\it D-branes wrapped on supersymmetric cycles} of
special holonomy manifolds. We will see that the non-trivial
topology of the world-volume allows getting a scale-anomalous theory
living at low energies on the flat part of the brane world-volume;
\item {\it Fractional D-branes} on orbifold or conifold
backgrounds. These branes can be thought of as D-branes wrapped on
cycles that, in the limit in which the Calabi--Yau manifold
degenerates into a metrically singular space, result to be wrapped
on shrinking cycles, effectively losing some world-volume directions
and being stuck at the singularity of the background.
\end{itemize}

Finally, another possibility involving NS5 branes consists in a
stack of D--branes stretched between two sets of parallel
NS5--branes. As a matter of fact, D--brane world-volumes can end on
NS5--branes, and this causes the freezing of some of the moduli of
the theory. As a result, the gauge theory living at low energies on
the intersection of the D--branes and NS5--branes can acquire a
scale anomaly. We will not discuss this scenario any further.

\subsection{The Topological Twist}

Consider a Dq--brane with worldvolume $\mathcal{W}$ living in a
ten-dimensional space-time $M$. There are $9-q$ scalars corresponding to
the fluctuations of the Dq--brane along directions transverse to
$\mathcal{W}$, which
are interpreted as collective coordinates. The tangent space $TM$
can be decomposed as $TM = T\mathcal{W} + N_{\mathcal{W}}$ where
$N_{\mathcal{W}}$ is the normal bundle of $\mathcal{W}$. Its
dimension is precisely $9-q$ and we can regard our transverse
scalars to be sections of this bundle: the {\it R-symmetry group}
$SO(9-q)_R$ of the world-volume gauge theory arises from the normal
bundle $N_{\mathcal{W}}$. Consider now the theory on a Calabi--Yau
$n$-fold, and denote a real $d$-dimensional cycle inside it as
$\Sigma_d$.

Now, take $q = p+d$ and take the D(p+d)--branes with $p$ flat
spatial directions and $d$ wrapped directions spanning the cycle
$\Sigma_d$. $T\mathcal{W}$ decomposes in two parts corresponding to
the ($p+1$ dimensional) flat and the ($d$ dimensional) wrapped
directions on the world-volume of the D(p+d)--branes. Similarly,
$N_{\mathcal{W}}$ is naturally split into two parts,
$N_{\Sigma_{d}}$ (which is the normal bundle within the Calabi--Yau
manifold) and the trivial transverse flat part. Correspondingly, the
full Lorentz group gets broken as follows:
$$ SO(1,9) \to SO(1,p) \times SO(d)_{\Sigma_{d}}
\times SO(2n-d)_R \times SO(9-2n-p)_R ~. $$
See, for example, Table 4. Now, the connection on $N_{\Sigma_{d}}$
and the (spin) connection on $\Sigma_{d}$ are to be related.
This amounts to picking $SO(d)_R
\subset SO(2n-d)_R$ and implementing the identification
$SO(d)_{\Sigma_{d}} = SO(d)_R$. What we actually identify is the
Lorentz connection of the former with the gauge connection
corresponding to the latter. This is a {\it topological twist},
since the behavior of all fields under Lorentz transformations,
determined by their spin, gets changed by this identification
\cite{mm}.

\TABLE[ph]{\centerline {\small \begin{tabular}{@{}cccccccccc@{}}
\multicolumn{1}{c}{ }
&\multicolumn{3}{c}{$\overbrace{\phantom{\qquad\qquad}}^{SO(1,3)}$}
&\multicolumn{2}{c}{$\overbrace{\phantom{\quad\qquad}}^{SO(2)_{\Sigma_2}}$}
&\multicolumn{2}{c}{$\overbrace{\phantom{\quad\qquad}}^{SO(2)_R}$}
&\multicolumn{2}{c}{$\overbrace{\phantom{\quad\qquad}}^{SO(2)_R}$}\\
\hline
{} &{} &{} &{} &{} &{}&{} &{} &{} &{}\\[-1.5ex]
Directions & 1 & 2 & 3 & 4 & 5 & 6 & 7 & 8 & 9 \\[1ex]
\hline
{} &{} &{} &{} &{} &{}&{} &{} &{} &{}\\[-1.5ex]
D5--brane & $\times$ & $\times$ & $\times$ & $\times$ & $\times$ &{} &{} &{} &{}\\[1ex]
$CY_2$ & {} & {} & {} & $\times$ & $\times$ & $\times$ & $\times$  &{} &{}\\[1ex]
$\Sigma_2$ & {} & {} & {} & $\times$ & $\times$ & {} & {}  &{} &{}\\[1ex]
\hline
\end{tabular}
\caption{A D5--brane wrapping a calibrated two cycle
inside a Calabi-Yau two-fold and the splitting of the full Lorentz
group ($p=3,d=2,n=2$).}
\label{wrapping}}}

The gravitino transformation law, in particular, is modified by the
presence of the additional external {\it gauge} field $A_\mu$
coupled to the R-symmetry, and becomes \be \left(\partial_\mu +
\half \omega_\mu + A_\mu\right)\epsilon = 0 ~, \label{ttwist} \ee
which admits a solution. The twist schematically imposes $A_\mu = -
\half \omega_\mu$, so that the condition that $\epsilon$ should be
covariantly constant boils down to $\epsilon$ being a constant
spinor and supersymmetry can be preserved. If we reduce along
$\Sigma_d$ the resulting field content on the flat $\IR^{1,p}$ is a
direct consequence of the twist: all fields with charges such that
$A_\mu = - \half \omega_\mu$, will result in massless fields of the
$(p+1)$-dimensional gauge theory. The resulting {\it low-energy
theory} on $\IR^{1,p}$ is a gauge theory with no topological twist.

It is clear from the analysis presented here how to realize what is
the field content of a given configuration of D--branes wrapping
supersymmetric cycles. But it is less clear how to obtain a
supergravity solution which includes the gravitational backreaction
of the D--branes. The question we now want to answer is: {\it How to
perform the topological twist at the level of the supergravity
solutions?}.

\subsection{The Uses of Gauged Supergravity}

There is a set of theories that have the right elements to do the
job \cite{mn1}: lower dimensional gauged supergravities. These theories are
obtained by Scherk--Schwarz reduction of 10d or 11d supergravities
after gauging the isometries of the compactifying manifold. Let us
illustrate the procedure in the case of D6--branes. We need an 8d
theory of gauged supergravity to have enough room to place the
D6--branes with a transverse coordinate. Luckily, Salam and Sezgin
\cite{ss} have built such a theory, by Kaluza--Klein reduction of
Cremmer--Julia--Scherk's 11d supergravity on $S^3$, and gauging its
isometries (which, as discussed above, give the {\it R-symmetry} of
the gauge theory living on the branes). It will be enough for our
discussion to consider a consistent truncation of the theory in
which all the forms that come from the 11d $F_{[4]}$ are set to zero
\footnote{The case in which some of the descendant of $F_{[4]}$ are
turned on will not be considered in these lectures. The interested
reader should take a look at \cite{epr1,hs}.}. We only have the
metric $g_{\mu\nu}$, the dilaton $\Phi$, five scalars $L_\a^i \in
SL(3,\IR)/SO(3)$, and an $SU(2)$ gauge potential $A^i$. The
Lagrangian governing the dynamics in this sector reads \be e^{-1}
\mathcal{L} = \frac{1}{4} R - \frac{1}{4} e^{2\Phi}
(F_{\mu\nu}^{~i})^2 - \frac{1}{4} (P_{\mu ij})^2 - \half
(\pa_\mu\Phi)^2 - V(\Phi,L_\a^i) ~, \label{sslag} \ee where $e =
\det e_\mu^a$, $F_{\mu\nu}^i$ is the Yang--Mills field strength and
$P_{\mu ij}$ is a symmetric and traceless quantity \be P_{\mu ij} +
Q_{\mu ij} \equiv L_i^\a (\pa_\mu \delta_\a^{~\b} - \e_{\a\b\g}
A_\mu^\g) L_{\b j} ~, \ee $Q_{\mu ij}$ being the antisymmetric
counterpart. The scalar potential can be written as \be
V(\Phi,L_\a^i) = \frac{1}{16} e^{-2\Phi} \left( T_{ij} T^{ij} -
\half T^2 \right) ~, \ee where \be T^{ij} \equiv L_\a^i L_\b^j
\delta^{\a\b} ~, ~~~~~~~ T = \delta_{ij} T^{ij} ~. \ee The
supersymmetry transformation laws for the fermions are given by \ba
\d_\e\psi_\g & = & \CD_\g \e + \frac{1}{24} e^{\Phi} F_{\mu\nu}^i
\hG_i
(\G_\g^{~\mu\nu} - 10 \d_\g^{~\mu} \G^\nu) \e \\
& & ~~~~~~~~~~~~~~~~~~~~~~~~~~~~~~ - \frac{1}{288}
e^{-\Phi} \e_{ijk} \hG^{ijk} \G_\g T \e ~, \cr ~\cr
\d_\e\chi_i & = & \half \left( P_{\mu ij} + \frac{2}{3} \d_{ij}
\partial_\mu\Phi \right) \hG^j \G^\mu \e - \frac{1}{4} e^{\Phi} F_{\mu\nu i}
\G^{\mu\nu} \e \\
& & ~~~~~~~~~~~~~~~~~~~~~~~~~~~~~~ - \frac{1}{8} e^{-\Phi} \left( T_{ij}
- \half \d_{ij} T \right) \e^{jkl} \hG_{kl} \e ~, \nonumber
\label{susytr2}
\ea
where we use, for the Clifford algebra, $\G^a = \g^a \times \II$, $\hG^i
= \g_9 \times \s^i$, $\g^a$ are eight dimensional gamma matrices, $\g_9 =
i \g^0 \g^1 \dots \g^7$, $\g_9^2 = 1$, and $\s^i$ are the Pauli matrices.
It is also convenient to introduce $\hG_9 \equiv -i \hG^{123} =
\g_9 \times \II$.

\subsubsection{Near Horizon of Flat D6--branes Revisited}

Let us now allow for a varying dilaton $\Phi$ and one scalar
$\varphi \in SL(3,\IR)/SO(3)$, and consider a domain wall ansatz
that captures the symmetries of a system of flat (parallel)
D6--branes placed at the origin $\rho = 0$, \be ds^2 = e^{2 f(\rho)}
dx^2_{1,6} + d\rho^2 ~. \label{ansd6} \ee The corresponding BPS
equations, emerging from $\d_\e\psi_\mu = \d_\e\chi_i = 0$, are \ba
\Phi'(\rho) & = & \frac{1}{8} e^{-\Phi} (e^{-4\varphi} + 2 e^{2\varphi}) ~, \\
\varphi'(\rho) & = & \frac{1}{6} e^{-\Phi} (e^{-4\varphi} -
e^{2\varphi}) ~, \label{bpsflatd6} \ea while the equations for $f$
and $\e$ can be easily integrated with the result \be f =
\frac{1}{3} \Phi ~, ~~~~~~~ \e = i \hG_9 \G_r \e = e^{\frac{1}{6}
\Phi} \e_0 ~, \label{solfd6} \ee $\e_0$ being a constant spinor. A
single projection is imposed on the Killing spinor. Thus, as
expected, one half of the supersymmetries are preserved. After a
change of variables $d\rho = e^{\Phi - 2\varphi} dt$, the BPS
equations decouple and can be integrated: \ba
\Phi(t) & = & \frac{3}{4} \left[ \varphi(t) + \half (t - t_0) \right] ~, \\
\varphi(t) & = & \frac{1}{6} \left[ \log (e^t - \xi_0) - t \right]
~, \label{sol2fd6} \ea where $t_0$ and $\xi_0$ are integration
constants. If we perform a further change of variables $e^t = r^4 -
a^4$, with $\xi_0 = - a^4$, and uplift the solution to eleven
dimensions, by applying the formulas in Salam and Sezgin's paper
\cite{ss} in reverse we obtain the form \be ds_{11d}^2 = dx^2_{1,6}
+ \frac{1}{{1 - \frac{a^4}{r^4}}} dr^2 + \frac{r^2}{4} \left[
(\tw^1)^2 + (\tw^2)^2 + \left( {1 - \frac{a^4}{r^4}} \right)
(\tw^3)^2 \right] ~, \label{egha} \ee where $\tw^i$ are the left
invariant Maurer--Cartan $SU(2)$ one--forms corresponding to $\tilde
S^3$ (see Appendix A). Besides the 7d Minkowskian factor, we get the
metric for a non--trivial asymptotically locally Euclidean (ALE)
four manifold with a $SU(2) \times U(1)$ isometry group, namely the
Eguchi--Hanson metric. This is precisely the well-known uplift of
the near-horizon solution corresponding to D6--branes in type IIA
supergravity. This is the solution whose physics was discussed
above.

This result extends quite naturally and the following statement can
be made: {\it domain wall solutions of gauged supergravities
correspond to the near-horizon limit of D--brane configurations}
\cite{bst}. This provides the gravity dual description of the gauge
theories living in their worldvolumes. However, we did not introduce
lower dimensional gauged supergravities simply to reobtain known
solutions. We shall now explore the uses of these theories to seek
more involved solutions representing wrapped D--branes.

\subsection{D6--Branes Wrapping a sLag $3$--cycle}

There is an entry in the table of special holonomy manifolds that
tells us that there is a calibrated homology special Lagrangian
(sLag) $3$--cycle in a Calabi--Yau threefold where we can wrap
D6--branes while maintaining supersymmetry. ~The configuration will
actually have $\frac{1}{2} \times \frac{1}{4} \times 32 = 4$
supercharges.

From the point of view of the world--volume, the twist acts as
follows. The fields on the D6--branes transform under $SO(1,6)
\times SO(3)_R$ as $({\bf 8},{\bf 2})$ for the fermions and $({\bf
1},{\bf 3})$ for the scalars, while the gauge field is a singlet
under $R$--symmetry. When we wrap the D6--branes on a three--cycle,
the symmetry group splits as $SO(1,3) \times SO(3) \times SO(3)_R$.
The corresponding representation for fermions and scalars become,
respectively, $({\bf 4},{\bf 2},{\bf 2})$ and $({\bf 1},{\bf 1},{\bf
3})$. The effect of the twist is to preserve those fields that are
singlets under a diagonal $SO(3)_D$ built up from the last two
factors. The gauge field survives (it remains singlet after the
twist is applied) but the scalars are transformed into a vector
since ${\bf 1} \times {\bf 3} = {\bf 3}$. So, we are left with a
theory with no scalar fields in the infrared; besides, since ${\bf
2} \times {\bf 2} = {\bf 1} + {\bf 3}$, one quarter of the
supersymmetries, {\it i.e.} four supercharges, are preserved. This
is the content of the vector multiplet in $\CN = 1$ supersymmetric
gauge theory in four dimensions.

Let us seek the corresponding supergravity solution. An ansatz that
describes such a deformation of the world--volume of the D6--branes
is \cite{en}
\be
ds^2 = e^{2f(r)} dx^2_{1,3} + \frac{1}{4} e^{2h(r)} \sum_{i=1}^3 (w^i)^2
+ dr^2 ~,
\label{anstz}
\ee
where $w^i$ are the left invariant one--forms corresponding to the $S^3$. The twist is achieved by turning on the non--Abelian gauge field $A_{\mu}^i = -
\frac{1}{2} ~w_{\mu}^i$. It is easy to see that in this case we can
get rid of the scalars $L_\a^i = \d_\a^i ~ \Rightarrow ~ P_{ij} = 0
~, ~Q_{ij} = - \e_{ijk} A^k$. By imposing the following projections
on the supersymmetric parameter $\e$: $\e = i \hG_9 \G_r \e$
~$\Gamma_{ij}\,\e\,=\,-\hat\Gamma_{ij}\,\e\, ~,~~ i \neq j = 1, 2,
3$, which leave unbroken $\frac{1}{8}$ of the original
supersymmetries, that is, $4$ supercharges, the first order BPS
equations are, \be f'(r) = \frac{1}{3} \Phi'(r) = - \frac{1}{2}
e^{\Phi - 2h} + \frac{1}{8} e^{-\Phi} ~, \label{buu} \ee \be h'(r) =
\frac{3}{2} e^{\Phi - 2h} + \frac{1}{8} e^{-\Phi} ~. \label{bud} \ee
When uplifted to eleven dimensions, the solution of the BPS
equations reads $ds_{11d}^2 = dx^2_{1,3} + ds_{7d}^2$, with
\be
ds_{7d}^2 = \frac{1}{\Bigl( 1- \frac{a^3}{\rho^3} \Bigr)} d\rho^2 +
\frac{\rho^2}{12} \sum_{i=1}^3 (\tw^i)^2 + \frac{\rho^2}{9} \Bigl( 1
- \frac{a^3}{\rho^3} \Bigr) \sum_{i=1}^3 \Bigl[ w^i - \frac{1}{2}
\tw^i \Bigr]^2 ~,
\ee
after a convenient change of the radial
variable \cite{en}. It is pure metric, as advanced in the previous discussion.
This is the metric of a $G_2$ holonomy manifold \cite{bsal,gpp}
which is topologically $\IR^4 \times S^3$. The radial coordinate
lies in the range $\rho \geq a$. Furthermore, close to $\rho = a$,
$\tS^3$ retains finite volume $a^3$ while $S^3$ shrinks to zero
size.

Several importants comments are in order.\footnote{The subject of M--theory
on $G_2$ manifolds is sufficiently vast and insightful as to be appropriately discussed in the present lectures. Some of the features that we discuss here
have been originally presented in \cite{Bobby}. An exhaustive discussion of 
the dynamics of M--theory on these manifolds was carried out in \cite{AtW}.
The appearance of chiral fermions from singular $G_2$ manifolds was also
explored \cite{AchW}. Besides, there is a quite complete review on the
subject \cite{AchGuk} whose reading we encourage.} As we discussed before,
the gauge coupling constant is related to the volume of the wrapped cycle,
\be
\frac{1}{\gym^2} \sim \frac{1}{l_s^2} \int_{\tS^3} d^3\xi\ \sqrt{\det G} ~, \label{cgtwo}
\ee
where $G_{ab}$ is the induced metric. If we start to decrease the value of
$a$, the geometry flows towards the singularity and, due to (\ref{cgtwo}),
the gauge theory is flowing to the IR. This seems worrying because we would
not trust this solution if we are forced to hit the singularity. However, remembering that $\mathcal{N} = 1$ theories have no real scalar fields, we immediately realize that there must be a partner of the quantity above such
that a complexified coupling arises. Clearly, we are talking about $\tym$,
that can be computed in terms of the three--form potential $C_{[3]}$ as
\be
\tym \sim \frac{1}{l_s^2} \int_{\tS^3} C_{[3]} ~.
\ee
Therefore, the singularity is really a point in a complex plane that can
consequently be avoided. Atiyah, Maldacena and Vafa argued \cite{amv} that
there is a {\it flop} (geometric, smooth) transition, the blown up $S^3$
being related to the gaugino condensate, where the gauge group disappears
and the theory confines.

The latter discussion opens an interesting avenue to think of gauge/gravity
duality as the result of geometric transitions in string theory. This idea
was put forward by Vafa \cite{vafa} and further elaborated in several
papers, beside \cite{amv}. It is an important observation for many reasons.
Most notably, there are supersymmetric gauge theories whose D--brane set up
is under control while their gravity duals are extremely difficult to find.
In these cases, the framework of geometric transitions might shed light,
at least, into the holomorphic sector of the gauge/gravity duality. As an
outstanding example, we would like to point out the case of $\CN = 1$ supersymmetric gauge theory with an arbitrary tree-level superpotential.
The effective superpotential at low energies, as a function of the gluino condensate composite superfields, can be computed by means of applying
geometric transitions \cite{civ,eot}.

There are two very different quotients of this $G_2$ manifold: a singular
one by $\IZ_N \subset U(1) \subset SU(2)$, and a non--singular
quotient if one instead chooses $\IZ_N \subset U(1) \subset
\tilde{SU}(2)$. They are pictorically represented in Fig.\ref{fig:compli2}.
The former results in an $A_{N-1}$ singularity fibered over $\tS^3$ so
that, after Kaluza--Klein (KK) reduction along the circle
corresponding to the $U(1)$, one ends with $N$ D6--branes wrapped on
a special Lagrangian $\tS^3$ in a Calabi--Yau three--fold. The
latter, instead, as there are no fixed points, leads to a smooth
manifold admitting no normalizable supergravity zero modes. Thus,
M--theory on the latter manifold has {\it no massless fields}
localized in the transverse four-dimensional spacetime. ~By a smooth
interpolation between these manifolds, M--theory realizes the {\it
mass gap of $\mathcal{N}=1$ supersymmetric four-dimensional gauge
theory}. After KK reduction of the smooth manifold one ends with a
non--singular type IIA configuration (without D6--branes) on a space
with the topology of $\mathcal{O}(-1) + \mathcal{O}(-1) \to \IP^1$,
and with $N$ units of RR flux through the finite radius $S^2$.

\FIGURE[ph]{\centerline{\epsfig{file=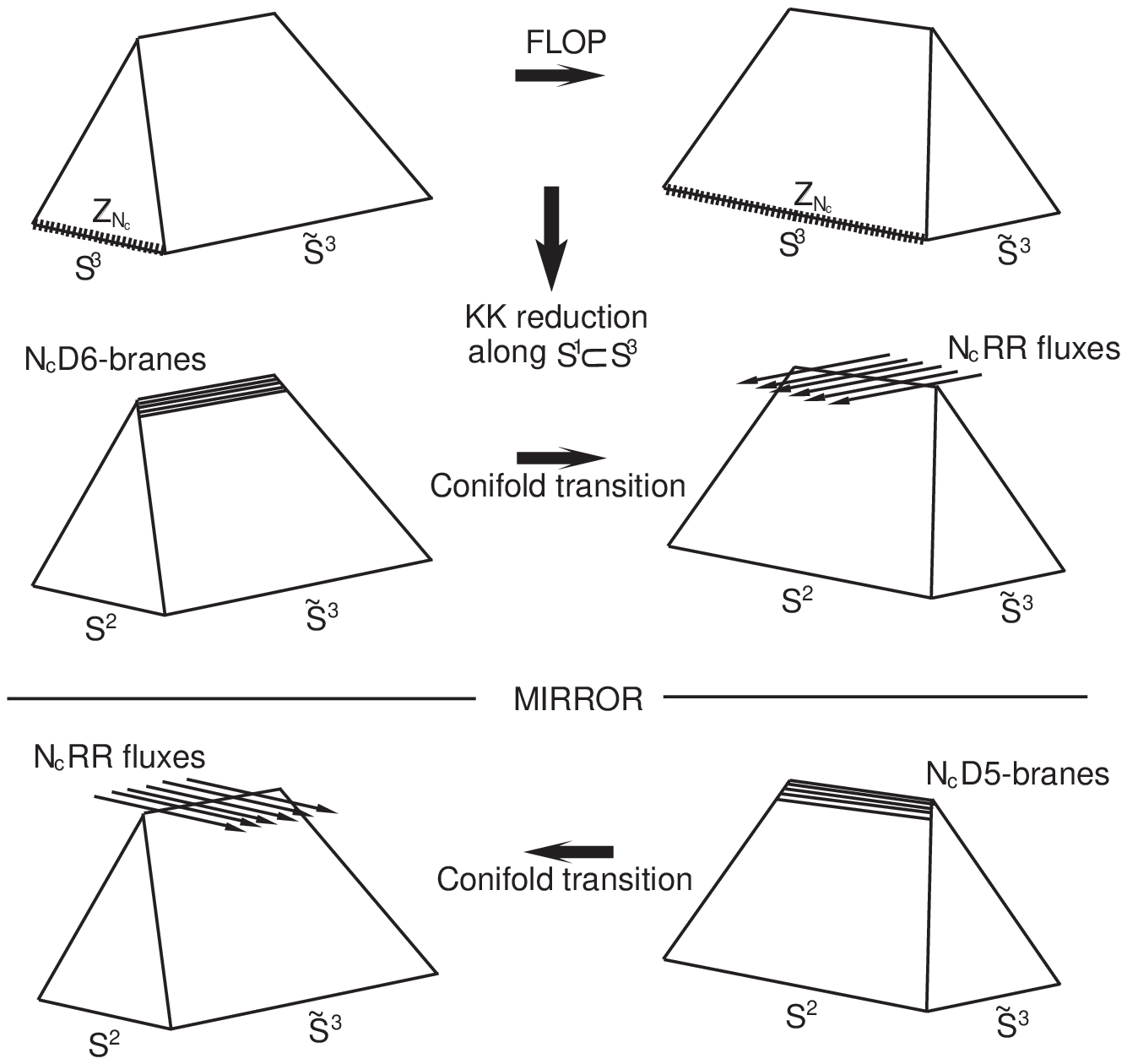,width=13cm}}
\caption{The flop transition in the $G_2$ manifold amounts to exchanging
the r\^ole of the spheres. The action of $\IZ_{N_c} \subset S^1 \subset
S^3$ is singular/regular on the left/right hand side of the upper figure.
Corresponding quotients of the $S^3 \in G_2$ give a singular manifold
(upper left) which is an $A_{N_c-1}$ singularity fibered over $\tS^3$ that,
when KK reduced along
the $S^1$ represents $N_c$ D6-branes wrapping a sLag 3-cycle in the cotangent
bundle $T^*\tS^3$ (middle left). RG flow towards the IR drives the system
into the flop transition to a smooth $G_2$ manifold (upper right) or, from
the 10d point of view, the system undergoes a conifold transition such that
D6-branes dissapear and are replaced by RR fluxes through the exceptional
$\IP^1$ of the manifold $\mathcal{O}(-1) + \mathcal{O}(-1) \to \IP^1$
(middle right). There is a mirror version of this story, in type IIB string
theory, where $N_c$ D5-branes wrapping an holomorphic 2-cycle in
$\mathcal{O}(-1) + \mathcal{O}(-1) \to \IP^1$ are dual to $N_c$ fluxes
through the sLag 3-cycle of the deformed conifold.}
\label{fig:compli2}}

The resulting geometry in eleven dimensions is asymptotically conical. As
such, it can be seen as the uplift of the system of wrapped D6--branes for
an {\it infinite value of the string coupling}. Remind that the latter is related to the asymptotical value of the exponential of the dilaton which,
in turn, is interpreted as the radius of the eleventh dimension. It should
be possible to find a supergravity dual of this system with a finite radius
circle at infinity, thus representing a string theory configuration at
finite coupling. Indeed, a solution with the desired behaviour was found,
and its uses in the context of the gauge/gravity duality were thoroughly
studied \cite{bggg}.

\subsection{The Maldacena--N\'u\~nez setup}

There is an analogous setup in type IIB superstring theory that was
proposed before the previous one by Maldacena and N\'u\~nez
\cite{mn2}. Roughly speaking, it can be thought of as the mirrored
version of the type IIA scenario discussed before, as schematically
displayed in Fig.\ref{fig:compli2}.
It is actually simpler to tackle the problem in type IIB
because it lacks the subtleties related to world-sheet instantons
that affect the calculus of $\CN = 1$ superpotential terms. Again,
we might start by drawing attention to an entry in the table of
special holonomy manifolds that says that there are holomorphic
$2$--cycles in a Calabi--Yau manifold where we can wrap D5--branes
while preserving some supersymmetry. Recalling the discussion about
flat D5--branes, we can already see that this system will be similar
to $\CN = 1$ supersymmetric Yang--Mills theory in the IR but its UV
completion will be given by a conformal 6d theory constructed on a
stack of flat NS5--branes. We will not discuss this point here.

Consider a system of D5--branes wrapped on a supersymmetric
$2$--cycle inside a Calabi--Yau {\it threefold}. The $R$--symmetry
group is $SO(4)_R = SU(2)_1 \times SU(2)_2$. We need to choose an
$SO(2)$ subgroup in order to perform the twist and cancel the
contribution of the spin connection in $S^2$. There are several
options. The right twist to get $\CN = 1$ supersymmetric Yang--Mills
is performed by identifying $SO(2)_{S^2}$ and $SO(2) \subset
SU(2)_1$ (notice that $SO(2) \subset SU(2)_2$ is the same; other
nontrivial options involve both $SU(2)$ factors). We can mimic the
arguments given in the previous section. The fields on the
D5--branes transform under $SO(1,5) \times SO(4)_R = SO(1,5) \times
SU(2)_1 \times SU(2)_2$ as $({\bf 4^+},{\bf 2})$ and $({\bf
4^-},{\bf 2})$ for the fermions (the $\pm$ superscripts correspond
to chirality), $({\bf 1},{\bf 4})$ for the scalars and the gauge
field remains a singlet under $R$--symmetry. When we wrap the
D5--branes on a two--cycle, the symmetry group splits as $SO(1,3)
\times SO(2) \times SU(2)_1 \times SU(2)_2$. The corresponding
representation for the fermions become $({\bf 2},{\bf +1},{\bf
2},{\bf 1})$, $({\bf \bar 2},{\bf -1},{\bf 2},{\bf 1})$, $({\bf \bar
2},{\bf +1},{\bf 1},{\bf 2})$ and $({\bf 2},{\bf -1},{\bf 1},{\bf
2})$, while the scalars transform in the $({\bf 1},{\bf 0},{\bf
2},{\bf 2})$. The effect of the twist is to preserve those fields
which are uncharged under the diagonal group $SO(2)_{\rm diag}$ or,
better, $U(1)_{\rm diag}$. Again, they are exactly the field content
of the vector multiplet in $\CN = 1$ supersymmetric gauge theory in
four dimensions: the gauge field and the fermions transform under
$SO(1,3) \times SO(2)_D$, respectively, in the
$(\mathbf{4},\mathbf{1})$ and $(\mathbf{2},+) \oplus
(\mathbf{\bar{2}},-)$ representations \footnote{As an exercise, the
reader can attempt to perform the twist when $SO(2)_{S^2}$ is
identified with the diagonal $SU(2)$ built out from $SU(2)_1 \times
SU(2)_2$. The result is the supermultiplet of $\CN = 2$
supersymmetric gauge theory.}.

It is possible to follow the same steps that we pursued in the case of
the D6--brane. The supergravity solution shall appear as an appropriate
domain wall solution of 7d gauged supergravity with the needed twist.
There is however a subtlety: by turning on the $U(1)$ gauge field
corresponding to $SO(2) \subset SU(2)_1$, we get a {\it singular
solution}. We can turn on other components of the $SU(2)$ gauge
fields by keeping their UV behavior, where the perturbative degrees
of freedom used to discuss the topological twist are the relevant ones.
This is a specific feature that, as we will see in the last lecture,
can be mapped to an extremely important IR phenomenon of $\mathcal{N}=1$
supersymmetric gauge theory: {\it chiral symmetry breaking due to gaugino
condensation}. From the point of view of gauged supergravity, turning on
other components of the $SU(2)$ gauge field amounts to a generalization
of the twisting prescription that has been discussed in \cite{epr2}. We
don't have time to comment on this point any further.

Instead of deriving the so-called Maldacena--N\'u\~nez (MN) solution
(which was actually originally obtained in a different context by
Chamseddine and Volkov \cite{chv}), we will just write it down
following a notation that makes contact with the gauged supergravity
approach. The gauge field is better presented in terms of the
triplet of Maurer--Cartan 1-forms on $S^2$ \be \s_1 = d\theta ~,
~~~~~ \s_2 = \sin\theta d\varphi ~, ~~~~~ \s_3 = \cos\theta d\varphi
~, \label{maucar} \ee that obey the conditions $d\s_i = - \half
\e_{ijk} \s_j\wedge \s_j$. It is given by: \be A^1 = - \half a(\rho)
~\s_1 ~, ~~~~~ \,\,\, A^{2} = \half a(\rho) ~\s_2 ~, ~~~~~ \,\,\,
A^{3} = - \half \s_3 ~, \label{gaugemn} \ee where, for future use,
we write the explicit dependence of $a(\rho)$, \be a(\rho) = \frac{2
\rho}{\sinh(2\rho)} ~. \label{arho} \ee If we take $a(\rho)$ to
zero, the solution goes to the previously mentioned singular case
(which is customarily called the Abelian MN solution). The string
frame metric is \be ds_{str}^2 = dx_{1,3}^2 + g_s N \alpha^\prime
\bigg( d\rho^2 + e^{2h(\rho)} (\s_1^2 + \s_2^2) + \frac{1}{4} \sum_i
(w^i - A^i)^2 \bigg) ~, \label{metricmn} \ee where \be e^{2h(\rho)}
= \rho \coth(2\rho) - \frac{\rho^2}{\sinh^2(2\rho)} - \frac{1}{4} ~.
\label{gmn} \ee As expected, the solution has a running dilaton \be
e^{2\Phi(\rho)} = \frac{\sinh 2\rho}{2 e^{h(\rho)}} ~, \label{rdil}
\ee that behaves linearly at infinity. There is a RR 3--form flux
sourced by the D5--branes whose explicit expression reads
\be
F_{[3]} = g_s N \alpha^\prime \left[ 2\ (w^1 - A^1) \wedge (w^2 - A^2)
\wedge (w^3 - A^3) - \sum_{i=1}^{3} F^i \wedge w^i \right] ~,
\label{RR3f}
\ee
where the physical radial distance is $r = \sqrt{g_s N \alpha^\prime} \rho$
and $F^i = dA^i + \epsilon^{ijk} A^j \wedge A^k$, and $w^i$ are left-invariant one-forms parameterizing $S^3$ (see the Appendix). If we perform the 7d gauged supergravity analysis, it becomes clear that we are left with {\it four supercharges}. We will use this supergravity solution in Lecture V and show
how it encapsulates field theory physics.

We wish to point out at this stage that several interesting scenarios
involving D5--branes and NS5--branes wrapping supersymmetric cycles have
been considered so far. Most notably, systems which are dual to pure
$\CN = 2$ supersymmetric gauge theories in four \cite{gkmw1,bcz} and three
\cite{gkmw2} spacetime dimensions, as well as to $\CN = 1$ theories in
3d \cite{st,mnas}.

\subsection{Decoupling}

There are two issues of decoupling when considering wrapped
Dp--branes. One is the decoupling of Kaluza-Klein modes on the
wrapped cycle and the other is the decoupling of higher worldvolume
modes from the field theory in the `field theory limit'.

Let us first consider the Kaluza-Klein modes.  The mass scale
associated with the Kaluza-Klein modes should be \be
\Lambda_{\text{KK}}^q \equiv M_{\text{KK}}^q \sim
\frac{1}{\text{Vol}\,\Sigma_q} , \ee where $\Sigma_q$ is the
$q$-cycle in the Calabi-Yau wrapped by the branes. There is no a
priori reason to identify this cycle with any particular cycle in
the dual (backreacted) geometry.  When probed at energies above
$\Lambda_{\text{KK}}$ the field theory will be $p+1$ dimensional.
These field theories typically have a positive $\beta$-function so
the dimensionless effective coupling will decrease as one decreases
the energy scale.  At energy scales comparable to
$\Lambda_{\text{KK}}$ the field theory will become $p-q+1$
dimensional; in the above we have considered $p-q+1=4$ in order to
make contact with QCD-like theories and we will continue to do so in
this section.  We know from (\ref{bef}) that the beta function is
now negative, and the dimensionless effective coupling will increase
until it hits the characteristic mass scale of the field theory
generated through dimensional transmutation $\Lambda_{\text{QCD}}$
(see Fig.4).
\FIGURE[ph]{\centerline{\epsfig{file=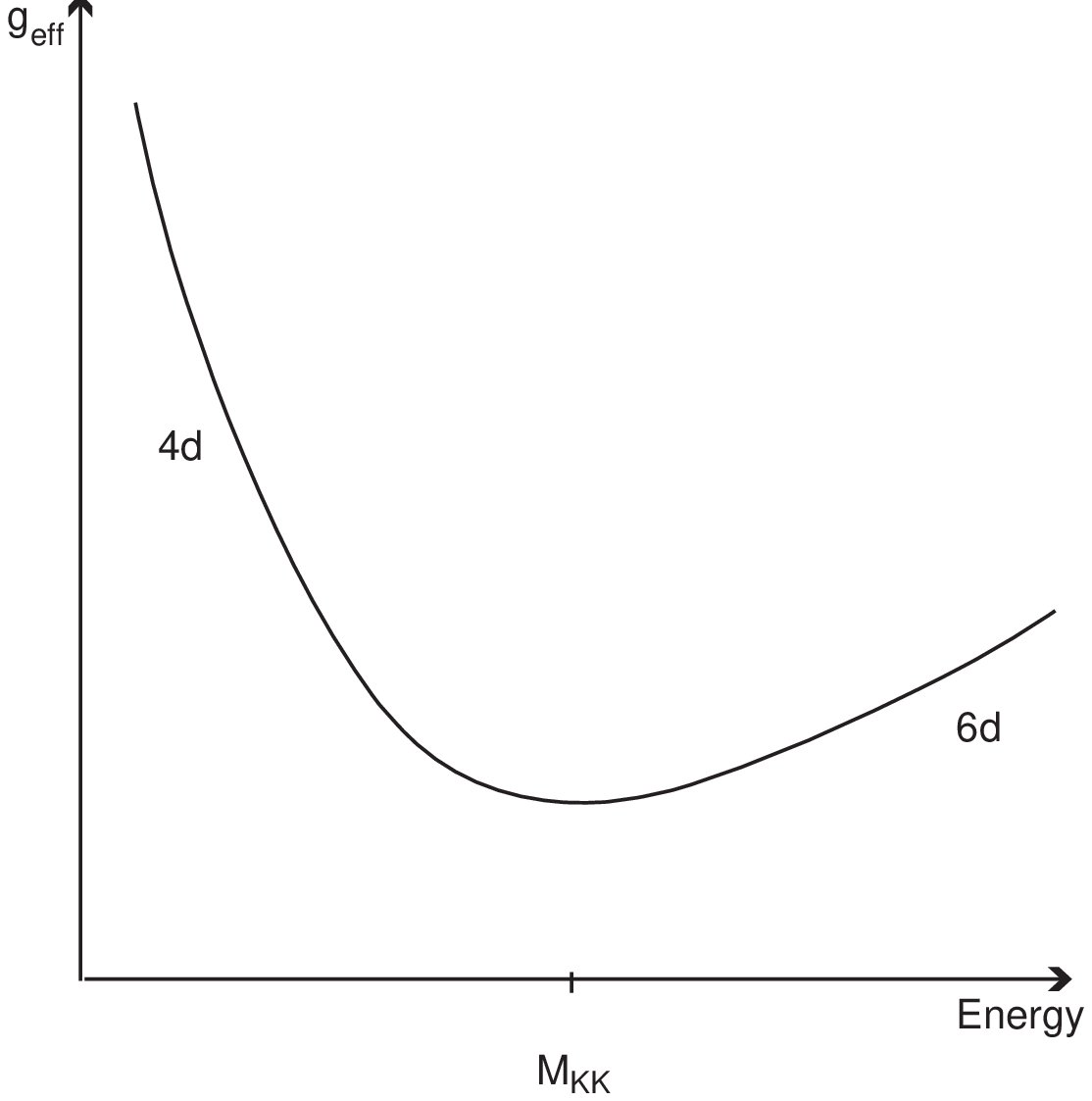,width=7cm}}
\caption{A plot of the effective
coupling and its dependence on energy for the MN setup. The
Kaluza-Klein scale $M_{KK}$ is the energy scale at which the wrapped
dimensions become visible.  Non-decoupling is the statement that
$\Lambda_{QCD}$ is very close to $M_{KK}$.}
\label{fig:running}}
This is the scale at which nonperturbative effects in the field theory become significant, and the scale at which we wish to study the gravity dual. It can
be shown that for the gravity duals involving wrapped branes, it is {\it not possible} to decouple the two scales $\Lambda_{\text{QCD}}$ and
$\Lambda_{\text{KK}}$. This can be shown in two ways, either by assuming
some relations between certain cycles within the gravitational background
with the initial cycle the D--branes wrap, as is often done in discussions
of the Maldacena-N\'u\~{n}ez background \cite{bcpz,Loewy:2001pq} or by
assuming that the gravitational background should not contain a regime dual
to a weakly coupled field theory \cite{Gursoy:2003hf}. The study of strong
coupling field theory via gravity duals would therefore, on general grounds,
be contaminated by KK modes. This issue would be overcome, of course, provided
suitable sigma models for a string theory in these background were found and their quantization carried out.

In spite of the precedent arguments, a further comment is in order at this point. In the context of supergravity duals corresponding to gauge theories
with a global $U(1) \times U(1)$ symmetry, it has been recently shown that
it is possible to perform a $\beta$--deformation (meaningly, an $SL(2,\IZ)$ transformation in the $\tau$ parameter of the corresponding torus in the
gravity side) on both sides of the gauge/gravity duality \cite{lumal}. In
the particular case of the MN background, it has been argued that the
resulting field theory is a dipole deformation of $\CN = 1$ supersymmetric Yang--Mills theory and that the deformation only affects the KK sector. It
is then possible to render these states very massive, this possibly allowing
to disentangle them from the dynamics \cite{gunu} (see also \cite{bodira}). 

Let us now turn to study the gravitational modes. Supersymmetric Yang--Mills
is not the complete field theory that describes the Dp--brane dynamics. The question at hand is whether there exists a limit, compatible with the other limits, that can be taken such that other degrees of freedom can be decoupled.  We will not describe the arguments discussed in \cite{imsy} and reviewed in
\cite{Gursoy:2003hf} here in detail, but just mention that in the case of wrapped D5 and D6--branes it is again {\it not possible} to decouple gravitational modes and little-string theory modes respectively from
$\Lambda_{QCD}$.

It is in a sense remarkable that, despite such problems with decoupling, dualities involving wrapped branes are able to reproduce so many features
of ${\mathcal{N}} = 1$ supersymmetric Yang--Mills theories. The study of
$\beta$--deformed systems opens an avenue to KK decoupling \cite{gunu}
which might provide a clue to puzzle out this conundrum.
\newpage

\section{Lecture IV: D-branes at Singularities}

In the previous lecture we introduced a framework to study the
gravity dual of non-maximal supersymmetric gauge theories by
wrapping Dp--branes, $p>3$ on calibrated cycles of special holonomy
manifolds. Among the difficulties posed by this approach, we have
seen that decoupling is far from being as clear as it is for
D3--branes. If we insist on keeping D3--branes and want to deform
the background, the simplest case amounts to considering orbifolds.
Take for example, $\IC^2/\IZ_2$, where $\IZ_2$ leaves invariant six
coordinates of spacetime and acts nontrivially on the last four
ones,
\be
z^1 = - z^1 ~, ~~~~~ z^2 = - z^2 ~,
\ee
with $z^1 = x^6 + i x^7$ and $z^2 = x^8 + i x^9$. Now, take $\IR^{1,9} \to
\IR^{1,5} \times \IR^4 \approx \IR^{1,5} \times \IC^2$ and consider how the
action with the $\IZ_2$ cyclic group affects the oscillators. Clearly,
\be
\alpha_n^r \to - \alpha_n^r ~, \qquad \tilde \alpha_n^r \to - \tilde
\alpha_n^r ~, \qquad \psi_n^r \to - \psi_n^r ~, \qquad
\tilde \psi_n^r \to - \tilde \psi_n^r ~.
\ee
We keep full control over perturbative string theory just by keeping states which are $\IZ_2$--invariant. We now put flat D3--branes along the directions
$0123$. If a D3--brane is not at the origin of the transverse space, it is
not invariant under the action of $\IZ_2$.
\FIGURE[ph]{\centerline{\epsfig{file=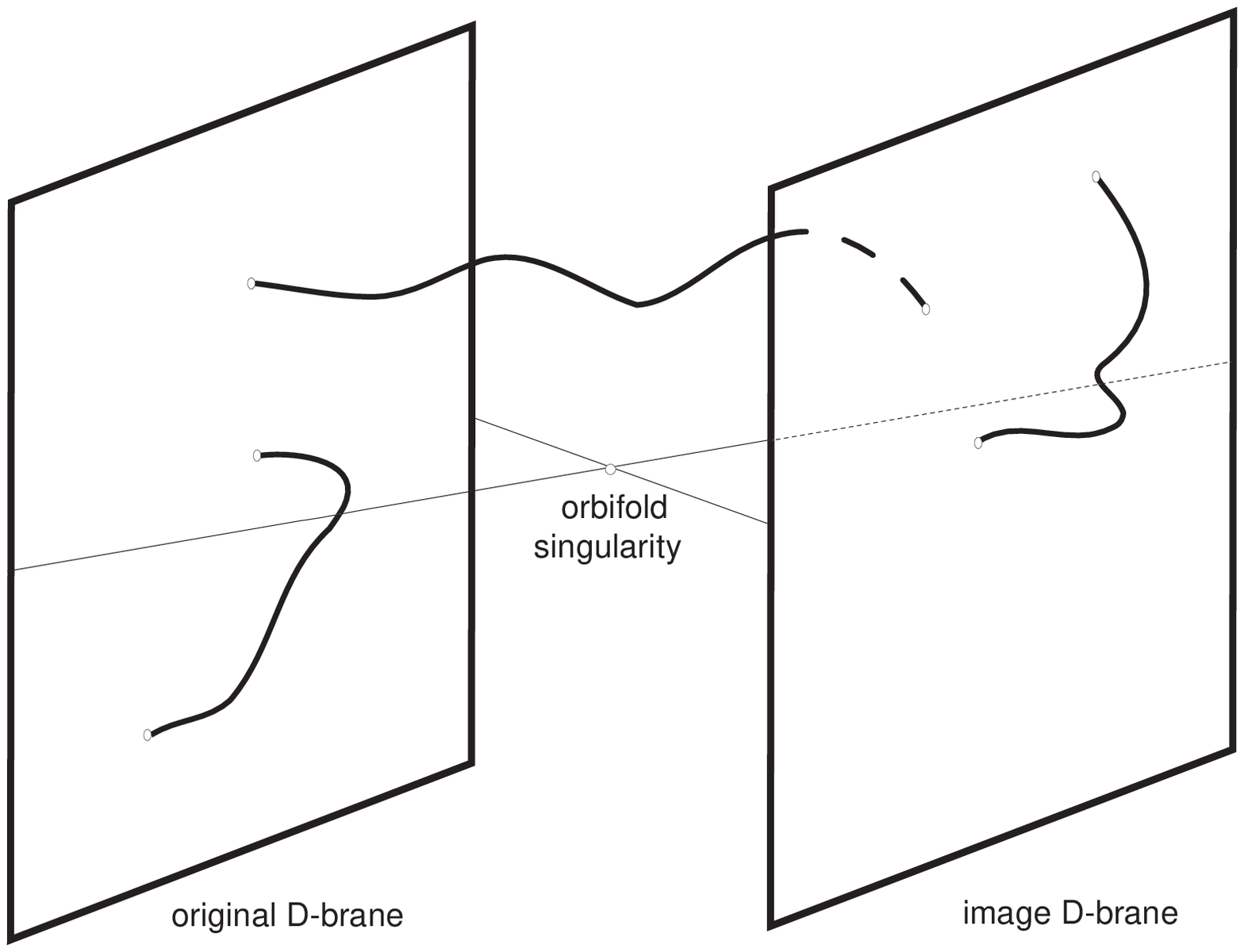,width=10cm}}
\caption{D--branes in presence of orbifold singularities and their image. There
are open strings whose end points are on the D3--brane, on its image, and on both. The resulting gauge group is $U(1) \times U(1)$}
\label{fig:orbifold}}
We need to introduce an accompanying image to obtain a $\IZ_2$--invariant
state. Therefore, there will be open strings whose end points are placed only
on the D3--brane, only on its image, and on both (see Fig.\ref{fig:orbifold}). Correspondingly, the massless spectrum of this orbifold is $\mathcal{N}=2$ supersymmetric $U(1) \times U(1)$ with matter multiplets in the bifundamental representation with charges $(+1,-1)$ and $(-1,+1)$. Let us call them, in
$\mathcal{N}=1$ notation, respectively $A_i$ and $B_i$. They are both doublets under the flavor (global) symmetry group $SU(2) \times SU(2)$. These theories
are called {\it quiver theories} and their field content is represented by diagrams such as Fig.\ref{fig:quiver}. The number of supersymmetries is compatible with the fact that $\IC^2/\IZ_2$ orbifolds arise in the singular limit of $K3$. If we take the number of D3--branes to be $N_c$, the resulting theory will be $\mathcal{N}=2$ supersymmetric $U(N_c) \times U(N_c)$ with
matter multiplets in the bifundamental with charges $(N_c, \bar N_c)$ and
$(\bar N_c, N_c)$. One can compute the one-loop $\beta$--function and, given
the fact that this theory has two $\CN=2$ vector multiplets and two $\CN=2$ matter hypermultiplets, it vanishes identically. The theory is conformal.

Consider now a D3--brane which is placed at the origin. There is no need to introduce an image. Such a brane is constrained to remain at the origin and,
as such, the corresponding representation of the orbifold group is one dimensional. There are two such representations, the trivial one and the
sign. The states that survive correspond to a single $\CN=2$ vector
multiplet: the scalars which parameterize the displacement away from the singularity and their superpartners are identically zero. In a sense, this
is one half of the D3--brane which sits outside the origin. For this
reason, it has been termed a {\it fractional brane}. It is immediate
to see that its field content provides a nontrivial contribution to
the $\beta$--function. Thus, when introducing fractional branes, we
are dealing with nonconformal field theories.

\FIGURE[ph]{\centerline{\epsfig{file=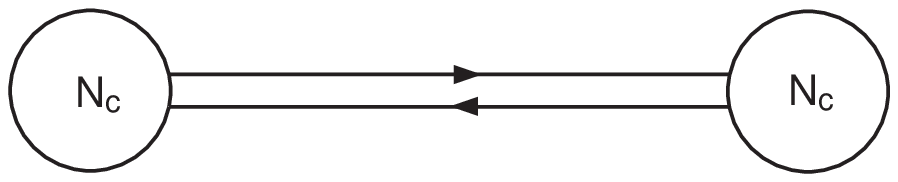,width=7cm}}
\caption{The quiver diagram representing a $SU(N_c) \times SU(N_c)$
supersymmetric gauge theory with adjoint vector fields and chiral
bifundamental multiplets in $(N_c,\bar N_c)$ and $(\bar N_c,N_c)$
representations.}
\label{fig:quiver}}

Let us make a couple of claims without discussing the details. It turns out
that the orbifold singularity can be seen as the end of a shrinking process
of a two cycle $\Sigma$: the inverse of a blowup. Indeed, a fractional brane
can be interpreted as a D(p+2)--brane with two directions wrapping the exceptional cycle $\Sigma$ in the orbifold limit, where its geometrical
volume vanishes. This would lead to a problem with the tensions if it wasn't
for the fact that the $B_{[2]}$ is such that $b = \int_\Sigma B = 1/2$ is
its only nonvanishing component and it is responsible for keeping the tension
finite. Further support to this interpretation comes from the fact that the fractional Dp--brane couples to both $B_{\mu\nu}$ and $C_{[p+3]}$, as a
D(p+2)--brane would do.

We would like to discuss in greater detail a case in which we have
$\CN=1$ supersymmetry. In order to do this we first need to
introduce the conifold.

\subsection{The Conifold}

\subsubsection{The Singular Conifold}

The {\it conifold}, $\CY_6$, is a six dimensional (three complex
dimensions) Calabi--Yau cone. It can be most easily introduced by
considering its embedding in $\IC^4$ given by the quadric
\be \CY_6
\equiv \sum_{A=1}^4 ~(z^A)^2 = 0 ~, ~~~~~~~ z^A = x^A + i y^A ~.
\label{quadric} \ee
This is a smooth surface everywhere except at
the point $z^A=0$ (which is a double point singularity: $\CY_6 = 0$,
$d\CY_6 = 0$). Notice, moreover, that if $z^A \in \CY_6$, then $\l
z^A \in \CY_6$ for any $\l \in \IC$. Thus, the surface is made up of
complex lines through the origin and, therefore, it is a cone. The
apex of the cone is precisely the point $z^A=0$. The base of the
cone, $\CX_5$, is given by the intersection of the quadric with a
sphere of radius $\rho$ in $\IC^4$, \be \CS_\rho \equiv \sum_{A=1}^4
~|z^A|^2 = \rho^2 ~. \label{sph} \ee In terms of $x^A$ and $y^A$,
this gives the following equations:
$$ \vec{x}\cdot\vec{x} = \half \rho^2 ~, ~~~~~ \vec{y}\cdot\vec{y} =
\half \rho^2 ~, ~~~~~ \vec{x}\cdot\vec{y}=0 ~. $$ The first equation
defines an $S^3$ while the other two define an $S^2$ fibered over
$S^3$. All such bundles are trivial, so the topology of the base of
the cone is $S^2 \times S^3$.

For some applications it is useful to define the matrix \be W =
\rhalf z^A \s_A =  \left(
\begin{array}{cc} u & z \\ w & v \end{array} \right) \equiv \rhalf \left(
\begin{array}{cc} z^3 + i z^4 & z^1 - i z^2 \\ z^1 + i z^2 & -z^3 +
i z^4 \end{array} \right) ~, \label{wmatrix} \ee where $\s_A =
(\s_k, i\II)$, $\s_k$ are the Pauli matrices.  In terms of $W$ the
quadric above can be immediately rewritten as \be \det W = u v - w z
= 0 ~. \label{quadbis} \ee
Notice that the sphere (\ref{sph}) can be cast in terms of $W$ as
\be Tr\, [W^{\dagger},W] = \rho^2 ~. \label{sphbis} \ee The
existence of a Ricci-flat K\"ahler metric on $\CY_6$ implies that
the base of the cone, $\CX_5$, admits an Einstein metric. The
simplest option (besides the trivial one given by $S^5$), and the only
one that was known until very recently, is $\CX_5 = T^{1,1}$
whose Einstein metric is \cite{cdlo}
\be
ds_5^2\,(T^{1,1}) = \frac{1}{9}\,\left( d\tilde{\psi} + \sum_{i=1}^2
\cos\t_i\,d\phi_i \right)^2 + \frac{1}{6} \sum_{i=1}^2 (d\t_i^2 +
\sin^2\t_i d\phi_i^2) ~,
\label{dsToo}
\ee
where $0 \leq \theta_i <\pi$, $0 \leq \phi_i < 2\pi$ and $0 \leq \tilde\psi
< 4\pi$. This is not the induced metric coming from the defining embedding
in $\IC^4$ (which is not, indeed, Ricci flat). The isometry group of
$T^{1,1}$ is $SU(2) \times SU(2) \times U(1)$. Moreover, the metric makes evident that $T^{1,1}$ is a $U(1)$ fibration over a K\"ahler--Einstein
manifold, $S^2 \times S^2$. It is compact and homogeneous,
\be
T^{1,1} = \frac{SU(2)\times SU(2)}{U(1)} ~.
\ee
The {\it singular conifold} (or simply conifold) metric is the metric of
a cone over $T^{1,1}$:
\be
ds_6^2(\CY_6)\,=\,d\rho^2\,+\,\rho^2\,ds_5^2\,(T^{1,1}) ~.
\label{metcfd}
\ee
We write in the Appendix a set of useful formulas that should clarify some
of the computations in the remaining of this section.

\subsubsection{The Deformed Conifold}

How can we {\it resolve} the singularity at the apex?. The most
natural way seems to add a non--vanishing r.h.s. to the quadric \be
\CY_6^\m \equiv \sum_{A=1}^4 (z^A)^2 = \mu^2 ~, \label{defcon} \ee
or, in terms of the matrix $W$, \be \det W = u v - w z = -\half \m^2
~. \label{defconw} \ee This is the {\it deformation of the
conifold}: at the apex, the $S^3$ remains finite while the $S^2$
shrinks. The appearance of $\mu \neq 0$ breaks a $U(1)$ invariance
$z^A \to e^{i\a} z^A$ to a discrete subgroup $\IZ_2$. This will be
interpreted as the realization of chiral symmetry breaking at the IR
as seen from geometry.

\subsubsection{The Resolved Conifold}

Another way of getting rid of the singular point is by the standard
blowup procedure. By definition, this amounts to considering the
same manifold and replacing a point, in this case the apex of the
cone, by a projective sphere. This is done as follows. Consider the
defining equation for the conifold (\ref{quadbis}). Now, replace
this equation by
\be
\begin{pmatrix} u & & z \cr w & & v \end{pmatrix} ~\begin{pmatrix}
\l_1 \cr \l_2 \end{pmatrix} = 0 ~,
\label{resconi}
\ee where
$\l_1,\l_2 \in \IC$ and they cannot both be zero. Only at the apex,
$\l_1/\l_2$ is unconstrained, so one has an entire ~$\IP^1 = S^2$.
This defines what is called the {\it small resolution of the conifold}: at
the apex, $S^2$ is finite while $S^3$ shrinks.

\FIGURE[ph]{\centerline{\epsfig{file=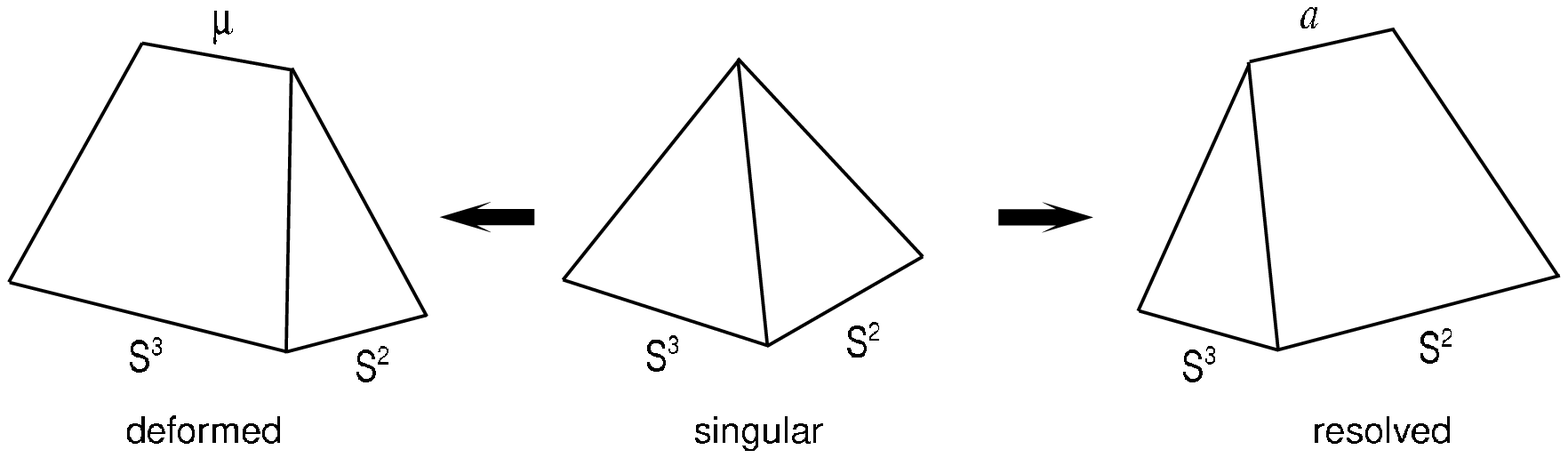,width=13cm}}
\caption{Different resolutions of the conifold correspond to either the $S^3$
(deformed) or the $S^2$ (resolved) being non-vanishing at the origin. The resolution parameter will appear as an additional parameter in the metrics
for these two cases.}
\label{fig:conifold}}

Ricci-flat K\"ahler metrics on these manifolds, compatible with $ds_6^2
(\CY_6)$ are known. They have the same asymptotics as the singular conifold
and differ only in the region near the tip of the cone (the IR of the dual
field theory). Both the metric of the conifold, as well as those of its resolution and its deformation, can be found by considering the uplift to
eleven dimensions of $D6$--branes wrapping a 2-cycle in a Calabi--Yau
manifold \cite{en,epr3}.

Let us finally comment on the fact that this manifold is also called $\CO(-1)
+ \CO(-1) \to \IP^1$. This means that it can be constructed by taking two
copies of $\IC^3$ with coordinates $(x,u,z)$ and $(x',u',z')$, identifying
$z$ (respectively $z'$) as the coordinate of the northern (southern)
hemisphere of the $\IP^1$, and finally gluing the two copies of $\IC^3$ with
the identification:
\be
z' = \frac{1}{z} ~, \qquad x' = x z ~, \qquad u' = u z ~.
\ee

\subsection{The Klebanov--Witten setup}

\vskip-7.06mm
\hskip70mm {\bf (dual to $\mathcal{N}=1$, $SU(N)\times
SU(N)$ superconformal Yang-Mills with bi-fundamental matter)}

~

\noindent
Consider now the addition of a stack of flat D3--branes at the apex
of the conifold. It is not difficult to write down the corresponding
supergravity solution \cite{kw}:
\be
ds_{10}^2 = h(r)^{-\frac{1}{2}} (dx_i)^2 + h(r)^{\frac{1}{2}} \left( dr^2 +
r^2 ds_{T^{1,1}}^2 \right) ~,
\label{kw}
\ee
where
\be
h(r) = 1 + \frac{L^4}{r^4} ~, \qquad L^4 = 4\pi g_s N (\alpha')^2 ~,
\ee
as
in the case of flat D3--branes presented in section 2 and the Einstein metric
on $T^{1,1}$ is given in the appendix. Clearly, the near horizon limit of
(\ref{kw}) is just given by the direct product $AdS_5 \times T^{1,1}$. This
is a particular example of the more general statement that the gravity dual
to the field theory living on D3--branes placed at a conifold singularity is $AdS_5\times X_5$, where $X_5$ is the Einstein manifold which is the base of
the cone \cite{Gubser:1998vd}. The $AdS_5$ factor in the metric is telling
us that the gauge theory has a conformal fixed point. The number of supersymmetries can be deduced from the holonomy of the background. It leads
to four supercharges which are doubled by the presence of a conformal
symmetry.

In summary, the resulting theory is $\CN=1$ superconformal gauge
theory. The $U(1)$ isometry corresponding to shifts in $\tilde\psi$
amounts to the $R$ symmetry while the $SU(2) \times SU(2)$ isometry
corresponds to a global flavor symmetry. If we look at the defining
equation of the conifold (\ref{quadbis}), we see that this equation
is automatically satisfied if we write \be u = A_1 B_1 ~, ~~~ v =
A_2 B_2 ~, ~~~ z = A_1 B_2 ~, ~~~ w = A_2 B_1 ~. \label{uvzw} \ee
The defining equation of the manifold is related to the moduli space
of the gauge theory. Thus, the theory can be thought of as having
constituents $A_i$ and $B_i$ (in presence of $N$ D3--branes they
have to be promoted to matrix-valued bifundamental chiral fields of
the gauge group $SU(N) \times SU(N)$ transforming, respectively, in
the $({\bf N}, {\bf \bar N})$ and $({\bf \bar N},{\bf N})$
representations. Besides, $A_i$ is a doublet of the first $SU(2)$
(and a singlet of the second) and, conversely, $B_i$ is the other
way around. The dual superconformal gauge theory can be represented
by a quiver diagram. The only ingredient that is still lacking is
the superpotential. Taking into account the fact that the
superpotential has $R$ charge equal to two and it is a singlet under
flavor symmetries, it has to be of the form \cite{kw}
\be
W = \epsilon^{ij} \epsilon^{kl} Tr A_i B_k A_j B_l ~.
\label{superkw}
\ee
Indeed, the $A_i$ and $B_i$ fields have $R$ charge equal to $\frac{1}{2}$,
conformal dimension $\frac{3}{4}$ and, respectively, baryon number equal
to $+1$ and $-1$.

\subsection{Klebanov--Tseytlin solution}

\vskip-7.2mm
\hskip67mm {\bf (the UV of $\mathcal{N}=1$, $SU(N+M)\times SU(N)$
supersymmetric Yang--Mills theory with bi-fundamental matter)}

~

\noindent
If we add fractional branes into the picture, conformal symmetry will be immediately spoiled. The solution to this system was originally developed
in Refs.\cite{kn,kt}. Let us consider a set up in which, in addition to
the metric and a RR 5-form, we switch on the RR 3-form $F_3$. We keep the
metric ansatz
\be
ds_{10}^2 = h(r)^{-\frac{1}{2}} (dx_i)^2 + h(r)^{\frac{1}{2}} \left( dr^2
+ r^2 ds_{T^{1,1}}^2 \right) ~,
\label{kt1}
\ee where $h(r)$ is now a function to
be determined.  In order to make an ansatz for $F_3$, we can make
use of the closed 3-form of the conifold $w_3$ (see Appendix B). We
can make the following ansatz, \be F_3 = f w_3 ~, \label{ansatzF3}
\ee where $f$ still has to be determined. We will fix $f$ by means
of the quantization condition corresponding to a stack of D5-branes.
Taking $\star F^{(7)} = F_3$, we have \be \int_{S^3} F_3 = 4 \pi^2
\alpha^\prime ~ M ~, ~~~~~ M \in  \IZ ~, \label{qconditD5} \ee $M$
being the number of D5-branes and $S^3$ the sphere which is given by
the set of $T^{1,1}$ with $\t_2 = \p_2 = 0$. These D5-branes are
wrapped on a (supersymmetric) 2-cycle and, thus, we are going to
have $N$ D3-branes and $M$ fractional D3-branes. From equations
(\ref{ansatzF3}) and (\ref{qconditD5}), we get $f = \frac{M}{2}
\alpha^\prime$, and \be F_3 = \frac{M}{2} \alpha^\prime ~w_3 ~,
\label{actualF3} \ee First of all, it is clear that $d F_3 = 0$ is
automatic. Since $d F_5 \neq 0$, we must also switch on a NSNS
3-form $H_3 = dB$. In order to construct the appropriate NSNS 2-form
potential $B$, we make use of the closed 2-form of the conifold
$w_2$. Let us take \be B = T(r)\, w_2 ~, ~~~ \Rightarrow ~~~ H_3 =
T'(r)\, dr \we w_2 ~, \label{ansatzB} \ee where $T(r)$ is to be
determined. Plugging in these expressions into $dF_5 = H_3 \wedge
F_3$, we can write $T(r)$ as \be (r^5 h'(r))' = - 27 M T'(r)
\alpha^\prime g_s ~. \ee Now, given that $F_1 = 0$ in this
background, the dilaton equation of motion reads \be \nabla^2 \Phi =
\frac{g_s}{12} \left( e^\Phi F_3^2 - e^{-\Phi} H_3^2 \right) ~. \ee
Thus, for a constant dilaton $e^\Phi = g_s$, we must have \be g_s^2
F_3^2 = H_3^2 ~. \ee From this equation we can derive an explicit
expression for $T'(r)$ such that \be H_3 = \frac{3 \gs M}{2r}
\alpha^\prime ~dr \we w_2 \,,\qquad B_2 = \frac{3 \gs M}{2}
\log\frac{r}{r_0} \alpha^\prime  w_2 ~. \label{actualH3} \ee Then
\be H_3 \we F_3 = \frac{3 \gs M^2}{4r} (\alpha^\prime)^2 ~dr \we w_2
\we w_3 ~. \label{H3F3} \ee Using Eq.(\ref{w32}), we can readily see
that \be F_3^2 = \frac{3^5 M^2}{r^6} (\alpha^\prime)^2 h^{-3/2} ~.
\label{F32} \ee Using this solution for $T'(r)$, we can compute
$h(r)$. A first integral of the equation above is easy to perform,
\be r^5 h'(r) = - \frac{81}{2} M^2 g_s^2 (\alpha^\prime)^2 \log r +
{\it const} ~. \ee This equation can be integrated to be \be h(r) =
\frac{81 M^2 g_s^2 (\alpha^\prime)^2}{2} \left( \frac{\log r}{4 r^4}
+ \frac{1}{16 r^4} \right) + \frac{\it const}{r^4} + {\it const'}
\ee In the near-horizon limit we can drop the additive constant.
Thus \be h(r) = \frac{27 \pi (\alpha^\prime)^2}{4 r^4} \left( g_s N
+ a (g_s M)^2 \log\frac{r}{r_0} + \frac{a}{4} (g_s M)^2 \right) ~,
\ee with $a = \frac{3}{2\pi}$. It is convenient to define at this
point \be N_{eff} = N + \frac{3}{2\pi} \gs M^2 \log \frac{r}{r_0} ~,
\label{Neff} \ee such that, for example, \be r^5 h'(r) = - 27 \pi
(\alpha^\prime)^2 g_s N_{eff} ~, \ee and the RR 5-form becomes \be
g_s F_5 = 27 \pi (\alpha^\prime)^2 g_s N_{eff} dVol(T^{1,1}) + {\rm
Hodge dual} ~. \ee From this formula it can be deduced \be
\int_{T^{1,1}} F_5 = (4 \pi^2 \alpha^\prime)^2 N_{eff} ~.
\label{fluxF5} \ee Notice that $N_{eff}$ is a {\it logarithmically
running number of colors}.

\subsubsection{The Running of $N_{eff}$ and  Seiberg Duality}

The gauge couplings of the two gauge groups are given by \cite{kw}
\be
\frac{1}{g_1^2} + \frac{1}{g_2^2}\approx e^{-\Phi} ~, \qquad
\frac{1}{g_1^2} - \frac{1}{g_2^2}\approx e^{-\Phi} \left( \int_{S^2}
B_2 - \frac{1}{2} \right) ~,
\label{rungcoup}
\ee
and we recall that we are considering a constant dilaton $\Phi$. Equation
(\ref{actualH3}) then shows that $1/g_1^2$ and $1/g_2^2$ run logarithmically
in different directions. If we think of the RG flow in the $SU(N+M) \times
SU(N)$ theory, it is not difficult to see that there exists a scale (value
of $r$, say, $\Lambda_{N+M}$) at which the gauge coupling corresponding to
the first factor, $g_1$, diverges. To go beyond this scale we must perform a Seiberg duality. This is an S--duality statement telling us that an $\CN = 1$ supersymmetric gauge theory with gauge group $SU(N_c)$ and $N_f$ flavors,
when strongly coupled, is dual to a weakly coupled $SU(N_f - N_c)$ theory
with $N_f$ flavors, provided $\frac{3}{2} N_c < N_f < 3 N_c$. Thus, at a
scale given by $\Lambda_{N+M}$, given that $N_c = N+M$ and $N_f = 2N$ for
the first factor, $SU(N+M) \to SU(N-M)$ under Seiberg duality. The solution
we presented above has a cascade of such dualities \footnote{This point is
exposed in this set of lectures avoiding some nice details. The reader may
quench his/her thirst for a more accurate explanation on these topics in a masterly review written by Strassler \cite{strass}.}
\be
SU(N+M) \times SU(N) \rightarrow SU(N) \times SU(N-M) \rightarrow SU(N-M)
\times SU(N-2M) \rightarrow \dots
\label{cascade}
\ee
in which, at decreasing energy scales, the coupling of the first gauge group diverges, and we can think of performing a Seiberg duality on that factor.

The cascade of Seiberg dualities is encoded in the supergravity background
that we have presented above in the logarithmic running of $N_{eff}$
(\ref{Neff}).  There becomes a point where the effective number of colors vanishes and the solution can be shown to be singular. The way to obtain a
well behaved supergravity solution valid in the infrared was presented in
\cite{ks}, and is related to the deformed conifold which we presented in
section 4.1.2. We come to this point in what follows.

\subsection{The Klebanov--Strassler setup}

\vskip-7.7mm
\hskip73mm {\bf (dual to $\mathcal{N}=1$, $SU(N+M)\times SU(N)$
supersymmetric Yang-Mills theory with bi-fundamental matter)}

~

\noindent
We would like to deform the conifold such that the RG flow leads to
a duality cascade that ends at the IR. Let us consider an ansatz for
the 10d line element similar to the ones above \be ds^2 =
h^{-1/2}(\tau) dx_{1,3}^2 + h^{1/2}(\tau) ds_6^2 ~, \label{linel}
\ee where now however $ds_6^2$ is the metric of the deformed
conifold (see the Appendix for further details),
\ba ds_6^2 & = &
\frac{1}{2} \e^{4/3} K(\tau) \bigg[ \frac{1}{3 K(\tau)^2} (d\tau^2 +
(g^5)^2) + \sinh^2(\tau/2) ((g^1)^2 + (g^2)^2) \cr &&
~~~~~~~~~~~~~~~~~~~~~~+ \cosh^2(\tau/2)\, ((g^3)^2 + (g^4)^2) \bigg]
~, \label{cydef}
\ea the function $K(\tau)$ being given by the
following expression \be K(\tau) = \frac{(\sinh 2\tau -
2\tau)^{1/3}}{2^{1/3} \sinh \tau} ~. \label{ktau} \ee Notice that
this function has a finite nonvanishing limiting value when $\tau$
goes to zero, \be \lim_{\tau \to 0} K(\tau) = \left( \frac{2}{3}
\right)^{1/3} ~. \ee For this reason, it is clear that there is a
$S^2$ (spanned by $g^1$ and $g^2$) that shrinks to zero volume at
$\tau = 0$ while there is a finite $S^3$.

This finite $S^3$ presents one argument in favor of this idea.  The
divergence in the KT solution could be shown to arise from an
infinite energy density in $F_3^2$, due to the fact that as the flux
of $F_3$ through the $S_3$ is constant and equal to $M$, when the
$S^3$ collapses, $F_3$ must diverge.  A more rigorous justification
involves studying the moduli space of the field theory and showing
that it is in fact given by the deformed conifold (\ref{defconw})
when the nonperturbative contributions to the superpotential are
considered \cite{Affleck:1983mk}.

The derivation of the solution is involved and follows similar steps
to the ones presented above for the KT case.  The solution is more
complicated and we present it now.

The fields are
\ba \gs F_5 & = & d^4x \we dh^{-1} + {\rm Hodge~dual}
=  \frac{\gs^2 M^2 (\alpha^\prime)^2}{4} \ell(\tau) g^1 \we \dots
\we g^5 + {\rm Hodge~dual} \nn \\ F_3 & = & \frac{M
\alpha^\prime}{2} \left[ (1 - F) g^3 \we g^4 \we g^5 + F g^1 \we g^2
\we g^5 + F' d\tau \we (g^1 \we g^3 + g^2 \we g^4) \right] \nn \\ B
& = & \frac{g_s M \alpha^\prime}{2} \left( f(\tau) g^1 \wedge g^2 +
K(\tau) g^3 \wedge g^4 \right) \nn \\ H_3 & = & \frac{\gs M
\alpha^\prime}{2} \left[ d\tau \we (f' g^1 \we g^2 + K' g^3 \we g^4)
+ \half (K - f) g^5 \we (g^1 \we g^3 + g^2 \we g^4) \right]\,.
\ea
The equations of motion can be solved and the above functions
are found to be
\ba F(\tau) & = & \frac{\sinh\tau -\tau}{2\sinh\tau} \,, \nn \\
h(\tau) & = & \frac{2^{2/3}\alpha}{4} \int_\tau^\infty dx
\frac{x\coth x - 1}{\sinh^2x} (\sinh 2x - 2x)^{1/3} \,, \nn \\
f(\tau) & = & \frac{\tau \coth\tau -1}{2\sinh\tau}(\cosh\tau -1) \,,
\nn \\
l(\tau) & = & \frac{\tau \coth\tau -1}{4\sinh^2 \tau}(\sinh 2\tau
-2\tau)\,, \ea where \be \alpha = 4 g_s^2 M^2 (\alpha^\prime)^2
\e^{- 8/3} \,. \ee This solution has the same UV (large $r$)
behavior as the KT solution presented above, but is non-singular in
the IR (small $r$).
\newpage

\section{Lecture V: Nonperturbative phenomena as seen from supergravity}

Let us end these lectures by providing a summary of several
nonperturbative phenomena in $\CN=1$ supersymmetric gauge theory as
seen from the supergravity side of the duality. We will go over
those phenomena presented in lecture I and discuss the way this
information is encoded in the dual supergravity description. We will
mostly work within the framework given by the MN solution in type
IIB.

\subsection{Chiral symmetry breaking}

Chiral symmetry breaking is the phenomenom by which the $U(1)_A$
mentioned in (\ref{symqcd}) gets broken to a discrete subgroup: \be
U(1)_A \rightarrow \IZ_{2N_c} \,.\ee In field theory this is known
to be a UV effect caused by instantons.

We know from probe brane analysis the specific relations between the gauge
coupling and $\theta$--angle in the gauge theory and supergravity fields.
In the case of $N_c$ D5--branes wrapping a two--cycle $\Sigma$ inside a
Calabi--Yau, they read:
\ba
\frac{4\pi}{\gym^2} & = & \frac{1}{(2\pi l_s)^2 g_s} \int_{\Sigma} d^2\xi\
e^{-\Phi} \sqrt{\det G} ~,\\
\frac{\tym}{2\pi} & = & \frac{1}{(2\pi l_s)^2 g_s} \int_{\Sigma}
C_{[2]} ~,
\ea
where $G$ and $C_{[2]}$ are space-time fields restricted to the cycle. The
two-cycle is non-trivially embedded in the geometry \cite{bm}. Using the
formulas corresponding to the MN background, we can explicitly compute
\ba
\frac{4\pi}{\gym^2} & = & \frac{N_c}{4\pi} \left(e^{2h(\rho)} + (a(\rho)
-1)^2\right) = \frac{N_c}{\pi}\ \rho \tanh \rho ~,\\
\frac{\tym}{2\pi} & = & \frac{N_c}{2\pi} ~\psi_0 ~,
\label{gauth2}
\ea
where $\psi_0$ is an integration constant that, due to the embedding,
gets fixed to $0$ or $2\pi$.

The $U(1)$ $R$--symmetry in the gauge theory has to do with shifts
in the angular variable $\psi$ in supergravity. At the UV, we can
approach the solution by the Abelian (singular) one because $a(\rho)
\to 0$ when $\rho \to \infty$. Then, using the value of $C_{[2]}$ in the
Abelian solution,
\be
\frac{\tym}{2\pi} = \frac{1}{(2\pi l_s)^2 g_s} \int_{\Sigma} C_{[2]}
\approx \frac{N_c}{2\pi} (\psi + \psi_0) ~,
\label{invun1}
\ee
which is clearly invariant under ($\tym \to \tym + 2\pi k$)
\be
\psi \to \psi + \frac{2\pi k}{N_c} ~.
\label{invun2}
\ee Since an $R$--symmetry transformation of parameter $\epsilon$
changes the $\theta$--angle by $\tym \to \tym - 2 N_c \epsilon$, it
translates to $\psi \to \psi + 2\epsilon$. This means that the
remaining UV symmetry corresponds to $U(1)_R \to \IZ_{2N_c}$ which
is a well-known instanton effect in the gauge theory: this is
nothing other than the chiral anomaly, a one--loop effect that is an
UV phenomenon. The holographic dual of this feature is well
understood \cite{kow}. The value of $\tym$ in the complete solution
tells us that $\IZ_{2N_c} \to \IZ_{2}$ in the IR, as can be seen,
for example, in equation (\ref{gauth2}). Notice that, in a sense, it
is the non-vanishing value of $a(\rho)$ which is somehow responsible
for encoding the IR effects of the gauge theory dual. Let us discuss
this aspect in greater detail in the following section.

Before proceeding though, let us mention the effect in the gravity
picture which is dual to chiral symmetry breaking in the field
theory.  This was first studied by \cite{kow} in the KS background.
Fluctuations $\psi\rightarrow \psi + \lambda$ about the $U(1)$
isometric direction of the metric with coordinate $\psi$ where
described by a vector in terms of a vector field $A_i$.  Even though
the direction is an isometry, when taking into account the
contribution from the RR potentials, it was found that the effective
action for the fluctuation corresponds to the action of a {\it
massive} (and gauge invariant) vector field $W_i = A_i +
\partial_i \lambda$.  These fluctuations are dual to the chiral
symmetry current, and hence the anomaly in the field theory arises
as a Higgs' effect in the gravity dual.  Correlators involving the
chiral current in this picture were studied in
\cite{Krasnitz:2000ir,Krasnitz:2002ct}.  The effect was also studied
in the IIA gravity duals and their M theory uplifts
\cite{Gursoy:2003hf}.  The subtlety in this context involves the
fact that as already mentioned, the IIA gravity duals lift to pure
geometry in M theory, therefore fluctuations dual to the chiral
current can no longer be around a $U(1)$ isometry.

More on chiral symmetry breaking in the context of AdS/CFT as applied to
confining theories can be found in \cite{BEEGK,BdVFLM,BdVFLMi}.

\subsection{Gaugino Condensation}

We have seen that the supergravity solution nicely accounts for the
chiral symmetry breaking phenomenon. The IR effects can be traced to
the appearance of the function $a(\rho)$ in the solution. In the
gauge theory side we know that the IR breaking $\IZ_{2N_c} \to
\IZ_{2}$ is a consequence of gaugino condensation. Thus, it seems
natural to identify $<\l\l>$ with $a(\rho)$ in a way that we would
explore in what follows. Let us first give further arguments to
support this claim. From the second order equation that is obeyed by
the funcion $a(\rho)$, \be \left( e^{-4h(\rho) - 2\Phi(\rho)}
a'(\rho) \right)' = e^{2h(\rho) - 6\Phi(\rho)} ~a (\rho)~(a(\rho)^2
- 1) ~, \ee we can extract the asymptotic behavior of the solution.
Following the discussion in the original Maldacena's setup, we need
the asymptotic value of $a(\rho)$ for both linearly independent
solutions, \be a(\rho) \to a_{d}(\rho) = \frac{1}{\sqrt{\rho}} ~,
~~~~~{\rm and}~~~~~ a(\rho) \to a_{sd}(\rho) = 2 \rho\, e^{-2\rho}
~. \label{arhoes} \ee where the subindices mean, respectively, {\it
dominant} and {\it sub-dominant}. The former is a normalizable
solution while the latter is non-normalizable. If we blindly apply
the AdS/CFT and holographic RG flow prescriptions, we can see that
the dominant solution corresponds to the insertion of the operator
$<\bar\l\l>$. This is a mass term for the gaugino which breaks
supersymmetry. In turn, the sub-dominant solution --which actually
corresponds to the behavior in the MN background--, is giving a
vacuum expectation value to the operator $<\l\l>$. Thus, the gaugino
condensate is related to $a(\rho)$ \cite{ABCPZ}.

The gaugino condensate is a protected operator in the gauge theory,
which is related to the dynamical scale via $\langle \Tr
~\lambda\lambda \rangle \approx \L_{QCD}^3$. Notice that, due to
this issue, the identification between $a(\rho)$ and the gaugino
condensate should be an exact equation. We still need to introduce
the subtraction scale $\mu$ of the gauge theory. Taking into account
mass units and the precedent arguments, it seems natural to identify
\cite{dVLM}
\be
\mu^3\ a(\rho) = \L_{QCD}^3 ~.
\label{gauco1}
\ee
Namely,
\be
\frac{\L_{QCD}^3}{\mu^3} = \frac{2\rho}{\sinh 2\rho} ~.
\label{gauco2}
\ee
This relation gives implicitly the {\it energy/radius relation} between supergravity coordinates and gauge theory scales. A final comment is in
order. The function $a(\rho)$ is responsible for regularizing the otherwise singular Abelian MN background. It is interesting to point out that this relation between the last step of chiral symmetry breaking and space-time
singularity resolution is also present in Klebanov-Strassler's setup
\cite{ks}. There, gaugino condensation and chiral symmetry breaking arise
as a consequence of the deformation of the conifold singularity.

\subsection{The Beta Function}

Recall the definition of the $\beta$--function. It encodes the
variation of the gauge coupling constant with energy, \be
\beta(\gym) = \frac{\partial \gym}{\partial \ln
\frac{\mu}{\L_{QCD}}} ~. \label{bf0} \ee Taking into account the
radial dependence of all quantities involved in this expression, we
can write \be \beta(\gym) = \frac{\partial \gym}{\partial \rho}
\frac{\partial \rho}{\partial \ln \frac{\mu}{\L_{QCD}}} ~.
\label{bf1} \ee Let us first use the asymptotic behavior, neglecting
subleading exponential corrections. The leading contribution to the
Yang--Mills coupling is linear, \be \frac{4\pi}{\gym^2} \approx
\frac{N_c}{\pi} ~\rho ~. \label{bf2} \ee We therefore get: \be
\frac{\partial \gym}{\partial \rho} \approx \frac{\pi}{\sqrt{N_c}}
\rho^{-3/2} \approx - \frac{N_c \gym^3}{8\pi^2} ~. \label{bf3} \ee
On the other hand, we can compute from equation (\ref{bf1}), \be
\frac{\partial \rho}{\partial \ln \frac{\mu}{\L_{QCD}}} \approx
\frac{3}{2} \left( 1 - \frac{1}{2\rho} \right)^{-1} \approx
\frac{3}{2} \left( 1 - \frac{N_c\gym^2}{8\pi^2} \right)^{-1} ~, \ee
where we used the asymptotic relation in the last step to trade
$\rho$ for $\gym$. Putting together these results, we obtain
\be
\beta(\gym) = - \frac{3N_c\gym^2}{16\pi^2} \left( 1 -
\frac{N_c\gym^2}{8\pi^2} \right)^{-1} ~,
\label{bnsvz}
\ee
which is the correct NSVZ $\beta$-function at all-loops computed in the
Pauli--Villars renormalization scheme for $\mathcal{N}=1$ supersymmetric
Yang--Mills theory \cite{nsvz}.

We can now work out the calculation using the complete solution,
this resulting in the following expression for the $\beta$--function
\cite{dVLM,bm,mer},
\be
\beta(\gym) = -\frac{3N_c\gym^3}{16\pi^2} \left( 1 - \frac{N_c\gym}{8\pi^2}
+ \frac{2e^{-16\pi^2/N_c\gym^2}}{1-e^{-16\pi^2/N_c\gym^2}} \right)^{-1} ~. \label{fullbeta}
\ee
The last contribution seems to be originated in non-perturbative {\it
fractional instantons} contributions to the running of the coupling. It is
not yet understood what the origin of these configurations might be, and
this formula for the $\beta$--function is still under debate. \footnote{In
a recent paper \cite{CaNuPa}, by studying a family of deformations of the
MN setup, it is suggested that the gravity computation is not trustable
after an appropriate cut-off related to $\Lambda_{\rm UV}$, this implying
that the full integration using the complete solution that gives raise to
the expression (\ref{fullbeta}) is possibly meaningless.} On the other
hand, the result must be taken with care as it is unclear how to decouple
the four-dimensional dynamics Kaluza--Klein modes on $\Sigma$ within the
region of validity of supergravity.

\subsection{Confinement and Screening of Magnetic Monopoles}

A fundamental charge is described by a fundamental string extended
along one of the gauge theory directions, say, $z$. The
quark-antiquark potential $V_{q\bar q}$, then, is given by its
Euclidean Nambu--Goto action, \be S_{NG} =
\frac{1}{2\pi\alpha^\prime} \int d\tau d\sigma \sqrt{-{\rm det}\,
g_{ab}} ~, \label{ngaction} \ee where $g_{ab}$ is the pull-back of
the string frame metric on the worldvolume of the string. It is an
immediate consequence of the form of the metric in the MN solution
that the string will prefer to stretch out sitting at $r=0$, where
the value of $e^{\phi}$ is the least. Thus, \be S_{NG} =
\frac{e^{\phi_0}}{2\pi\alpha^\prime}\int dx \,dt \qquad \Rightarrow
\qquad T_{q\bar q} = \frac{e^{\phi_0}}{2\pi\alpha^\prime} \neq 0 ~.
\label{ngf1} \ee The string tension does not vanish and the theory
is confining. If the same computation is performed in the $AdS_5
\times S^5$ background, the result is zero, this reflecting the
non-confining nature of $\CN = 4$ SYM theory. We will discuss
confining strings in more detail in the next section. Let us
conclude the present one by computing the monopole-antimonopole
potential. Magnetic monopole sources correspond to D3--branes
wrapping $S^2$ and extending along the radial direction. The
potential $V_{m\bar m}$ then corresponds to the energy of a D3--brane
wrapping $S^2$ and extending along a gauge theory direction. The D3--brane worldvolume coordinates are $(\tau,\sigma,\Theta,\Phi)$ and the embedding
into the background geometry reads $t = \tau$, $x = \sigma$, $\theta = \Theta$,
$\phi = \Phi$, and $r = r(\sigma)$. Then,
\ba
S_{NG} & = & \frac{1}{2\pi\alpha^\prime} \int d\tau\, d\sigma\, d\Theta\,
d\Phi\, \sqrt{\det (-g)} \cr
& = & 4 \pi \gs N_c \int d\sigma\, e^{2\phi(r(\sigma))} \bigg(
e^{2h(r(\sigma))} + \frac{(a(r(\sigma))-1)^2}{4} \bigg) \sqrt{1 +
\alpha^\prime \gs N_c (r'(\sigma))^2} ~.
\label{monop}
\ea
But, as we have seen, the $S^2$ shrinks in the IR ($r \to
0$), this implying that the corresponding wrapped D3--branes become
{\it tensionless},
\be
T_{m \bar m} = 0 ~;
\label{tensmon}
\ee
{\it Magnetic monopoles are screened, not confined}.

\subsection{Confining Strings}

Let us discuss in greater detail the tension of confining strings.
The natural candidate for a confining string in the gravity dual is
a fundamental string.  So we will study the energetics of $k$
infinite fundamental strings extending in the "physical" spacetime
directions in the gravity duals. Confinement takes place at the IR,
where the $S^2$ shrinks, and the geometry is essentially given by
the blown-up $S^3$ times Minkowski four-space, which we called
"physical" spacetime. The metric of the $S^3$ is \be ds_{\tS^3}^2 =
c\, \gs N_c \alpha^\prime \left( d\psi^2 + \sin^2\psi (d\t^2 +
\sin^2\t d\phi^2) \right) ~, \label{blup} \ee where $c=1$ in the MN
solution and $c \approx .933$ for KS. A salient feature of these
solutions is the existence of $N_c$ units of RR 3-form flux through
this $S^3$,
\be
F_{[3]} = dC_{[2]} = 2 N_c \alpha^\prime \sin^2\psi \sin\t\, d\psi \we d\t
\we d\phi ~,
\label{rrfthree}
\ee
with $\psi \in [0,\pi]$ the azimuthal angle. Fundamental strings couple to
the RR potential $C_{[2]}$ through the Wess-Zumino term in their action. This
coupling causes the strings to expand into a D3--brane wrapped over an $S^2$
at fixed azimuthal angle $\psi$ on the $S^3$ via the so-called Myers' effect \cite{Myers:1999ps}.  In fact, it is energetically favorable for the strings
to expand in this way. The tension $T_k$ of the $k$-strings (they are still strings as viewed from the "physical" spacetime) is given by the linear
energy density of the brane:
\be
T \sim \sqrt{c^2 \sin^4\psi + \left( \psi - \half \sin 2\psi -
\frac{k\pi}{N_c} \right)^2} ~.
\label{kstension}
\ee
Minimizing with respect to $\psi$, we obtain
\be
\psi - \frac{k\pi}{N_c} = \frac{1 - c^2}{2} \sin 2\psi ~.
\label{minpsi}
\ee
Notice that $c$ is either one or close to it. If $c=1$, we get
\be
T_k \sim \sin \left( \frac{k\pi}{N_c} \right) ~.
\label{dshformula}
\ee
This formula was first derived in this context in \cite{hk2} that appears in several field theory approaches to QCD as softly broken $\mathcal{N}=2$ supersymmetric gauge theory \cite{ds} and MQCD \cite{hsz}.  Notice that it is
invariant under the symmetry $k\rightarrow N_c - k$ as required by charge conjugation symmetry. It is possible in the gravity dual to study finite
length $k$-strings, using the same method of probe branes \cite{HaPo}, and derive finite-length corrections to the sine formula for the tension
(\ref{dshformula}). Notice that this formula is also inaccurate in the KS
setup. Furthermore, a deformation of the MN configuration has been considered
very recently \cite{CaNuPa} that somehow captures information of the KK sector,
and it is rewarding to confirm that the sine formula does not hold, precisely
matching the field theory expectations \cite{EGFMMG}. Finally, let us mention
for completeness that confining strings were also discussed in the framework
of M--theory in $G_2$ manifolds \cite{Bobby2}.

\subsection{Domain Walls}

In $\mathcal{N} = 1$ SYM theory there are no central charges \cite{hls}. Consequently, particles cannot be BPS saturated. However, if the Lorentz symmetry $SO(3,1)$ is broken as it would be, for example, in the case of
domain walls (SYM theory does not have a central charge for the QCD string),
a central extension might be induced upon compactification \cite{DvSh}. Thus,
a BPS domain wall can be realized that interpolates two vacua with different values of a non-perturbative condensate (for example the gluino or the
monopole/dyon condensate). In this perspective, a domain wall can be thought
of as a sort of homogeneous distribution of baryon vertices connected by fundamental strings \cite{witten}. Thus, we expect the domain wall to be an extended object where the confining strings can end, and given the association of the latter with the fundamental string, we can argue that a domain wall corresponds to a D5--brane wrapping $S^3$ \cite{witten}.

Another piece of evidence comes from the fact that these domain walls are
known to be BPS objects preserving one half of the supersymmetries and, furthermore, the tension of these domain walls in the large $N_c$ limit
scales differently to that of a soliton in an effective glueball theory.
Thus, if $k$ domain walls sit on top of each other we would expect a $2+1$ dimensional $U(k)$ gauge theory on their worldvolume, $\CM^6 \sim \IR^3
\times S^3$. Let us write the worldvolume action for these D5--branes. It
is given by (\ref{SwvDp})
\be
S_{D5} = \tau_5 \int_{\CM^6}  \Big( dt\, d^2x\, d\Omega_3\; e^{-\phi}
\sqrt{\det \left(-g + 2\pi\alpha^\prime F \right)} + (2\pi\alpha^\prime)^2
\; C_{[2]} \we F \we F \Big) ~,
\label{wvactd5}
\ee
where $F$ is the world-volume gauge field strength and the last term is the Wess--Zumino contribution arising from the non-vanishing $C_{[2]}$ field in
the MN solution (\ref{RR3f}). Expanding in powers of $\alpha^\prime$ and
computing both terms in (\ref{wvactd5}) explicitly we have, besides the
usual Yang--Mills and scalar terms (\ref{dbiym}) --and, of course, their
supersymmetric partners-- \footnote{The theory on the domain wall has $\CN =
1$ supersymmetry in three dimensions, {\it i.e.} two supercharges, and
the field content is organized accordingly in vector ($A,\lambda$) and
scalar ($\Phi,\Psi$) multiplets \cite{av}.}
\be
S_{D5} \sim \frac{1}{(2\pi\alpha^\prime)^2}\, \int_{S^3} d\Omega_3\;
\int_{\IR^3} dt\, d^2x\; e^{-\phi} \sqrt{\det \left( -g \right)} +
\int_{S^3} F_{[3]}\; \int_{\IR^3} A \we F + \cdots ~,
\label{actdw}
\ee
where we split the integral taking into account the natural factorization
in the wrapped and flat parts of the worldvolume of the D5--branes. Also, the dots amount to the previously alluded terms and quantities are evaluated at
$\rho = 0$, where the volume of the wrapped $S^3$ is minimum. The first term
gives the tension of the domain wall \cite{Loewy:2001pq},
\be
{\rm T}_{\rm domain ~wall} = \left( \alpha^\prime \right)^{-\frac{3}{2}}\,
N_c\; \left( \gs N_c \right)^{-\frac{1}{2}} ~.
\label{dwtens}
\ee
On the other hand, recalling that the $F_{[3]}$ form has $N_c$ units of flux threading the sphere, the latter term in (\ref{actdw}) gives nothing but a
level $N_c$ {\it Chern--Simons} gauge theory dynamics induced on the domain
wall \cite{av}.

It is also possible to show that if we cross a domain wall built out of
$k$ wrapped D5--branes, the corresponding shift (\ref{invun2}) takes place
due to a change in the flux of the $C_{[2]}$ field on the sphere. This
indicates that we have tunneled from a given (say, the $n$-th) vacuum, to
the ($n+k$)-th one. Let us close this subsection by stating, without further
comments, that in the M--theory approach a domain wall corresponds to an
M5--brane wrapping on $S^3/\IZ_N$.

\subsection{Wilson Loops}

Consider a Wilson loop (\ref{wilson1}) with a rectangle contour $\CC$. A
natural prescription to compute this quantity in string theory, since the
fundamental strings is associated to the QCD--string, would be to sum over
all world-sheets having $\CC$ as a boundary. The leading contribution then
implies the identification of the Wilson loop in the Maldacena conjecture as
\be
<W[\CC]> \sim e^{- S_{NG}} ~,
\label{wilson3}
\ee
where $S_{NG}$ is the Nambu-Goto action for a fundamental string going from
one quark to the other \cite{Rey:1998ik,maldaw}. The general analysis was performed in \cite{Kinar:1998vq}. Consider a 10d background
\be
ds^2 = - g_{tt}(r) dt^2 + g_{xx}(r) d\vec x^2 + g_{rr}(r) dr^2 +
g_{ij}(r) d\theta^i d\theta^j ~,
\ee
where the spatial coordinates on the D3--brane are the $x^a$ and the
internal compact coordinates are $\theta^i$.  Let us consider the quark
anti-quark to be separated in the $x$ direction. If we parameterize the fundamental string as $t = \tau$, $x = \sigma$, and $r = r(\sigma)$, and
we define
\be
f^2(r(\sigma)) = g_{tt}(r(\sigma))\, g_{xx}(r(\sigma)) ~, \qquad
g^2(r(\sigma)) = g_{tt}(r(\sigma))\, g_{rr}(r(\sigma)) ~,
\ee
the Nambu--Goto action reads
\ba
S_{NG} & = & \int d\tau\, d\sigma\; \sqrt{\det (-g)} \cr
& = & T \int d\sigma\; \sqrt{f^2(r(\sigma)) + g^2(r(\sigma))\,
(r'(\sigma))^2} ~,
\label{wilson4}
\ea
where $T$ is the time interval. This action implies an equation of motion
that dictates the embedding $r(\sigma)$ of the fundamental string to be
\be
\frac{dr}{d\sigma} = \pm \frac{f(r)}{g(r)}\; \frac{f^2(r) - f^2(0)}{f(0)} ~.
\label{wilson5}
\ee
The distance $L$ separating the static pair of quark anti-quark is
\be
L = \int d\sigma = 2 \int_0^{r_f} \frac{g(r)}{f(r)}\; \frac{f(0)}{f^2(r)
- f^2(0)}\; dr ~,
\label{wilson6}
\ee
where $r_f$ is the maximal radial distance from the fundamental string to
the D3--branes (see Fig.\ref{fig:wilson}).
\FIGURE[ph]{\centerline{\epsfig{file=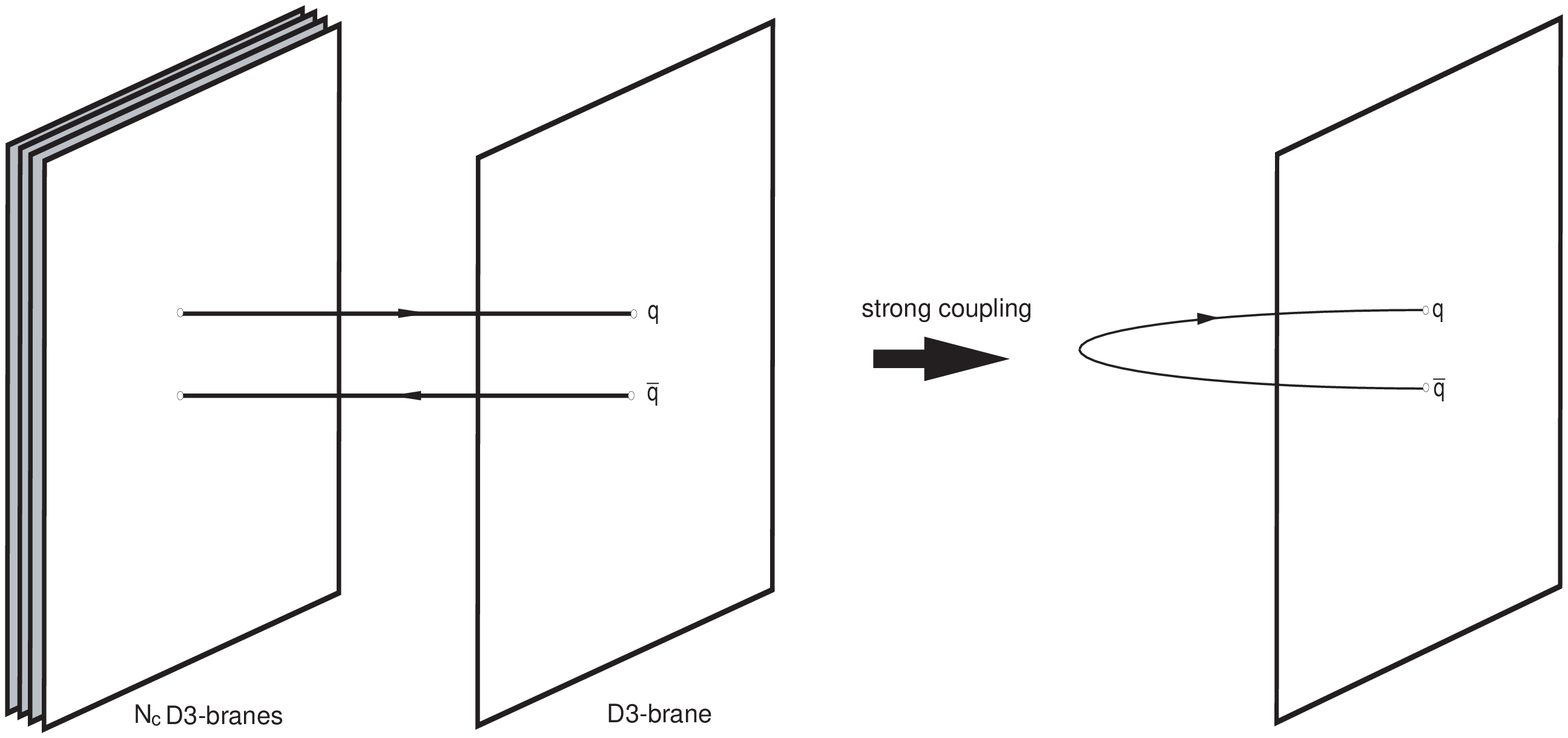,width=14cm}}
\caption{Wilson loop and confinement. We introduce the quark pair in terms
of the W bosons associated to the symmetry breaking $U(N_c+1) \to U(N_c)
\times U(1)$, which we perform by separating a D3--brane. At strong coupling,
as discussed earlier, the $N_c$ D3--branes are replaced by the background.}
\label{fig:wilson}}
The quark anti-quark potential
is read off from (\ref{wilson2}) and (\ref{wilson3}), after a suitable
renormalization \cite{cobi} --whose details we skip-- of the otherwise
divergent Nambu--Goto action. It is given by
\be
E = f(0) L + 2 \int_0^{r_f}  \frac{g(r)}{f(r)}\; \bigg( \sqrt{f^2(r) -
f^2(0)} - f(r) \bigg)\; dr ~,
\label{wilson7}
\ee
In this way, we can determine whether a supersymmetric gauge theory is
confining or not from its gravity dual. It is possible to extract from the
above formulas a sufficient condition for confinement \cite{cobi}: If
$f(r)$ is analytic, $f'(r)$ is positive and $g(r)$ is smooth for $r \in
(0,\infty)$, and if the integral of $g(r)/f^2(r)$ does not diverge, then
there is confinement if either $f(0) \neq 0$ or $g(\tilde r) \to \infty$
for some $r = \tilde r$, with $f(\tilde r) \neq 0$. It is not hard to check
that the first condition is accomplished in the MN solution. Notice that
confinement reduces to the behaviour of the gravity dual at small radius
which, from the point of view of the gauge theory, is the IR. The tension
of the {\it QCD string} in the MN set up is finite (actually, proportional
to $f(0)$): the theory {\it confines}.

\subsection{Instantons}

As discussed earlier, instantons in $\CN = 1$ SYM theory are
responsible for breaking $U(1)_R$ to $\IZ_{2N_c}$. The action of an
instanton, from the field theory point of view, is given by \be
S_{\rm inst} = \frac{8\pi^2}{\gym^2} + i \tym ~. \ee In the large
$N_c$ limit, this breaking pattern is not apparent in the isometries
of the MN background. Instantons can be identified with Euclidean
D1--branes wrapping an $S^2$ which is orthogonal to the gauge theory
directions \cite{mn2}. The precise supersymmetric 2-cycle is defined
by the equations
\be
\t = \tilde\t ~, \quad\quad\quad \tilde\varphi = 2\pi - \varphi ~,
\quad\quad\quad \psi = (2n+1) \pi ~.
\ee
In the UV we have to use the S-dual background, that is, the D1--brane should
be replaced by a fundamental string (the instantons in the gauge theory
being related to worldsheet instantons). As the MN configuration asymptotes
to the so-called Abelian MN solution in the UV, we should use the latter to
compute instanton effects. The upshot being that the Nambu--Goto action (plus
the Wess--Zumino contribution),
\be
S_{\rm NG} = \frac{1}{2\pi\alpha^\prime\gs} \int_{\cal W} d^2\xi
~e^{-\phi} ~\sqrt{e^{2h} + \frac{1}{4} (a-1)^2} - i \int_{\cal W}
~B_2 ~,
\ee
must be identified with the action of an instanton, $S_{\rm NG} = S_{\rm
inst}$. When computing explicitly, we get
\ba
\frac{8\pi^2}{\gym^2} & = & \frac{1}{2\pi\alpha^\prime\gs} \int_{\cal W}
d^2\xi ~e^{-\phi} ~\sqrt{e^{2h} + \frac{1}{4} (a-1)^2}\,, \\
&& \cr \tym & = & 4 N_c \psi_0 ~. \ea
The latter expression tells us that only $\IZ_{2N_c}$ rotations of the phase
$\psi \to \psi + \frac{2\pi k}{N_c}$ leave the path integral invariant.

\subsection{Finite Temperature Effects}

We would like to shortly mention an interesting application of the
gauge/gravity duality to confining theories at finite temperature. There is
a conjectured deconfining phase transition in these theories, at a given
temperature $T_c$, where a new state of matter appears; namely, the {\it
quark--gluon plasma}. In recent years, the traditional view on quark--gluon plasma properties has changed. Instead of
being thought of as a weakly interacting gas of quasiparticles, it is now
viewed as a strongly coupled nearly perfect fluid \cite{RHIC}. As such, it
is better described in terms of hydrodynamics. It is then a vital issue
to characterize its transport properties such as the shear and bulk
viscosities, $\eta$ and $\zeta$, the thermal and electric conductivity
$\kappa_T$ and $\sigma$, the charge diffusion constant $D_Q$, the speed of
sound $v_s$ and the entropy density $s$. Transport coefficients are hard to compute from first principles, even in perturbation theory, and are not
amenable to lattice gauge theory techniques either (see, however,
\cite{SakNa}). The uses of the gauge/gravity correspondence in this
respect are of sound relevance.

The shear viscosity represents the transferred momentum between different
layers of a fluid. It is computed in the high energy regime of field theories
as a correlation function of the stress-energy tensor by the so-called Kubo formula
\be
\eta = \lim_{w \to 0} ~\frac{1}{2w} \int dt d^3\vec x ~e^{iwt}
~<[T_{xy}(\vec x,t), T_{xy}(\vec 0,0)]> = - \lim_{w \to 0}
~\frac{{\rm Im} ~G_{xy,xy}^{\rm R}(w,\vec 0)}{w} ~,
\label{Kubo}
\ee
where $G_{xy,xy}^{\rm R}(w,\vec 0)$ is the retarded Green function. This has been studied in detail when $T >> T_c$ in perturbative field theory. {\it How can this be computed on the gravity side?}

The gravity duals of finite temperature gauge theories involve black holes
in asymptotically AdS backgrounds \cite{wittent}. Hawking's temperature and Bekenstein's entropy of the gravitational background are identified with
those of the gauge theory in thermal equilibrium. Given that transport properties correspond to near-equilibrium processes, and taking into
account that $g_{\mu\nu}$ couples to the energy--momentum tensor and we
wish to compute correlators of the kind given in (\ref{Kubo}), they can be studied by considering (metric) perturbations of the appropriate thermal
supergravity background \cite{pss}. We will not explain this technically involved approach here. Instead, we would like to comment on a misterious feature that, even having being developed some twenty years ago, makes more
sense under the {\it a posteriori} light of AdS/CFT. This goes under the
name of {\it the membrane paradigm} \cite{HaRu}.

The black hole membrane paradigm can be simply recasted as stating that the
event horizon can be thought of as a high temperature classically radiating surface (see \cite{ThPrMc} for details). Being out of equilibrium, it
presents all kinds of characteristic features of a dissipative system.
Consider, for example, the shear viscosity. Let us start from a generic
black hole background of the form
\be
ds^2 = - g_{tt}(r) dt^2 + g_{xx}(r) d\vec x^2 + g_{rr}(r) dr^2 +
Z(r) \tilde g_{ij}(y) dy^i dy^j ~,
\label{mempdg}
\ee
where $r_0$ stands for the position of the black hole horizon. Without
entering into details that are well out of the scope of these lectures, we
would like to comment on a striking result. It turns out that the shear viscosity per unit of specific entropy of this configuration that follows
from the membrane paradigm, fully coincides with the expression obtained by
computing the lowest quasinormal frequency in this background \cite{kss},
\be
\frac{\eta}{s} = T \left.\sqrt{\frac{\det g}{g_{tt} g_{rr}}}\;
\right|_{r=r_0} ~\int_{r_0}^\infty dr ~\frac{g_{tt} g_{rr}}{g_{xx}
\sqrt{\det g}} ~.
\label{ratioes}
\ee
Something similar happens with other transport coefficients. Furthermore (and remarkably enough), the above formula gives a {\it universal} value for backgrounds corresponding to duals of a huge class of supersymmetric gauge theories of the black brane form (\ref{mempdg}) \cite{BuLiu},
\be
\frac{\eta}{s} = \frac{1}{4\pi} ~,
\label{univratio}
\ee
in the regime in which it is well described by its gravity dual, {\it i.e.},
at strong 't Hooft coupling. \footnote{It is timely to add that it was
found very recently that the same universal ratio is attained in gauge
theories with a chemical potential \cite{Mas,SoSt,Saremi,MaNaOk} whose
gravity dual background --an $R$-charged black hole-- cannot be written as
in (\ref{mempdg}). This may suggest that there is a generalization of the membrane paradigm that encompasses these systems.} This universality is
related to that of the low energy absortion cross section for gravitons
\cite{DaGiMa,Emparan,KoSoSt}. Furthermore, and most important to us, this ubiquitous value has been explicitely derived by direct computation of
Kubo's formula within the AdS/CFT framework for a wide class of relevant
supergravity backgrounds \cite{Buchel}. This is quite a small value. For example, the same quantity for water under normal pressure and temperature
is about 400 times higher. In good accord with this prediction, measurements
at the Relativistic Heavy Ion Collider (RHIC) also suggest a small value for this ratio \cite{RHIC}.

Let us quickly check how this result emerges within the membrane paradigm
in the case of a finite temperature version of the Maldacena--N\'u\~nez background. It is given by the following 10d metric in the Einstein frame
\cite{Buch},
\be
ds^2 = C^2(r) \bigg[ - \Delta_1^2(r)\, dt^2 + d\vec x_2^2 + \alpha^\prime
\gs N_c\, \bigg( \frac{dr^2}{r^2~\Delta_2(r)^2} + e^{2 \hat h(r)}\,
d\Omega^2_2 + \frac{1}{4}\, d\tilde\Omega^2_2 + \frac{1}{4}\, e_\psi^2
\bigg) \bigg] ~,
\label{thermn}
\ee
where $e_\psi = d\psi + \cos\t\, d\phi + \cos\tilde\t\, d\tilde\phi$, and
the set of functions $C(r)$, $\Delta_1(r)$, $\Delta_2(r)$ and $\hat h(r)$,
which are defined in \cite{Buch}, satisfy the identity
\be
\Delta_1'(r) \Delta_2(r) = \frac{\mu}{4 r C(r)^8 e^{2 \hat h(r)}} ~,
\label{Deltas}
\ee
$\mu$ being the near extremal parameter. This solution has a regular horizon $r_0$. One can compute the temperature of the black hole to be
\be
2\pi T = \left. \frac{\mu}{\sqrt{\alpha^\prime\gs N}} ~\frac{1}{C^2(r)
\hat h(r)} \right|_{r=r_0} ~.
\ee
Inserting this expression in (\ref{ratioes}), and using (\ref{Deltas}), it
is not hard to verify that the universal value $1/4\pi$ is attained
\cite{BuLiu}.

Plugging in units into (\ref{univratio}), it is amusing to check that neither the Newton constant $G$ nor the speed of light $c$ appear in this expression. This seems to imply that it is not of relativistic or gravitational origin.
The constants entering (\ref{univratio}) are Planck's $\hbar$ and Boltzmann's $k_B$. It actually looks more quantum mechanical. Like the saturation of the uncertainty principle. Indeed, it has been conjectured that this value is a lower bound for any system in Nature \cite{kss,KoSoSt}.

\subsection{Adding Flavors}

In order to introduce matter in the fundamental representation of the gauge group, it is necessary to incorporate an open string sector in the dual supergravity background. To do so, it seems natural to consider the addition
of D--brane probes at a characteristic distance. Let us call them {\it flavor branes}. This distance will give a lower bound for the energy of a string stretched between both branes which is to be related with the mass of the quarks. It is interesting to point out that this is also
suggested by 't Hooft's large $N_c$ arguments, since flavor corresponds to
the presence of boundaries in the worldsheet and these amount to D--branes.
It was originally proposed in the $AdS_5 \times S^5$ case that addition of flavor corresponds to spacetime filling D7--branes \cite{KaKa}. Open strings coming into the stack of D3--branes where the gauge theory is defined,
represent quarks and the fluctuations of these D7--branes provide the meson spectrum \cite{KMMW}.

The addition of matter multiplets in supersymmetric gauge theories does not
reduce the amount of supersymmetry. Therefore, the flavor branes have to be
introduced into the configuration in a symmetry preserving way. At the probe
approximation, the supersymmetric embedding of a D--brane probe is obtained
by imposing the kappa-symmetry condition:
\beq
\Gamma_{\kappa}\,\epsilon\,=\,\epsilon\,\,,
\label{kappacondition}
\eeq
where $\epsilon$ is a Killing spinor of the background and $\Gamma_{\kappa}$
is a matrix that depends on the embedding. In the absence of worldvolume gauge
fields living on the probe D--brane,
\beq
\Gamma_{\kappa}\,=\,\frac{1}{(p+1)!\sqrt{-g}}\,\epsilon^{\mu_1\cdots\mu_{p+1}}\,
(\tau_3)^{\frac{p-3}{2}}\,i\tau_2\,\otimes\,
\gamma_{\mu_1\cdots\mu_{p+1}}\,\,,
\label{gammakappa}
\eeq
where $g$ is the determinant of the induced metric $g_{\mu\nu}$ and
$\gamma_{\mu_1\cdots\mu_{p+1}}$ denotes the antisymmetrized product of
the induced Dirac matrices. In order that the probe approximation remains
valid and backreaction of the flavor branes do not need to be considered, we
shall assume that we are in the so-called quenched approximation, $N_f << N_c$.
\footnote{By the time of submitting these notes, it came to our notice that
there was an upcoming paper dealing with the supergravity dual of flavored
supersymmetric gauge theories for any $N_f \leq 2 N_c$ \cite{CaNuPa}.}
In the case of MN, flavor is introduced through D5--branes wrapping a
two-cycle different from the one wrapped by the bunch of $N_c$ D5--branes
defining the gauge theory \cite{npr}. There are no solutions at a fixed distance. This is related to the lack of moduli space in $\CN = 1$ theory.
Other aspects that can be studied within this framework include $U(1)_R$ symmetry breaking by the formation of a squark condensate, non-smoothness
of the massless quark limit, and $U(1)_B$ baryonic symmetry preservation.
We have unfortunately no time to discuss these in the present lectures.
Further work on the gauge/gravity duality in presence of flavors include
\cite{AEEGG}. Aspects of the meson spectrum were considered in \cite{ErKi},
while meson decay at strong coupling has been recently discussed in
\cite{PeSoZa,CoMaTr}.

Many additional aspects of the gauge/gravity duality in confining theories
would have deserved to be discussed in these lectures. For example, the mass
spectrum of glueballs. These are composite objects whose constituents are
nothing but gluons. The mass spectrum of these states is quite amenable to
lattice studies \cite{tep}. By these means, it has been shown that they have
a bound state spectrum, the lightest glueball being a scalar with a mass of
1630 MeV. The examination of this problem in the dual gravity background
demands the study of fluctuations of supergravity fields. The linearized
system is reduced to a Schr\"odinger problem whose eigenvalues are nothing
but the desired spectrum. Remind that we have seen in the second lecture that
fluctuations of the fields in the gravity side correspond to operators in
the dual gauge theory. Thus, what this approach computes is a correlation
function from which we can actually read the masses of the glueballs. In
the MN setup, this issue has been addressed quite recently in
\cite{CaNun,BeHaMu}.

\section*{Acknowledgments}

\noindent
JDE wishes to thank the organizers of the school, Leandro de
Paula, Victor Rivelles, and Rogerio Rosenfeld, for their invitation to deliver
this series of lectures in a great event and express his gratefulness to the
participants of the School who contributed with their enthusiasm and the
follow-up discussions after the lectures, at the piano bar and even when
playing soccer each afternoon. He is also glad to express his gratitude to
Alfonso Ramallo for comments and for lending him one of his mythical
notebooks that was really helpful to prepare the lectures. Both authors
wish to thank Carlos N\'u\~nez for discussions and a critical reading of
the manuscript. We would also like to thank Juan Maldacena for pointing out
a goof in the first version and Javier Mas for valuable comments. JDE has
been supported in part by ANPCyT (Argentina) under grant PICT 2002 03-11624, MCyT, FEDER and Xunta de Galicia under grant FPA2005-00188, by the FCT (Portugal) grant POCTI/FNU/38004/2001, and by the EC Commission under the
grants HPRN-CT-2002-00325 and MRTN-CT-2004-005104. RP thankfully acknowledges support from FONDECYT grant 3050086.
We further acknowledge support from the binational SECyT (Argentina)--CONICYT (Chile) grant CH/PA03-EIII/014.
Institutional support to the Centro de Estudios Cient\'\i ficos (CECS) from Empresas CMPC is gratefully acknowledged. CECS is a Millennium Science Institute and is funded in part by grants from Fundaci\'on Andes and the Tinker Foundation.

\appendix
\section{Useful formulas}

\subsection{Conifold}

The metric of the conifold and its resolved and deformed cousins are better
described in terms of the following vielbein:
\ba
g^1 & = & \rhalf \left[ - \sin\t_1\, d\p_1 - \cos\psi\, \sin\t_2\, d\p_2
+ \sin\psi\, d\t_2 \right] \cr
g^2 & = & \rhalf \left[ d\t_1 - \sin\psi\, \sin\t_2\, d\p_2
- \cos\psi\, d\t_2 \right] \cr
g^3 & = & \rhalf \left[ - \sin\t_1\, d\p_1 + \cos\psi\, \sin\t_2\, d\p_2
- \sin\psi\, d\t_2 \right] \cr
g^4 & = & \rhalf \left[ d\t_1 + \sin\psi\, \sin\t_2\, d\p_2
+ \cos\psi\, d\t_2 \right] \cr
g^5 & = & d\psi + \cos\t_1\, d\p_1 + \cos\t_2\, d\p_2
\label{gframe} \ea
The metric of $T^{1,1}$, for example, is then written as
\be
ds^2(T^{1,1}) = \frac{1}{6} \sum_{i=1}^4 (g^i)^2 + \frac{1}{9} (g^5)^2
\ee
Check, as a useful exercise, the following identities:
\be
d(g^1 \wedge g^2 + g^3 \wedge g^4) = g^5 \wedge (g^1 \wedge g^2 -
g^3 \wedge g^4)
\ee
\be
d(g^1 \wedge g^2 - g^3 \wedge g^4) = - g^5 \wedge (g^1 \wedge g^3 +
g^2 \wedge g^4)
\ee
Remind that $T^{1,1}$ is a $U(1)$ fibration over a K\"ahler--Einstein
manifold, $S^2 \times S^2$. The symplectic form of the 4d base manifold
reads
\be
J_4 = \frac{1}{6}\, \bigg( \sin\t_1\, d\t_1 \wedge d\p_1 + \sin\t_2\,
d\t_2 \wedge d\p_2 \bigg) ~.
\ee
Given the conical structure of the variety, we can write the symplectic
form of the conifold $J$ by means of $J_4$ and $g^5$
\be
J = r^2\, J_4 + \frac{r}{3}\, dr \we g^5 ~.
\ee
The reader should check that $dJ = 0$.
We can also define the $2$-form
\be
w_2 = \half \bigg( g^1 \we g^2 + g^3 \we g^4 \bigg)  = \half \bigg(
\sin\t_1 d\t_1 \we d\p_1 - \sin\t_2 d\t_2 \we d\p_2 \bigg) ~,
\label{w2} \ee
and the $3$-form $w_3$
\be
w_3 = g^5 \we w_2 ~.
\label{w3} \ee
Both are closed
\be
d w_2 = d w_3 = 0 ~.
\label{dw3} \ee
Furthermore, notice that
\ba
w_2 \we w_3 & = & \half \bigg( g^1 \we g^2 + g^3 \we g^4 \bigg) \we g^5
\we \half \bigg( g^1 \we g^2 + g^3 \we g^4 \bigg) \cr
& = & \half g^1 \we g^2 \we g^3 \we g^4 \we g^5 = 54\, d\Vol(T^{1,1}) ~.
\label{w2w3id} \ea
The following results are also useful throughout the text (in section 4):
\ba
w_2^2 & \equiv & w_{2,ab} w_2^{ab} = g^{ac} g^{bd} w_{2,ab} w_{2,cd}
= \frac{36}{r^4 h} ~, \cr
w_3^2 & \equiv & w_{3,abc} w_3^{abc} = \frac{3^5 4}{r^6 h^{3/2}} ~,
\label{w32} \ea
$h = h(r)$ being the function appearing as a warp factor in the metric
(\ref{kt1}). Using the metric (\ref{dsToo}), it is straightforward to
compute the volume of the base
\be
\Vol(T^{1,1}) = \frac{16}{27} \pi^3 ~.
\label{volToo}
\ee
Thus,
\be
L^4 = \frac{27\pi}{4} \gs N (\alpha^\prime)^2 ~.
\label{l4}
\ee

\subsection{Left invariant one forms}

Let $L$ be an element of $SU(2)$. The left invariant one forms
$w^k$, $k=1,2,3$, are defined by the expression
\be
L^{\dagger} dL = \ihalf w^k \s_k ~,
\label{elemsu2} \ee
where $\s_k$ are the Pauli matrices. The canonical basis is
\ba
w^1 & = & \cos\phi d\theta + \sin\theta \sin\phi d\psi ~, \cr
w^2 & = & \sin\phi d\theta - \sin\theta \cos\phi d\psi ~, \cr
w^3 & = & d\phi + \cos\theta d\psi ~. 
\label{canbasis}
\ea

\noindent {\sc Exercise}: Check that parametrization of $L$ in terms
of Euler angles,
$$ L = e^{\ihalf\phi\s_3} e^{\ihalf\t\s_1} e^{\ihalf\psi\s_3} ~,
~~~~~~~ 0 \leq \phi, \psi \leq 2\pi ~, ~~~ 0 \leq \t \leq \pi ~, $$
leads to the canonical basis. It is useful to notice that
$w^k = -i \Tr (\s_k L^{\dagger} dL)$.

~

\subsection{Useful type IIB formulas}

Let us recall the Lagrangian and equations of motion of type IIB supergravity.
In the string frame, the action reads \cite{schw}:
\ba
S_{IIB} & = & \frac{1}{2\k^2} \int e^{-2\Phi} \left( d^{10}x \sqrt{-g}\, R
+ 4 ~d \Phi \wedge \star d \Phi - \frac{1}{12} H_{[3]} \wedge \star H_{[3]} \right) \cr
& - & \frac{1}{4\k^2} \int \bigg( F_{[1]} \wedge \star F_{[1]}  +
\frac{1}{6} F_{[3]} \wedge \star F_{[3]} + \frac{1}{240} F_{[5]} \wedge
\star F_{[5]} - C_{[4]} \wedge F_{[3]} \wedge H_{[3]}  \bigg) ~,
\label{iibst}
\ea
where
\ba
F_{[1]} & = & d C_{[0]} ~, \nn \\
F_{[3]} & = & d C_{[2]} - C_{[0]} \, H_{[3]} ~, \nn \\
H_{[3]} & = & d B_{[2]} ~, \nn \\
F_{[5]} & = & d C_{[4]} - \half\, C_{[2]} \wedge H_{[3]} + \half\, B_{[2]} \wedge F_{[3]} ~,
\label{efes}
\ea
supplemented by the additional on-shell constraint $F_{[5]} = \star F_{[5]}$. Let us recall the change to the Einstein frame by keeping careful track of
the $\gs$ factors (following, for example, Ref.\cite{hk}). We start from
\be
(g_{\m\n})_{\rm string} = \gs^{-1/2}\, e^{\P/2}\, (g_{\m\n})_{\rm Einstein} ~, \label{chframes}
\ee
where $\gs = e^{\Phi(r\to\infty)}$ is the string coupling constant. Then,
\be
\sqrt{|g|}_{\rm string} = (\gs^{-1/2}\, e^{\P/2})^5\, \sqrt{|g|}_{\rm
Einstein} = \gs^{-5/2}\, e^{5\P/2}\, \sqrt{|g|}_{\rm Einstein} ~.
\ee
In general, for any $p$--form
\be
(F_{[p]} \we \star F_{[p]})_{\rm string} = \gs^{p/2}\, e^{-p\P/2}\,
(F_{[p]} \we \star F_{[p]})_{\rm Einstein} ~.
\ee
Reabsorbing the factor $g_s^2$ into the constant $\k^2$ by defining
$\bar\k^2 = g_s^2 \k^2$, we can write the action in the Einstein frame as
\ba S_{IIB} & = & \frac{1}{2\bar\k^2} \int \, \left( d^{10}x \sqrt{-g} R -
\frac{1}{2} ~d \Phi \wedge \star d \Phi - \frac{g_s}{12}\, e^{-\Phi}
H_{[3]} \wedge \star H_{[3]} \right. \\
& - & \frac{1}{2}\, \bigg[ e^{2\Phi} F_{[1]} \wedge \star F_{[1]} +
\frac{g_s}{6}\, e^{\Phi} F_{[3]} \wedge \star F_{[3]} + \frac{g_s^2}{240}\,
F_{[5]} \wedge \star F_{[5]} - \, g_s^2 \, C_{[4]} \wedge
F_{[3]} \wedge H_{[3]} \bigg] \bigg) ~, \nn
\label{iibei}
\ea
where all quantities are given in the Einstein frame. Beside the Einstein equation,
\ba
R_{MN} & = & \frac{1}{2} \partial_M \phi \partial_N \phi + \frac{1}{2}
e^{2\Phi} \partial_M C_{[0]} \partial_N C_{[0]} + \frac{g_s}{4}\,
e^{-\Phi} \bigg( {H_{[3]}}_{MPQ} {H_{[3]}}_N^{~PQ} - \frac{1}{12} g_{MN}\,
{H_{[3]}}^2 \bigg) \cr
& & + \frac{g_s^2}{96} {F_{[5]}}_{MPQRS} {F_{[5]}}_N^{~PQRS} +
\frac{g_s}{4}\, e^{\Phi} \bigg( {F_{[3]}}_{MPQ} {F_{[3]}}_N^{~PQ} -
\frac{1}{12} g_{MN} {F_{[3]}}^2 \bigg) ~,
\label{einiib}
\ea
where we used $F_{[5]}^2 = 0$, the equations of motion include
\be
\nabla^2 \phi = F_{[1]}^2 - \frac{g_s}{12}\, e^{-\Phi}\, H_{[3]}^2 +
\frac{g_s}{12}\, e^{\Phi}\, F_{[3]}^2 ~,
\ee
\be
d (e^{2\Phi} \star F_{[1]}) = - g_s\, e^{\Phi}\, H_{[3]} \wedge \star
F_{[3]} ~,
\ee
\be
d (e^{\Phi} \star F_{[3]}) = g_s\, F_{[5]} \wedge H_{[3]} ~,
\ee
\be
d (e^{-\Phi} \star H_{[3]} - C_{[0]}\, e^{\Phi} \star F_{[3]}) =
- g_s\, F_{[5]} \wedge \left( F_{[3]} + C_{[0]}\, H_{[3]} \right) ~,
\ee
\be
d (\star F_{[5]}) = - F_{[3]} \wedge H_{[3]} ~,
\ee
and the Bianchi identities
\be
d F_{[1]} = d H_{[3]} = 0 ~, \qquad d F_{[3]} = - F_{[1]} \wedge H_{[3]} ~,
\qquad d F_{[5]} = - F_{[3]} \wedge H_{[3]} ~.
\ee

\end{document}